\documentstyle[prl,aps,epsf]{revtex}

 \newcommand{\insertplot}[5]{\begin{figure}
 \hfill\hbox to 0.05in{\vbox to #5in{\vfill
 \inputplot{#1}{#4}{#5}}\hfill}
 \hfill\vspace{-.1in}
 \caption{#2}\label{#3}
 \end{figure}}
 \newcommand{\inputplot}[3]{
 \special{ps: plotfile #1}
}

\begin{document}
\draft
\title{STATIC AXIALLY SYMMETRIC EINSTEIN-YANG-MILLS-DILATON SOLUTIONS:
II.BLACK HOLE SOLUTIONS}
\author{
{\bf Burkhard Kleihaus, and Jutta Kunz}}
\address{Fachbereich Physik, Universit\"at Oldenburg, Postfach 2503\\
D-26111 Oldenburg, Germany}
\date{\today}
\maketitle
%
%
%
%
\begin{abstract}
We discuss the new class of static axially symmetric black hole solutions
obtained recently in Einstein-Yang-Mills and 
Einstein-Yang-Mills-dilaton theory.
These black hole solutions are asymptotically flat 
and they possess a regular event horizon.
The event horizon is almost spherically symmetric 
with a slight elongation along the symmetry axis.
The energy density of the matter fields is angle-dependent
at the horizon.
The static axially symmetric black hole solutions
satisfy a simple relation between mass, dilaton charge, entropy and
temperature.
The black hole solutions are characterized by two integers,
the winding number $n$ 
and the node number $k$ of the purely magnetic gauge field.
With increasing node number the magnetically neutral black hole solutions 
form sequences tending to limiting solutions with magnetic charge $n$,
corresponding to Einstein-Maxwell-dilaton black hole solutions
for finite dilaton coupling constant and
to Reissner-Nordstr\o m black hole solutions
for vanishing dilaton coupling constant.
\end{abstract}
 \vfill
 \noindent {Preprint gr-qc/9712086} \hfill\break
 \vfill\eject

\section{Introduction}

The ``no hair'' conjecture for black holes states,
that black holes are completely characterized 
by their mass $M$, their charge $Q$ and their angular momentum $J$.
This conjecture presents a generalization of
rigorous results obtained for scalar fields coupled to gravity
\cite{bek} as well as 
for Einstein-Maxwell (EM) theory \cite{hair-em}.
In EM theory,
the unique family of stationary Kerr-Newman black holes
with nontrivial values of $M$, $Q$, and $J$
contains the stationary Kerr black holes for $Q=0$,
the static Reissner-Nordstr\o m black holes for $J=0$
and the static Schwarzschild black holes for $J=Q=0$.
Notably, the static black hole solutions in EM theory
are spherically symmetric, and the stationary black hole solutions
are axially symmetric.

In recent years counterexamples to the ``no hair'' conjecture
were established in various theories with non-abelian fields,
including Einstein-Yang-Mills (EYM) theory,
Einstein-Yang-Mills-dilaton (EYMD) theory,
Einstein-Yang-Mills-Higgs (EYMH) theory,
and Einstein-Skyrme (ES) theory \cite{review}.
Possessing non-trivial matter fields outside
their regular event horizon,
these non-abelian black hole solutions
are no more completely determined by their global charges.
And they hold more surprises. 
There are static black hole solutions
with only axial symmetry \cite{kk3},
static black hole solutions
with only discrete symmetries \cite{ewein},
and there are non-static non-rotational black hole solutions
\cite{brod}.
Unlike the static spherically symmetric \cite{review}
and axially symmetric static black hole solutions \cite{kk3},
many of the new types of non-abelian
black hole solutions are only perturbative solutions
\cite{ewein,brod,slomo}.

The static axially symmetric black hole solutions 
in EYM and EYMD theory \cite{kk3} have many properties in common
with the globally regular static axially symmetric solutions
constructed previously \cite{kk2,kk4}.
Representing generalizations of
the static spherically symmetric globally regular and black hole solutions
\cite{bm,su2,eymd},
these static axially symmetric solutions
are characterized by two integers.
These are the node number $k$ of the gauge field functions
and the winding number $n$ 
with respect to the azimuthal angle $\phi$.
While $\phi$ covers the full trigonometric circle once,
the fields wind $n$ times around.
The static spherically symmetric solutions have winding number $n=1$.
Winding number $n>1$ leads to axially symmetric solutions.

The static axially symmetric EYM and EYMD black hole solutions
\cite{kk3} are asymptotically flat and possess a regular event horizon.
The event horizon of these solutions resides at a surface
of constant isotropic radial coordinate.
The energy density of the matter fields
is not constant at the horizon but angle-dependent.
Outside their regular event horizon, 
the static axially symmetric black hole solutions
possess non-trivial magnetic gauge field configurations,
but they carry no global magnetic charge.
For fixed winding number $n$ and increasing node number $k$ 
the solutions form sequences, tending to limiting solutions.
These limiting solutions are spherically symmetric
and abelian, representing Einstein-Maxwell-dilaton (EMD) \cite{emd}
and Reissner-Nordstr\o m (RN) black hole solutions for EYMD and EYM theory,
respectively \cite{kk2,kk3,kk4}, which carry magnetic charge $n$.

Having given a brief account of the properties of these
static axially symmetric black hole solutions in \cite{kk3}, 
we here present these solutions in detail.
The paper thus represents the second paper of our sequel
on static axially symmetric EYMD solutions, following \cite{kk4},
where a detailed presentation of the globally regular solutions
was given.
In section II of this paper we recall the action, we present the 
static axially symmetric ansatz in isotropic coordinates
and we discuss the boundary conditions.
Introducing temperature and entropy of the black hole solutions,
we derive a relation between these thermodynamic quantities
and the mass and the dilaton charge.
In section III we recall the static spherically symmetric black hole 
solutions and present them in isotropic coordinates.
In section IV we present the static axially symmetric black hole solutions,
discuss their properties, 
and analyse the properties of their event horizon.
We show that the sequences of neutral non-abelian solutions tend 
to limiting charged abelian solutions.
We present our conclusions in section V.
In Appendix A we present the expansion of the functions
at the horizon.
We show that the surface gravity is 
constant at the horizon and that the Kretschmann scalar
is finite there.
In Appendix B we discuss 
the final choice of functions for the numerical integration.

\section{\bf Static axially symmetric ansatz}

We consider the SU(2) Einstein-Yang-Mills-dilaton action
\begin{equation}
S=\int \left ( \frac{R}{16\pi G} + L_M \right ) \sqrt{-g} d^4x
\ \label{action} \end{equation}
with the matter Lagrangian
\begin{equation}
L_M=-\frac{1}{2}\partial_\mu \Phi \partial^\mu \Phi
 -e^{2 \kappa \Phi }\frac{1}{2} {\rm Tr} (F_{\mu\nu} F^{\mu\nu})
\ , \label{lagm} \end{equation}
the field strength tensor
\begin{equation}
F_{\mu \nu} = 
\partial_\mu A_\nu -\partial_\nu A_\mu + i e \left[A_\mu , A_\nu \right] 
\ , \label{fmn} \end{equation}
the gauge field
\begin{equation}
A_{\mu} = \frac{1}{2} \tau^a A_\mu^a
\ , \label{amu} \end{equation}
the dilaton field $\Phi$,
and the Yang-Mills and dilaton coupling constants
$e$ and $\kappa$, respectively.

Variation of the action (\ref{action}) with respect to the metric
$g^{\mu\nu}$ leads to the Einstein equations,
variation with respect to the gauge field $A_\mu$ 
and the dilaton field $\Phi$ 
leads to the matter field equations \cite{kk4}.

\subsection{\bf Static axially symmetric ansatz}

As for the globally regular static 
axially symmetric solutions \cite{kk2,kk4},
we adopt isotropic coordinates
to construct the static axially symmetric black hole solutions.
In terms of the coordinates $r$, $\theta$ and $\phi$
the isotropic metric reads
\begin{equation}
ds^2=
  - f dt^2 +  \frac{m}{f} d r^2 + \frac{m r^2}{f} d \theta^2 
           +  \frac{l r^2 \sin^2 \theta}{f} d\phi^2
\ , \label{metric2} \end{equation}
where the metric functions
$f$, $m$ and $l$ are only functions of 
the coordinates $r$ and $\theta$.
Regularity on the $z$-axis ($z=r \cos \theta$) requires \cite{book}
\begin{equation}
m|_{\theta=0}=l|_{\theta=0}
\ . \label{lm} \end{equation}

Again, we consider a purely magnetic gauge field, $A_0=0$,
and choose for the gauge field the ansatz \cite{rr,kk,kk1,kk2,kk3,kk4}
\begin{equation}
A_\mu dx^\mu =
\frac{1}{2er} \left[ \tau^n_\phi 
 \left( H_1 dr + \left(1-H_2\right) r d\theta \right)
 -n \left( \tau^n_r H_3 + \tau^n_\theta \left(1-H_4\right) \right)
  r \sin \theta d\phi \right]
\ , \label{gf1} \end{equation}
where the symbols $\tau^n_r$, $\tau^n_\theta$ and $\tau^n_\phi$
denote the dot products of the cartesian vector
of Pauli matrices, $\vec \tau = ( \tau_x, \tau_y, \tau_z) $,
with the spatial unit vectors
\begin{eqnarray}
\vec e_r^{\, n}      &=& 
(\sin \theta \cos n \phi, \sin \theta \sin n \phi, \cos \theta)
\ , \nonumber \\
\vec e_\theta^{\, n} &=& 
(\cos \theta \cos n \phi, \cos \theta \sin n \phi,-\sin \theta)
\ , \nonumber \\
\vec e_\phi^{\, n}   &=& (-\sin n \phi, \cos n \phi,0) 
\ , \label{rtp} \end{eqnarray}
respectively.
Since the fields wind $n$ times around, while the
azimuthal angle $\phi$ covers the full trigonometric circle once,
we refer to the integer $n$ as the winding number of the solutions.
The four gauge field functions $H_i$ 
and the dilaton function $\Phi$ depend only on 
the coordinates $r$ and $\theta$.
The spherically symmetric ansatz \cite{eymd} is recovered
for $n=1$ and $H_1=H_3=0$, $H_2=H_4=w(r)$ and $\Phi=\Phi(r)$.

The ansatz (\ref{gf1}) is axially symmetric in the sense,
that a rotation around the $z$-axis can be compensated
by a gauge rotation.
Besides being axially symmetric the ansatz
respects the discrete mirror symmetry $M_{xz}\otimes C$,
where the first factor represents reflection through the
$xz$-plane and the second factor denotes charge conjugation
\cite{kkb,kk,bk}.

The ansatz is form-invariant under the abelian gauge transformation
\cite{kkb,kk,kk1}
\begin{equation}
 U= \exp \left({\frac{i}{2} \tau^n_\phi \Gamma(r,\theta)} \right)
\ .\label{gauge} \end{equation}
The functions $H_1$ and $H_2$ transform inhomogeneously
under this gauge transformation,
\begin{eqnarray}
  H_1 & \rightarrow & H_1 -  r \partial_r \Gamma \ , \nonumber \\
  H_2 & \rightarrow & H_2 +   \partial_\theta \Gamma
\ , \label{gt1} \end{eqnarray}
like a 2-dimensional gauge field.
The functions $H_3$ and $H_4$ combine to form a scalar doublet,
$(H_3+{\rm ctg} \theta, H_4)$.
We choose the same gauge condition
as previously \cite{kkb,kk,kk1,kk2,kk3,kk4},
\begin{equation}
 r \partial_r H_1 - \partial_\theta H_2 = 0 
\ . \label{gc1} \end{equation}

With the ansatz (\ref{metric2})-(\ref{gf1})
and the gauge condition (\ref{gc1}) 
we obtain the set of EYMD field equations,
given in \cite{kk4}.

The energy density of the matter fields $\epsilon =-T_0^0=-L_M$ reads
\begin{eqnarray}
-T_0^0 &= & \frac{f}{2m} \left[
 (\partial_r \Phi )^2 + \frac{1}{r^2} (\partial_\theta \Phi )^2 \right]
       + e^{2 \kappa \Phi} \frac{f^2}{2 e^2 r^4 m} \left\{
 \frac{1}{m} \left(r \partial_r H_2 + \partial_\theta H_1\right)^2 
 \right.
\nonumber \\
      & +&  \left.
   \frac{n^2}{l} \left [
  \left(  r \partial_r H_3 - H_1 H_4 \right)^2
+ \left(r \partial_r H_4 + H_1 \left( H_3 + {\rm ctg} \theta \right)
    \right)^2 \right. \right.
\nonumber \\
      & + & \left. \left.
  \left(\partial_\theta H_3 - 1 + {\rm ctg} \theta H_3 + H_2 H_4
     \right)^2 +
  \left(\partial_\theta H_4 + {\rm ctg} \theta \left( H_4-H_2 \right) 
   - H_2 H_3 \right)^2 \right] \right\}
\ . \label{edens} \end{eqnarray}
Here the first gauge field term derives from $F_{r\theta}$, 
the second and third derive from $F_{r\phi}$ 
and the fourth and fifth from $F_{\theta\phi}$.
As seen from eq.~(\ref{edens}),
regularity on the $z$-axis requires 
\begin{equation}
H_2|_{\theta=0}=H_4|_{\theta=0}
\ . \label{h2h4} \end{equation}

\subsection{Boundary conditions}

To obtain asymptotically flat solutions
with a regular event horizon and with the proper symmetries,
we must impose the appropriate boundary conditions,
the boundaries being the horizon and radial infinity,
the $z$-axis and, because of parity reflection symmetry,
the $\rho$-axis.
The presence of a regular event horizon is the essential
new feature of the static axially symmetric black hole solutions,
as compared to the globally regular solutions.
We therefore begin with a detailed discussion
of the boundary conditions at the horizon.
The boundary conditions at infinity and along the $\rho$- and the $z$-axis
agree with those of the globally regular solutions \cite{kk2,kk4}.
They are only briefly recalled for completeness.

{\sl Boundary conditions at the horizon}

The event horizon of the static black hole solutions
is characterized by $g_{tt}=-f=0$,
$g_{rr}$ is finite at the horizon in isotropic coordinates.
We impose that the horizon of the black hole solutions
resides at a surface of constant $r$, $r=r_{\rm H}$
\cite{rh2}.
This ansatz for the event horizon is justified a posteriori,
since it leads to consistent solutions, possessing
a regular event horizon.

Requiring the horizon to be regular, we obtain
the boundary conditions at the horizon $r=r_{\rm H}$.
The metric functions must satisfy
\begin{equation}
f|_{r=r_{\rm H}}=
m|_{r=r_{\rm H}}=
l|_{r=r_{\rm H}}=0 
\ , \label{bc2a} \end{equation}
and the dilaton function 
\begin{equation}
\partial_r \Phi |_{r=r_{\rm H}}
=0
\ . \label{bc2b} \end{equation}
The conditions for the gauge field functions are
\begin{eqnarray}
& &(\partial_\theta H_1 + r \partial_r H_2) |_{r=r_{\rm H}} = 0 \ ,
\nonumber \\
& &(r \partial_r H_3-H_1 H_4) |_{r=r_{\rm H}} =0,   \ \ \ 
 \left( r \partial_r H_4+H_1( H_3 + {\rm ctg} \theta) \right)
|_{r=r_{\rm H}} =0
\ , \label{bc2c} \end{eqnarray}
which imply,
$F_{r\theta}=0$ and $F_{r\phi}=0$, respectively.

Thus the equations of motion yield only three boundary conditions
for the four gauge field functions $H_i$;
one gauge field boundary condition is left indeterminate.
However, for the black hole solutions
precisely one free boundary condition at the horizon is necessary
to completely fix the gauge.
The reason is, that in contrast to the case 
of the globally regular solutions \cite{kk2}, 
for the black hole solutions
the gauge condition (\ref{gc1})
still allows non-trivial gauge transformations satisfying
\begin{equation}
r^2 \partial^2_r \Gamma
+r \partial_r \Gamma
+  \partial^2_\theta \Gamma = 0
\ . \label{gfree} \end{equation}
To fix the gauge,
we have implemented various gauge conditions at the horizon, such as
\begin{equation}
(\partial_r H_1) |_{r=r_{\rm H}} = 0
\ , \label{gfree1} \end{equation}
or
\begin{equation}
(\partial_\theta H_1) |_{r=r_{\rm H}} = 0
\ , \label{gfree2} \end{equation}
obtaining the same results for the gauge invariant quantities.

The expansions of the functions at the regular horizon
are given in Appendix A.A.

{\sl Boundary conditions at infinity}

At infinity ($r=\infty$) we require 
the boundary conditions \cite{kk2,kk3,kk4}
\begin{equation}
f|_{r=\infty}= m|_{r=\infty}= l|_{r=\infty}=1
\ , \label{bc1a} \end{equation}
\begin{equation}
\Phi|_{r=\infty}=0 
\ , \label{bc1b} \end{equation}
(since any finite value of the dilaton field at infinity
can be transformed to zero via
$\Phi \rightarrow \Phi - \Phi(\infty)$, 
$r \rightarrow r e^{-\kappa \Phi(\infty)} $)
and
\begin{equation}
H_2|_{r=\infty}=H_4|_{r=\infty}=\pm 1, \ \ \ 
H_1|_{r=\infty}=H_3|_{r=\infty}=0
\ , \label{bc1c} \end{equation}
to obtain magnetically neutral solutions.
The expansion of the functions at infinity is given in \cite{kk4}.

The node number $k$ is defined by the number of nodes of
the gauge field functions $H_2$ and $H_4$ \cite{kk4}.
Because of the symmetry
with respect to the transformation $H_i \rightarrow -H_i$,
we can choose these gauge field functions 
to be positive at the horizon.
Solutions with an even number of nodes then
have $H_2(\infty)=H_4(\infty)=1$, whereas solutions
with an odd number of nodes 
have $H_2(\infty)=H_4(\infty)=-1$.

{\sl Boundary conditions along the axes}

The symmetries determine the boundary conditions 
along the $\rho$-axis and the $z$-axis \cite{kk2,kk3,kk4},
\begin{equation}
\begin{array}{lllllll}
\partial_\theta f|_{\theta=0} &=& \partial_\theta m|_{\theta=0} &=&
\partial_\theta l|_{\theta=0} &=&0 
\ , \\
\partial_\theta f|_{\theta=\pi/2} &=&
\partial_\theta m|_{\theta=\pi/2} &=&
\partial_\theta l|_{\theta=\pi/2} &=&0 \ ,
\end{array}
\   \label{bc4a} \end{equation}
\begin{equation}
\begin{array}{lll}
\partial_\theta \Phi|_{\theta=0} &=& 0
\ , \\
\partial_\theta \Phi|_{\theta=\pi/2} &=& 0 \ ,
\end{array}
\   \label{bc4b} \end{equation}
and
\begin{equation}
\begin{array}{lllllllllll}
H_1|_{\theta=0}&=&H_3|_{\theta=0}&=&0 &\ , \ \ \ &
\partial_\theta H_2|_{\theta=0} &=& \partial_\theta H_4|_{\theta=0} 
 &=& 0 \ ,
\\
H_1|_{\theta=\pi/2}&=&H_3|_{\theta=\pi/2}&=&0 &\ , \ \ \ &
\partial_\theta H_2|_{\theta=\pi/2} &=& 
\partial_\theta H_4|_{\theta=\pi/2} &=& 0 \ .
\end{array}
\   \label{bc4c} \end{equation}
In addition, regularity on the $z$-axis requires condition (\ref{lm})
for the metric functions to be satisfied (see Appendix B)
and condition (\ref{h2h4}) for the gauge field functions.
The expansion of the functions on the positive $z$-axis is given
in \cite{kk4}.

\subsection{Dimensionless Quantities}

As previously,
we introduce the dimensionless coordinate $x$,
\begin{equation}
x=\frac{e}{\sqrt{4\pi G}} r
\ , \label{dimx} \end{equation}
the dimensionless dilaton function $\varphi$,
\begin{equation}
\varphi = \sqrt{4\pi G} \Phi
\ , \label{dimp} \end{equation}
and the dimensionless dilaton coupling constant $\gamma$,
\begin{equation}
\gamma =\frac{1}{\sqrt{4\pi G}} \kappa
\ . \label{dimg} \end{equation}
The dilaton coupling constant $\gamma$ represents a parameter;
for $\gamma = 1$ contact with the low energy effective action
of string theory is made,
whereas in the limit $\gamma \rightarrow 0$ 
the dilaton decouples and EYM theory is obtained.

The dimensionless mass $\mu$ is related to 
the mass $M$ via
\begin{equation}
\mu = \frac{e G}{\sqrt{4 \pi G }} M
\ . \label{ms0} \end{equation}

\subsection{\bf Mass, temperature and entropy}

The mass $M$ of the black hole solutions can be obtained directly from
the total energy-momentum ``tensor'' $\tau^{\mu\nu}$
of matter and gravitation \cite{wein} 
\begin{equation}
M=\int \tau^{00} d^3r 
\ . \label{mwein} \end{equation}
As shown in \cite{kk4}, the dimensionless mass $\mu$
is then determined by the derivative of the metric function $f$
at infinity 
\begin{equation}
\mu = \frac{1}{2} \lim_{x \rightarrow \infty} x^2 \partial_x  f
\ . \label{mass} \end{equation}

The zeroth law of black hole physics states, that 
the surface gravity $\kappa_{\rm sg}$ 
is constant at the horizon of the black hole solutions,
here \cite{wald,ewein}
\begin{equation}
\kappa^2_{\rm sg}=-(1/4)g^{tt}g^{ij}(\partial_i g_{tt})(\partial_j g_{tt})
\ . \label{sg} \end{equation}
To evaluate $\kappa_{\rm sg}$, we need to
consider the metric functions at the horizon.
Expanding the equations in the vicinity of the horizon
in powers of the dimensionless coordinate $(x-x_{\rm H})/x_{\rm H}$,
we observe, that the metric functions are quadratic
in $x-x_{\rm H}$,
\begin{equation}
f(x,\theta)=f_2(\theta)\left(\frac{x-x_{\rm H}}{x_{\rm H}}\right)^2 
 + O\left(\frac{x-x_{\rm H}}{x_{\rm H}}\right)^3 \ , \ \ \ 
m(x,\theta)=m_2(\theta)\left(\frac{x-x_{\rm H}}{x_{\rm H}}\right)^2 
 + O\left(\frac{x-x_{\rm H}}{x_{\rm H}}\right)^3
\ , \label{expan} \end{equation}
and likewise for $l(x,\theta)$ (see Appendix A.A).
Since the temperature $T$ is proportional to the surface gravity
$\kappa_{\rm sg}$ \cite{wald},
\begin{equation}
T=\kappa_{\rm sg} /(2 \pi)
\ , \label{temp1} \end{equation}
we obtain for the dimensionless temperature
\begin{equation}
T=\frac{f_2(\theta)}{2 \pi x_{\rm H} \sqrt{m_2(\theta)} } 
\ . \label{temp} \end{equation}
With help of the Einstein equations we show in Appendix A.B, that
the temperature $T$, as given in (\ref{temp}), is indeed constant.

The dimensionless area $A$ 
of the event horizon of the black hole solutions is given by 
\begin{equation}
A = 2 \pi \int_0^\pi  d\theta \sin \theta
\frac{\sqrt{l_2 m_2}}{f_2} x_{\rm H}^2
\ . \label{area} \end{equation}
The entropy $S$ is proportional to the area $A$ \cite{wald}
\begin{equation}
S = \frac{A}{4}
\ , \label{area1} \end{equation}
leading to the dimensionless product
\begin{equation}
TS = \frac{x_{\rm H}}{4} \int_0^\pi  d\theta \sin \theta
{\sqrt{l_2 }}
\ . \label{area2} \end{equation}

Having defined temperature and entropy, we now derive
a second expression for the mass \cite{wald}.
As shown before \cite{kk4}, the equations of motion yield
\begin{equation}
\frac{1}{8 \pi G}
\partial_\mu \left( \sqrt{-g} \partial^\mu {\rm ln} f \right) =
  - \sqrt{-g} \left( 2 T_0^{\ 0} - T_\mu^{\ \mu} \right)
\ . \label{area3} \end{equation}
Integrating both sides over $r$, $\theta$ and $\phi$
from the horizon to infinity, we obtain
\begin{equation}
\frac{1}{4 G}
 \int_0^\pi d\theta \sin \theta  \left.
 \left[ r^2 \sqrt{l} \frac{\partial_r f}{f} \right]
 \right|_{r_{\rm H}}^\infty =
 - \int_0^{2 \pi} \int_0^\pi \int_{r_{\rm H}}^\infty
  d \phi d \theta d r
  \sqrt{-g} \left( 2 T_0^{\ 0} - T_\mu^{\ \mu} \right)
 = M_{\rm o}
\ . \label{area4} \end{equation}
Changing to dimensionless coordinates, we express the l.h.s.~with help
of the dimensionless mass $\mu$ and the product 
of temperature and entropy, $TS$, obtaining
\begin{equation}
\mu  =\mu_{\rm o} + 2 TS
\ , \label{mass2} \end{equation}
with $\mu_{\rm o}= (e/\sqrt{4 \pi G}) G M_{\rm o}$,
in agreement with the general mass formula
for static black hole solutions \cite{wald}.

Considering the dilaton field equation in the form \cite{kk4}
\begin{equation}
 \frac{1}{\kappa}
 \partial_\mu \left( \sqrt{-g} \partial^\mu \Phi \right) =
  \frac{1}{8 \pi G} 
 \partial_\mu \left( \sqrt{-g} \partial^\mu \ln f \right) 
\ , \label{eqdil3} \end{equation}
it is straightforward to derive a relation 
between mass the $\mu$ and the dilaton charge $D$,
\begin{equation}
D  =  \lim_{x \rightarrow \infty} x^2 \partial_x  \varphi
\ . \label{dilch} \end{equation}
We replace the l.h.s.~of eq.~(\ref{area3})
by the l.h.s.~of eq.~(\ref{eqdil3}) and integrate as above.
Recalling the boundary condition $\partial_r \Phi|_{r=r_{\rm H}}=0$ and
changing to dimensionless coordinates, we obtain
\begin{equation}
 D = \gamma \mu_{\rm o} = \gamma ( \mu - 2 TS ) 
\ , \label{der4} \end{equation}
where the second equality follows from eq.~(\ref{mass2}).

Previously we obtained for static spherically symmetric black holes
the relation \cite{kks3}
\begin{equation}
 D = \gamma \left( \mu - \frac{\mu_{\rm S}}{T_{\rm S}} T \right)
\ , \label{area5} \end{equation}
 where $\mu_{\rm S}= \tilde x_{\rm H}/2$ and
$T_{\rm S} = 1/(4 \pi \tilde x_{\rm H})$ represent the
Schwarzschild mass and the Schwarzschild temperature
of a black hole with horizon $\tilde x_{\rm H}$
in Schwarzschild-like coordinates.
This relation represents a special case of the more general relation
(\ref{der4}), since for a spherical black hole
$A = 4 \pi \tilde x_{\rm H}^2$.

\section{\bf Spherically symmetric solutions}

The static spherically symmetric EYM and EYMD solutions 
were obtained previously in Schwarzschild-like coordinates
with metric \cite{bm,eymd,kks3}
\begin{equation}
ds^2=
  - A^2 N dt^2 +  \frac{1}{N} d \tilde r^2 
  + \tilde r^2 \left( d \theta^2 + \sin^2 \theta d\phi^2 \right)
\   \label{metric3} \end{equation}
and metric functions $A(\tilde r)$ and $N(\tilde r)$,
\begin{equation}
N(\tilde r) = 1 - \frac{2 \tilde m(\tilde r)}{\tilde r}
\ . \label{N} \end{equation}
We here briefly present the spherical solutions
in isotropic coordinates. We exhibit the coordinate transformation between
the radial coordinates $r$ and $\tilde r$,
and we discuss the limiting solutions.

\subsection{Coordinate transformation}

The static spherically symmetric isotropic metric reads 
\begin{equation}
ds^2=
  - f dt^2 +  \frac{m}{f} \left ( d r^2 + r^2 \left( d \theta^2 
           +  \sin^2 \theta d\phi^2 \right) \right)
\ . \label{metric4} \end{equation}
The coordinates $r$ and $\tilde r$ are related by
\begin{equation}
\frac{dr}{r} = \frac{1}{\sqrt{N(\tilde r)}} \frac{d \tilde r}{\tilde r}
\ . \label{co2} \end{equation}

Changing to dimensionless coordinates $x$ and $\tilde x$,
the coordinate function $x(\tilde x)$ must be
obtained numerically from eq.~(\ref{co2}),
since the function $N(\tilde x)$ of the non-abelian solutions
is only known numerically.
To avoid the divergence of $1/N(\tilde x)$ at the horizon
in the numerical integration,
we introduce the function $\Delta(\tilde x)$
\begin{equation}
\Delta(\tilde x) = \frac{1}{\sqrt{N(\tilde x)}}
 - \frac{1}{c\sqrt{N_{\rm S}(\tilde x)}}
\ , \label{co3} \end{equation}
where 
\begin{equation}
c^2 = \left. \tilde x \frac{d N}{d \tilde x} 
 \right|_{\tilde x=\tilde x_{\rm H}}
\   \label{co4} \end{equation}
and $N_{\rm S}$ is the Schwarzschild metric function
\begin{equation}
N_{\rm S}(\tilde x) = 1 - \frac{\tilde x_{\rm H}}{\tilde x}
\ . \label{co4s} \end{equation}
After replacing on the r.h.s.~of eq.~(\ref{co2}) 
$1/\sqrt{N}$ by $1/c\sqrt{N_{\rm S}} + \Delta$,
we integrate the first term analytically and
the finite $\Delta$-term numerically.
This yields the coordinate transformation
\begin{equation}
x = \left( \frac{2 \sqrt{\tilde x(\tilde x-\tilde x_{\rm H})}
 + 2\tilde x - \tilde x_{\rm H}}{ 4}
\right)^{\frac{1}{c}}
 \exp \left[ \int_{\tilde x_{\rm H}}^{\tilde x}
 \frac{\Delta(\tilde x \,')}{\tilde x \,'} d \tilde x \,' \right]
\ , \label{co5} \end{equation}
where
\begin{equation}
 x_{\rm H} = \left( \frac{\tilde x_{\rm H}}{4} \right)^{\frac{1}{c}}
\   \label{co6} \end{equation}
is determined by the asymptotic requirement, $x/\tilde x \rightarrow 1$.
For the Schwarzschild solution $c=1$,
and $x_{\rm H} = \tilde x_{\rm H}/4$.

Fig.~1 demonstrates the coordinate transformation
for the static spherically symmetric EYMD solutions with $x_{\rm H}=0.02$,
$\gamma=1$ and $k=1-4$.
Figs.~2a,b show the metric functions $f$ and $m$,
and Figs.~3a,b show the gauge field function $w$
and the dilaton function $\varphi$.

Fig.~4 demonstrates the coordinate transformation
for the static spherically symmetric EYM solutions with $x_{\rm H}=0.02$
and $k=1-4$.
Figs.~5a,b show the metric functions $f$ and $m$.

\subsection{Limiting solutions}

For fixed dilaton coupling constant and horizon,
the sequences of neutral static 
spherically symmetric EYMD black hole solutions 
converge to limiting solutions \cite{kks3}.
These limiting solutions are EMD black hole solutions \cite{emd}
with the same dilaton coupling constant, the same horizon
and charge $P=1$ \cite{kks3}.

We now consider the limiting EMD solution for $\gamma=1$
and a general charge $P$.
Defining \cite{emd}
\begin{equation}
X_+ = \sqrt{\tilde x^2_{\rm H} + 2 P^2} \ , \ \ \
X_- = \frac{2 P^2}{X_+}
\ , \label{emd1} \end{equation}
and
\begin{equation}
X = \frac { X_- + \sqrt{4 \tilde x^2 + X_-^2}}{2}
\ , \label{emd2} \end{equation}
i.e.~$X_{\rm H}=X_+$,
the coordinate transformation for the limiting EMD solution reads
\begin{equation}
x= \frac{2\sqrt{(X-X_-)(X-X_+)} + 2 X - (X_-+X_+)}{4}
\ , \label{emd3} \end{equation}
with 
\begin{equation}
x_{\rm H} =  \frac{X_+-X_-}{4}
\ . \label{emd6} \end{equation}
The metric functions of the limiting solution read
in isotropic coordinates
\begin{equation}
f_\infty = \frac{ \left( 1 - \frac{x_{\rm H}}{x} \right)^2 }
 { \left( 1 + 2 \frac{ x_{\rm H}}{x}\sqrt{1 
          + \frac{1}{2} \left(\frac{P}{x_{\rm H}}\right)^2 }
  + \left( \frac{x_{\rm H}}{x} \right)^2 \right) }
\   \label{emd4} \end{equation}
and
\begin{equation}
m_\infty =  \left( 1 - \left( \frac{x_{\rm H}}{x}\right)^2 \right)^2 
\ . \label{emd5} \end{equation}
The dilaton function of the limiting solution reads
\begin{equation}
e^{2 \varphi_\infty} = 
 \frac{ \left( 1 + \frac{x_{\rm H}}{x} \right)^2 }
 { \left( 1 + 2 \frac{ x_{\rm H}}{x}\sqrt{1 
          + \frac{1}{2} \left(\frac{P}{x_{\rm H}}\right)^2 }
  + \left( \frac{x_{\rm H}}{x} \right)^2 \right) }
\ . \label{emd7} \end{equation}
The gauge field function of the limiting solution is trivial, 
$w_\infty=0$.
For $P=1$ and $x_{\rm H}=0.02$ the coordinate transformation
for the limiting solution is shown in Fig.~1, 
the metric functions of the limiting solution are shown
in Figs.~2 and the dilaton function in Fig.~3.
For the horizon $x_{\rm H}=0.02$ the convergence is rapid.
The solution with $k=3$ is already very close to the limiting
solution, and the solution with $k=4$ is almost indistinguishable,
for all functions except for the gauge field function $w$.
This function approaches its limiting function $w_\infty=0$
non-uniformly, since the boundary conditions require $w(\infty) \ne 0$.
A detailed discussion on the convergence of these solutions
is given in \cite{kks3}.

We now turn to the limiting solutions of the
the sequences of static spherically symmetric EYM black hole solutions.
In Schwarzschild-like coordinates, the limiting solution 
of the sequence with fixed horizon $\tilde x_{\rm H}$,
is a RN solution with the same horizon
and with charge $P=1$, if $\tilde x_{\rm H}>1$ \cite{lim,kks3}.
If $\tilde x_{\rm H}<1$, the limiting solution
consists of two parts, an exterior part 
covering the interval $1 < \tilde x < \infty$, which
represents the exterior of an extremal RN solution with mass $\mu=1$, 
horizon $\tilde x_{\rm H}=1$ and charge $P=1$,
and an oscillating interior part covering the interval
$\tilde x_{\rm H} < \tilde x < 1$ \cite{lim,kks3,fn1}.
In isotropic coordinates with fixed horizon $x_{\rm H}$,
the limiting solution corresponds to the exterior
of a RN solution with the same horizon and with charge $P=1$.

The coordinate transformation for the limiting RN solution reads
\begin{equation}
x = \frac{ \sqrt{ \tilde x^2 - 2 \mu \tilde x + P^2}
  + \tilde x - \mu }{ 2}
\ , \label{rn2} \end{equation}
with
\begin{equation}
\mu = \frac{\tilde x^2_{\rm H} + P^2}{2 \tilde x_{\rm H}}
\ . \label{rn2a} \end{equation}
The metric functions are given by 
\begin{equation}
f_\infty = \frac{ \left( 1 - \left( \frac{x_{\rm H}}{x}\right)^2 \right)^2 }
 { \left( 1 + 2 \frac{ x_{\rm H}}{x}\sqrt{1 
          + \frac{1}{4} \left(\frac{P}{x_{\rm H}}\right)^2 }
  + \left( \frac{x_{\rm H}}{x} \right)^2 \right)^2 }
\   \label{rn1a} \end{equation}
and
\begin{equation}
m_\infty =  \left( 1 - \left( \frac{x_{\rm H}}{x}\right)^2 \right)^2 
\ , \label{rn1b} \end{equation}
i.e.~$m_\infty$ is identical for the EMD and RN solutions.
Again, the gauge field function of the limiting solution is trivial,
$w_\infty=0$.
For $P=1$ and $x_{\rm H}=0.02$ the coordinate transformation
of the limiting solution is shown in Fig.~4
and the metric functions in Figs.~5.

\boldmath
\subsection{Limit $x_{\rm H} \rightarrow 0$}
\unboldmath

Let us now consider the limit $x_{\rm H} \rightarrow 0$
for the black hole solutions.
In this limit the solutions tend towards
the corresponding globally regular solutions \cite{su2,eymd,kks3}.
For several quantities of interest, however,
the limit $x_{\rm H} \rightarrow 0$ is not smooth.
For instance the energy density of the matter fields
of the black hole solutions approaches the energy density 
of the globally regular solutions with a discontinuity at the origin.
The reason is, that the magnetic field of the black hole solutions
is purely radial at the horizon 
\begin{equation}
\vec B= B_r \vec e_r
\ , \label{br} \end{equation}
with $B_r=F_{\theta\phi}$,
because the boundary conditions (\ref{bc2c}) require
\begin{equation}
 B_\theta=0\ , \ \ \ B_\phi=0
\ , \label{btp} \end{equation}
with $B_\theta=-F_{r\phi}$ and $B_\phi=F_{r\theta}$,
whereas the magnetic field of the globally regular solutions 
has non-vanishing $B_\theta$ at the origin.
We demonstrate this discontinuous behaviour of the energy density
of the matter fields in Figs.~6a,b for the EYMD ($\gamma=1$) solutions with 
$k=1$ and 2 and $x_{\rm H}=0.001,\ 0.01, \ 0.1$ and 1.

We finally consider the Kretschmann scalar $K$,
\begin{equation}
 K= R^{\mu\nu\alpha\beta}R_{\mu\nu\alpha\beta}
\ . \label{ks} \end{equation}
An analytical expression for $K$ is given in Appendix A.C.
We show $K$ in Fig.~7 for the same set of EYMD solutions.
Again, the limit $x_{\rm H} \rightarrow 0$ is not smooth.
However, as required for a regular horizon,
$K$ is finite at the horizon for finite $x_{\rm H}$.

\section{\bf Axially symmetric solutions}

Subject to the above boundary conditions,
we solve the equations for the static axially symmetric
black hole solutions numerically.
We employ the same numerical algorithm \cite{schoen}
as for the static axially symmetric 
globally regular solutions \cite{kk2,kk3,kk4}.
To map spatial infinity to the finite value $\bar{x}=1$,
we here employ the radial coordinate 
\begin{equation}
\bar{x} = 1-\frac{x_{\rm H}}{x}
\ . \label{barx} \end{equation}
The equations are then discretized on a non-equidistant
grid in $\bar{x}$ and $\theta$,
where typical grids used have sizes $150 \times 30$, 
covering the integration region 
$0\leq\bar{x}\leq 1$, $0\leq\theta\leq\pi/2$.
The numerical error for the functions is estimated to be 
on the order of $10^{-3}$.

The solutions depend on two continuous parameters,
the ``isotropic radius'' $x_{\rm H}$ of the horizon
and the dilaton coupling constant $\gamma$,
as well as on two integers,
the winding number $n$ and the node number $k$.

\subsection{Energy density and horizon}

We begin our discussion of the static axially symmetric
black hole solutions
by considering the energy density of the matter fields.
As an example we show 
in Figs.~8 the energy density of the matter fields
for the black hole solution with $x_{\rm H}=1$, $\gamma=1$,
$n=2$ and $k=1$.
Fig.~8a shows a 3-dimensional plot of the energy density
as a function of the coordinates $\rho = x \sin \theta$ and
$z= x \cos \theta$ together with a contour plot,
and Figs.~8b-e show surfaces of constant energy density.
For small values of $\epsilon$ the energy density appears ellipsoidal,
being flatter at the poles than in the equatorial plane.
With increasing values of $\epsilon$ a toruslike shape appears
with two additional ellipsoids covering the poles.
The ellipsoids covering the poles 
persist up to the largest values of the energy density, 
showing that the maximum of the energy density resides at the poles.
Furthermore these black hole solutions have the remarkable property,
that the energy density is not constant at the horizon but angle-dependent.

The static axially symmetric black hole solutions are
self-consistent solutions arising from the interplay 
of gravity with the non-abelian gauge fields.
In isotropic coordinates the horizon 
of the static axially symmetric black hole solutions
resides at a surface of constant radial coordinate $x$, $x=x_{\rm H}$.
Since the energy density of the matter fields 
of the static axially symmetric black hole solutions
is angle-dependent at the horizon, this suggests that
the horizon is deformed.
We therefore 
measure the circumference of the horizon along the equator, $L_e$, 
\begin{equation}
L_e = \int_0^{2 \pi} { d \phi \left.
 \sqrt{ \frac{l}{f}} x \sin\theta
 \right|_{x=x_{\rm H}, \theta=\pi/2} }
 = 2 \pi x_{\rm H} \left. \sqrt{\frac{l_2}{f_2}} 
  \right|_{\theta=\pi/2} 
\ , \label{le} \end{equation}
and the circumference of the horizon along the poles, $L_p$,
\begin{equation}
L_p = 2 \int_0^{ \pi} { d \theta \left.
 \sqrt{ \frac{m  }{f}} x
 \right|_{x=x_{\rm H}, \phi=const.} }
 = 2 x_{\rm H} \int_0^\pi { d \theta 
 \sqrt{\frac{m_2(\theta)}{f_2(\theta)}} }
\ . \label{lp} \end{equation}
A spherical horizon would require $L_e=L_p$. 
For the static axially symmetric black hole solutions
we observe, however, 
\begin{equation}
L_p > L_e
\ . \label{lelp} \end{equation}
Thus the horizon itself possesses only axial symmetry.
The deviation from spherical symmetry is small, though.
For instance for the solution of Figs.~8 we find
$L_e/L_p=0.998$.

\boldmath
\subsection{$x_{\rm H}$ dependence of the solutions}
\unboldmath

Let us now study the dependence of the static axially symmetric solutions
on the isotropic black hole horizon radius $x_{\rm H}$.
To be specific, we consider black hole solutions with $\gamma=1$,
$n=2$ and $k=1$, the parameters also employed in the solution shown in
Figs.~8.
In Fig.~9 the energy density of the matter fields
is shown for several black hole solutions,
with values of the isotropic black hole horizon radius
$x_{\rm H} = 0.01$, 0.1, 1 and 10,
as well as for the globally regular solution.
As noted above, for larger values of $x_{\rm H}$
the global maximum of the energy density of the black
hole solutions resides on the $z$-axis at the horizon,
while a local maximum is located
on the $\rho$-axis away from the horizon.
With decreasing $x_{\rm H}$,
the maximum on the $\rho$-axis away from the horizon increases
and becomes the global maximum,
while the maximum on the $z$-axis at the horizon diminishes.
At the same time, a pronounced minimum develops
on the $\rho$-axis at the horizon,
aggravating the angle-dependence of the energy density at the horizon.

With decreasing $x_{\rm H}$
the energy density of the matter fields of the black hole solutions
tends increasingly towards the energy density of the
globally regular solution, which possesses a toroidal shape
because of the strong global maximum
on the $\rho$-axis \cite{kk2,kk3,kk4}.
However, the limit $x_{\rm H} \rightarrow 0$ is not smooth,
as already observed in section III
for the static spherically symmetric solutions.
The reason is, that
the magnetic field of the black hole solutions
is purely radial at the horizon,
while the magnetic field of the globally regular solutions
also has non-vanishing $B_\theta$ at the origin.
In the static axially symmetric solutions,
both $B_r$ and $B_\theta$ are angle-dependent.
For the globally regular solutions
the contributions from both $B_r$ and $B_\theta$
precisely add to an angle-independent density at the origin.
In contrast, the black hole solutions possess an angle-dependent
density at the horizon.

In Figs.~10-12
we show the functions of the black hole solutions
with $n=2$, $k=1$, $\gamma=1$ and 
$x_{\rm H}=0.01,\ 0.1, \ 1$ and 10 for three angles.
Figs.~10 show the metric functions,
Figs.~11 the gauge field functions and
Fig.~12 the dilaton function.
For small values of $x_{\rm H}$,
the angle-dependence of the metric functions is strongest around $x=1$. 
With increasing $x_{\rm H}$, the angle-dependence decreases strongly,
and the metric becomes increasingly spherical.
At the same time, with increasing $x_{\rm H}$
the matter fields become less important.
This is seen for instance in Fig.~13, where we present
relation (\ref{mass2}), $\mu= \mu_o +2TS$ (with $\mu_o=D/\gamma$),
and observe that $\mu_o \gg 2TS$ for small $x_{\rm H}$, 
whereas $\mu_o \ll 2TS$ for large $x_{\rm H}$.
Considering the gauge field functions,
we recall, that for the globally regular solutions
the functions $H_2$ and $H_4$ have precisely $k$ nodes,
the function $H_1$ has $k-1$ non-trivial nodes,
and the function $H_3$ has one non-trivial node \cite{kk4}.
For the black hole solutions we here observe accordingly,
that $H_1$ has no node and that
$H_2$ and $H_4$ each have one node. 
However, we observe two non-trivial nodes for the function $H_3$
of the black hole solutions of Fig.~11c,
which may be due to the choice of gauge.
We note, that the gauge field functions of Figs.~11
are obtained with gauge condition (\ref{gfree2}).
With increasing $x_{\rm H}$ the gauge field functions
retain a considerable angle-dependence. 
They approach a limiting shape, shifting towards larger values of $x$.
The dilaton function is slightly angle-dependent
at the horizon, as seen in Fig.~12.
Analogously to the metric functions,
for small values of $x_{\rm H}$
the angle-dependence of the dilaton function is strongest around $x=1$.
With increasing $x_{\rm H}$ the angle-dependence decreases strongly,
and at the same time the magnitude of the dilaton function
diminishes strongly.

In Fig.~14 we exhibit the Kretschmann scalar for these solutions.
With decreasing $x_{\rm H}$ the Kretschmann scalar
of the black hole solutions
tends to the Kretschmann scalar
of the globally regular solution except close to the horizon,
where it increases dramatically.
For finite $x_{\rm H}$, however,
the Kretschmann scalar remains finite at the horizon,
indicating that the black hole solutions indeed possess
a regular horizon, as required (see Appendix A.C).

In general the EYM solutions are very similar to the EYMD solutions.
We therefore do not exhibit these here. 
For instance,
the energy density of the matter fields and the metric functions 
$f$ and $m$ for the EYM black hole solutions with
$n=2$, $k=1$ and $x_{\rm H}=0.02$, 0.1, 0.5 and 1 are
shown in \cite{kk3}.

Let us now inspect the metric functions at the horizon more closely.
In Figs.~15a-c we exhibit the expansion coefficients
$f_2(\theta)$, $m_2(\theta)$ and $l_2(\theta)$ 
of the metric functions
for the above set of EYMD solutions.
With increasing $x_{\rm H}$ the angle-dependence
of the expansion coefficients first increases and then decreases again.
The shape of the horizon changes with $x_{\rm H}$ in a similar way.
This is seen in Fig.~16, where the ratio of the
circumference at the equator and the circumference at the poles,
$L_e/L_p$, is shown as a function of the mass
for the EYMD ($\gamma=1$) and EYM black hole solutions
with $n=2$ and $k=1$.
The maximal deformation of the horizon
occurs for an isotropic horizon radius of $x_{\rm H}=0.295$ for $\gamma=1$
and $x_{\rm H}=0.195$ for $\gamma=0$.
For $\gamma=0$, the maximal deformation of the horizon is greater
than for $\gamma=1$.
Fig.~17 shows the area of the horizon
as a function of the mass for the corresponding solutions.
For comparison, the area of a spherical horizon with
circumference $L_e$ is also shown,
deviating only little from the area of the deformed horizon.

Fig.~18 shows the inverse temperature as a function of the mass
for the static axially symmetric
EYMD ($\gamma=1$) and EYM black hole solutions
with $n=2$ and $k=1$.
These curves are very similar to those of the corresponding
static spherically symmetric black hole solutions \cite{kks3}.
Table~1 presents the dimensionless mass, temperature, entropy
and dilaton charge as well as the ratio $L_e/L_p$ 
of the above set of static axially symmetric black hole solutions
of EYMD theory ($\gamma=1$).

\subsection{Winding number dependence}

To illustrate the winding number dependence
of the black hole solutions we show in Figs.~19
the solutions with $n=4$, $k=1$, $\gamma=1$ and 
$x_{\rm H}=0.01,\ 0.1, \ 1$ and 10 for three angles.
In Fig.~19a we see the energy density of the matter fields.
We recall, that with increasing winding number $n$
the maximum on the $\rho$-axis 
of the energy density of the globally regular solutions 
shifts outward and decreases in height \cite{kk4}.
As compared to the $n=2$ black hole solutions, shown in Fig.~9, 
we here observe
that with decreasing $x_{\rm H}$ the globally regular solution is
approached faster for the greater winding number $n=4$,
while with increasing $x_{\rm H}$
the angle-dependence of the energy density of
the black hole solutions remains stronger.
Most strikingly, however, we observe, that the global maximum
always resides on the $\rho$-axis. The maximum on the $z$-axis
remains a local one also for large $x_{\rm H}$.

In Figs.~19b-d we show the metric function $f$, the gauge field function
$H_2$ and the dilaton function $\varphi$, respectively.
For the globally regular solutions
the angle-dependence of the metric and matter
functions increases strongly with $n$ and
the location of the biggest angular splitting moves further outward.
This dependence is reflected in the black hole solutions.
In particular we observe, that with increasing $n$
the angular dependence remains stronger for larger $x_{\rm H}$.

Let us now turn to the shape of the horizon.
As for $n=2$, with increasing $x_{\rm H}$ the ratio of the
circumference at the equator and the circumference at the poles,
$L_e/L_p$, first decreases and then increases again.
In Fig.~20 the ratio $L_e/L_p$ is shown as a function of the mass
for the EYMD ($\gamma=1$) and EYM black hole solutions with $n=4$ and
$k=1$.
The maximal deformation of the horizon
occurs for an isotropic horizon radius of $x_{\rm H}=0.95$
for $\gamma=1$
and $x_{\rm H}=0.65$ for $\gamma=0$.
Again, for $\gamma=0$ the maximal deformation of the horizon is greater
than for $\gamma=1$, but for $n=4$ it is smaller than for $n=2$.
Table~1 presents the ratio $L_e/L_p$ as well as
the dimensionless mass, temperature, entropy and dilaton charge
for the above set of static axially symmetric black hole solutions
of EYMD theory ($\gamma=1$).

\subsection{Node number dependence}

To illustrate the node number dependence
of the black hole solutions we show in Figs.~21
the solutions with $n=2$, $k=2$, $\gamma=1$ and 
the same set of radii $x_{\rm H}$ as above.
The energy density of the matter fields is shown in Fig.~21a.
We recall that
the maximum of the energy density of the globally regular solutions 
is located on the $\rho$-axis.
With increasing node number $k$ it
shifts inward and increases strongly in height \cite{kk4}.
For the black hole solutions we observe
that with decreasing $x_{\rm H}$ the globally regular solution is
approached more slowly for greater node number $k$,
while with increasing $x_{\rm H}$
the angle-dependence of the energy density of
the black hole solutions diminishes faster.
In particular, for large $x_{\rm H}$
the maximum on the $z$-axis
becomes the global maximum,
while the maximum on the $\rho$-axis disappears.

In Figs.~21b-d we show the metric function $f$, the gauge field function
$H_2$ and the dilaton function $\varphi$, respectively.
We recall, that for the globally regular solutions
the location of the biggest angular splitting moves inward with $k$
while its size stays roughly constant for most functions \cite{kk4}.
For the black hole solutions we here observe, that
as compared to the $k=1$ solutions, shown in Figs.~10-12,
the angular dependence of the metric and dilaton functions
diminishes faster with increasing $x_{\rm H}$.

Table~1 again shows the dimensionless mass, temperature, entropy
and dilaton charge as well as the ratio $L_e/L_p$
for the above set of static axially symmetric black hole solutions
of EYMD theory ($\gamma=1$).

\subsection{Limiting solutions}

For fixed $n$, $\gamma$ and $x_{\rm H}$ and increasing $k$,
the static axially symmetric black hole solutions form sequences,
tending to limiting solutions.
Whereas the solutions of the sequences are magnetically neutral,
axially symmetric and non-abelian, the limiting solutions possess
magnetic charge $n$ and they are spherically symmetric and abelian.
For finite $\gamma$ the limiting solutions are
EMD black hole solutions \cite{emd},
while for $\gamma=0$ the limiting solutions are 
Reissner-Nordstr\o m solutions \cite{kk3,kk4}.
The convergence of the global properties
is seen for instance in Table~1 of \cite{kk3}.

To illustrate the convergence of the metric and matter functions,
we exhibit in Figs.~22a-c the EYMD solutions
for $n=2$, $\gamma=1$ and $x_{\rm H}=1$ as an example.
With increasing $k$,
the metric functions converge rapidly
to the metric functions of the limiting EMD solution.
In Fig.~22a this is seen for the metric function $f$.
With increasing $k$,
the gauge field functions tend to their (vanishing)
limiting functions in an exponentially increasing interval,
but because of the boundary conditions
the convergence is not uniform for $H_2$ and $H_4$.
This is illustrated in Fig.~22b for the gauge field function $H_2$.
The dilaton function shown in Fig.~22c again converges rapidly
and uniformly.

\section{Conclusions}

We have constructed numerically a new class of black hole solutions
in EYM and EYMD theory \cite{kk3}.
These black hole solutions are asymptotically flat, static 
and possess a regular event horizon.
However, they are not spherically symmetric but only axially symmetric
with angle-dependent fields at the horizon.

The event horizon of the static axially symmetric black hole solutions
resides at a surface of constant isotropic radial
coordinate, $x=x_{\rm H}$.
However, the horizon is not spherical.
Evaluating the circumference of the horizon along the equator, $L_e$, 
and the circumference of the horizon along the poles, $L_p$, we observe
that the ratio $L_e/L_p$ is slightly smaller than one, i.e.~the horizon
is slightly elongated along the symmetry axis,
the maximal elongation occurring for small values of $x_{\rm H}$.

Like their globally regular counterparts,
the static axially symmetric black hole solutions
are characterized by two integers, the winding number $n>1$ and the node
number $k$ of the purely magnetic gauge field.
Whereas the energy density of the globally regular solutions
has a torus-like shape, due to a strong peak on the $\rho$-axis
away from the origin \cite{kk2},
the energy density of the black hole solutions has a more
complicated shape,
depending on the winding number $n$, the node number $k$ and 
the horizon radius $x_{\rm H}$.

The static spherically symmetric EYM and EYMD black hole solutions 
are unstable \cite{strau,eymd},
and there is all reason to believe, that
the static axially symmetric black hole solutions are unstable, too.
But we expect analogous black hole solutions in EYMH theory 
\cite{ewein,mon} and ES theory \cite{bc},
corresponding to black holes inside axially symmetric multimonopoles
and multiskyrmions, respectively,
and for $n=2$ these static axially symmetric solutions
should be stable \cite{ewein,mon,bc}.
In contrast, stable black hole solutions
with higher magnetic charges (EYMH) or higher baryon numbers (ES)
should not correspond to static axially symmetric solutions with $n>2$.
Instead these stable black hole solutions 
should exhibit only discrete crystal-like symmetries \cite{ewein,mon,bc}.
We expect analogous but unstable black hole solutions with crystal-like
symmetries also in EYM and EYMD theory.

We would like to thank the RRZN in Hannover for computing time.

\newpage

\section{Appendix A}
\subsection{Expansion at the horizon}

Here we present the expansion of the functions of the static axially symmetric
black hole solutions at the horizon $x_{\rm H}$ in powers of 
$\delta$,
\begin{equation}
\delta = \frac{x}{x_{\rm H}}-1
\ . \label{epsil} \end{equation}
The expansion of the functions at the horizon can be obtained from
the regularity conditions imposed on the Einstein equations and
the matter field equations. 
\begin{eqnarray}
f(\delta,\theta ) & = & \delta^2 f_{2} 
    \left\{  1-\delta+\frac{\delta^2}{24}
     \left[ \frac{f_{2}}{l_{2}} \left(\frac{n}{x_{\rm H}}\right)^2 e^{2 \gamma \varphi_0}
     \left[  \rule{0mm}{6mm} 24\cot \theta \left( \left( H_{30,\theta}+1-H_{20}^2 \right) H_{30}
     -H_{20} H_{40,\theta}
     +H_{40} H_{40,\theta} \right)
     \right. \right. \right.
\nonumber\ \\   & & 
     \left. \left. \left.
     +12 \left( H_{20}^2 \left( H_{30}^2+H_{40}^2-1 \right)
     +\frac{\left( H_{20}-H_{40}\right)^2 + H_{30}^2}{\sin^2 \theta}
     -\left( H_{30}^2+H_{40}^2 \right)
     +2 H_{20} \left( H_{30} H_{40,\theta}-H_{40} H_{30,\theta} \right) +1
     \right. \right. \right. \right.
\nonumber\ \\   & &  
     \left. \left. \left. \left.  
     \rule{0mm}{6mm}  
     +2 H_{30,\theta}+H_{30,\theta}^2+H_{40,\theta}^2 \right) \right] \right. \right.
\nonumber\ \\   & &  
     \left. \left.
     -2 \cot \theta \left( 3 \frac{f_{2,\theta}}{f_{2}}
     -2 \frac{l_{2,\theta}}{l_{2}} \right)
     -\left( 3 \frac{f_{2,\theta}}{f_{2}}\frac{l_{2,\theta}}{l_{2}} 
     +6 \frac{f_{2,\theta \theta}}{f_{2}}+\left( \frac{l_{2,\theta}}{l_{2}} \right)^2
     -2\frac{l_{2,\theta \theta}}{l_{2}} 
     -18-6 \left( \frac{f_{2,\theta}}{f_{2}} \right)^2 \right) \right] \right\}
     +O(\delta^5) \nonumber\ , \\
& & \nonumber \\    
m(\delta,\theta ) & = & \delta^2  m_{2}
\left\{1-3\delta+\frac{\delta^2}{24} \left[
150-4 \frac{l_{2,\theta ,\theta}}{l_{2}}
+2 \left(\frac{l_{2,\theta}}{l_{2}}\right)^2
+3\frac{l_{2,\theta}}{l_{2}}\frac{m_{2,\theta}}{m_{2}}
-6 \frac{m_{2,\theta ,\theta}}{m_{2}}
+6 \left(\frac{m_{2,\theta}}{m_{2}}\right)^2
-6 \left(\frac{f_{2,\theta}}{f_{2}}\right)^2
\right. \right. 
 \nonumber\ , \\   
& & 
\left. \left.
+2\cot \theta \left( 3\frac{m_{2,\theta}}{m_{2}}
                    -4\frac{l_{2,\theta}}{l_{2}}\right)
-24 \varphi_{0,\theta}^2  \right] \right\} +O(\delta^5) 
 \nonumber\ , \\                 
& & \nonumber \\    
l(\delta,\theta ) & = & \delta^2  l_{2} 
\left\{ 1-3\delta+\frac{\delta^2}{12} \left[ \left(\frac{l_{2,\theta}}{l_{2}}\right)^2
-2 \frac{l_{2,\theta \theta}}{l_{2}}
+75-4 \cot \theta 
\frac{l_{2,\theta}}{l_{2}}\right]\right\} +O(\delta^5) \nonumber\ , \\
& & \nonumber \\    
H_1(\delta,\theta ) & = & \delta \left( 1- \frac{\delta}{2} \right) H_{11}
 +O(\delta^3) \nonumber\ , \\
& & \nonumber \\    
H_2(\delta,\theta ) & = & 
H_{20}+\frac{ \delta^2}{4} \left[ \frac{m_2}{l_2} n^2 \left(\rule{0mm}{6mm}
 H_{20} \left(H_{30}^2+H_{40}^2-1 \right)
 +H_{30} H_{40, \theta }-H_{40} H_{30, \theta }
   +\frac{H_{20}-H_{40}}{\sin^2 \theta}
\right. \right.  
\nonumber\ \\   & &
\left. \left. 
   -\rule{0mm}{6mm}\cot \theta \left( 2 H_{20} H_{30}+H_{40, \theta } \right) \right)
   +\left( H_{11,\theta }-H_{20, \theta \theta }\right)\right]
 +O(\delta^3) \nonumber\ , \\
& & \nonumber \\    
H_3(\delta,\theta ) 
& = & H_{30}-\frac{\delta^2}{8} 
\left[\left( 4 \gamma \varphi_{0,\theta}
+2 \frac{f_{2,\theta}}{f_{2}}-\frac{l_{2,\theta}}{l_{2}}\right) 
\left( 1-H_{40} H_{20}+H_{30, \theta }+\cot \theta H_{30} \right)
    +2\cot \theta  H_{20} \left(H_{20}-H_{40}\right)
\right.
\nonumber\ \\   & &  
\left.
    +2 H_{30, \theta \theta }
    -4 H_{20} H_{40, \theta }
    -2 \left( \frac{H_{30}}{\sin^2 \theta}-\cot \theta  H_{30, \theta } \right)
    -2 H_{30} H_{20}^2    
    -2 H_{40} \left( 2 H_{11}+H_{20, \theta }\right)\right]  
 +O(\delta^3) \nonumber\ , \\ 
& & \nonumber \\    
H_4(\delta,\theta ) & = &
 H_{40}-\frac{\delta^2}{8} 
\left[\left( 4 \gamma \varphi_{0,\theta}
+2 \frac{f_{2,\theta}}{f_{2}}-\frac{l_{2,\theta}}{l_{2}}\right) 
 \left(H_{40, \theta }+H_{30} H_{20}
 -\cot \theta \left(H_{20}-H_{40}\right)\right) 
    +H_{20} \left(4 H_{30, \theta }+2\right)
\right.
\nonumber\ \\   & &  
\left.
    +2 \left(H_{30} \left(2 H_{11}+H_{20, \theta }\right)
    +H_{40, \theta \theta }-H_{40} H_{20}^2\right)
    +2 \frac{H_{20}-H_{40}}{\sin^2 \theta}
    -2 \cot \theta \left(2 H_{11}-H_{40, \theta }-H_{20} H_{30}+H_{20, \theta }\right)
    \right] \nonumber \\ & &
 +O(\delta^3) \nonumber\ , \\
& & \nonumber \\    
\varphi(\delta,\theta ) & = & 
\varphi_0 -\frac{\delta^2}{8}
\left[\rule{0mm}{6mm}
     2 \varphi_{0,\theta} \cot \theta 
     +\varphi_{0,\theta} \frac{l_{2,\theta}}{l_{2}}+2 \varphi_{0,\theta \theta}
\right.
\nonumber\ \\   & &
\left.  
-\frac{f_{2}}{l_{2}} \frac{n^2}{x_{\rm H}^2} \gamma e^{2 \gamma \varphi_0} 
     \left[
     \rule{0mm}{6mm}4 \cot \theta                \left(
     H_{30} \left( H_{30, \theta }+1 \right)
     -H_{20} \left( H_{20} H_{30}+H_{40, \theta } \right)
     +H_{40} H_{40, \theta }      \right) 
\right. \right.
\nonumber\ \\   & &  
\left. \left.  
     +2\left(\rule{0mm}{6mm}
     \left( H_{30}^2+H_{40}^2-1 \right) H_{20}^2 
     +2 H_{20} \left( H_{30} H_{40, \theta }-H_{40} H_{30, \theta } \right)
     +H_{30, \theta }^2+2 H_{30, \theta }
     +H_{40, \theta }^2+1 
\right. \right. \right.
\nonumber\ \\   & &  
\left. \left.  \left. 
     +\frac{\left(H_{20}-H_{40} \right)^2}{\sin^2 \theta }
     +\frac{H_{30}^2}{\sin^2 \theta }-\left(H_{30}^2+H_{40}^2\right)   \right )
     \right]
\right]
 +O(\delta^3)  \nonumber\ . \\  
\label{exp_hall} \end{eqnarray} 
The expansion coefficients
$f_2$, $m_2$, $l_2$, $H_{11}$, $H_{20}$, $H_{30}$, $H_{40}$
and $\varphi_0$ are functions of the variable $\theta$.
The expansion depends on the gauge condition imposed on the
gauge field functions at the horizon. For the expansion above we 
employed the gauge condition (\ref{gfree2}),
$\left. (\partial_\theta H_1)\right|_{x=x_{\rm H}} = 0$.

The expansion of the functions at infinity is given in \cite{kk4}.

\subsection{Surface gravity}

Here we show that the surface gravity $\kappa_{\rm sg}$
of the static axially symmetric black hole solutions
is indeed constant at the horizon.
We employ the expansion of the metric functions
of the static axially symmetric black hole solutions 
at the horizon $x_{\rm H}$ in powers of 
$\delta$, eq.~(\ref{epsil}),
\begin{eqnarray}
f(x,\theta) & = & 
\delta^2 (1- \ \delta) \ f_2(\theta ) \ +O(\delta^4) \nonumber\ , \\
m(x,\theta) & = & 
\delta^2 (1-3\delta) \ m_2(\theta ) +O(\delta^4)\nonumber\ , \\
l(x,\theta) & = & 
\delta^2 (1-3\delta) \ l_2(\theta ) \ +O(\delta^4)\nonumber \ .\\
\label{exp_horz1} \end{eqnarray}  
Expanding the $r \theta$ component of the Einstein equations 
at the horizon, we obtain
a relation between the expansion coefficients
$f_2(\theta)$ and $m_2(\theta)$,
\begin{eqnarray}
0 & = & \frac{\partial_\theta m_2}{m_2} - 2 \frac{\partial_\theta f_2}{f_2}
\ . \nonumber\\
\label{exp_horz2}
\end{eqnarray}
Comparison with eqs.~(\ref{sg}), (\ref{temp}) then shows 
that the surface gravity is constant at the horizon.
 
\subsection{Kretschmann scalar}

Here we derive the expression for the Kretschmann scalar
$K$, eq.~(\ref{ks}),
for the static axially symmetric black hole solutions
and show that it is finite at the horizon.

The nonvanishing components of the Riemann tensor are 
\begin{eqnarray} 
R_{0r0r} & = & \frac{f}{4 r^2} 
\left[ \frac{f_{,\theta}}{f} \left(\frac{m_{,\theta}}{m}-\frac{f_{,\theta}}{f}\right) 
 -r^2 \left(\frac{f_{,r}}{f} \frac{m_{,r}}{m} - 2\frac{f_{,r,r}}{f}\right) \right] 
\nonumber \\ & &
\nonumber\\
 & =&- R_{0rr0} = -R_{r00r} =  R_{r0r0} \ \ ,
\nonumber \\ & &
\nonumber\\
R_{0r0\theta}  & = & -\frac{f}{4 r} 
\left[ \left( 2 + r \frac{m_{,r}}{m} - r \frac{f_{,r}}{f} \right) \frac{f_{,\theta}}{f} 
+ r \frac{f_{,r}}{f} \frac{m_{,\theta}}{m}-2 r \frac{f_{,r,\theta}}{f} \right]
\nonumber \\ & &
\nonumber\\
 & = &           - R_{0r\theta 0} = -R_{r00\theta } = R_{r0\theta 0}=
R_{0\theta 0r} =- R_{0\theta r0} = -R_{\theta 00r} = R_{\theta 0 r0} \ \ ,
\nonumber \\ & &
\nonumber\\
R_{0\theta 0\theta} & = & \frac{f}{4}
\left[ r \frac{f_{,r}}{f} \left( 2 -r \frac{f_{,r}}{f}+ r \frac{m_{,r}}{m } \right)
-\frac{f_{,\theta}}{f} \frac{m_{,\theta}}{m} + 2 \frac{f_{,\theta ,\theta}}{f}\right]
\nonumber \\ & &
\nonumber\\
 & = &- R_{0\theta \theta 0}= - R_{\theta 0 0\theta}= R_{\theta 0 \theta 0} \ \ ,
\nonumber \\ & &
\nonumber\\
R_{0\phi 0\phi} & = &-\frac{f l }{4 m}  \sin^2 \theta 
\left[ r\frac{f_{,r}}{f} \left( r \frac{f_{,r}}{f} - r \frac{l_{,r}}{l} - 2\right)
+\frac{f_{,\theta}}{f} 
\left(\frac{f_{,\theta}}{f} -\frac{l_{,\theta}}{l} -2 \cot \theta\right)\right]
\nonumber \\ & &
\nonumber\\
 & = &- R_{0\phi \phi 0}= - R_{\phi 0 0\phi}= R_{\phi 0 \phi0} \ \ ,
\nonumber \\ & &
\nonumber\\
R_{r\theta r\theta} & = &-\frac{m}{2 f} 
\left[r \frac{m_{,r}}{m} - r \frac{f_{,r}}{f}  
+ r^2 \left(\frac{m_{,r,r}}{m} - \frac{f_{,r,r}}{f}\right)
+ \frac{m_{,\theta ,\theta}}{m} - \frac{f_{,\theta ,\theta}}{f} \right.
\nonumber\\
& & 
\left.
-\left( \left(r\frac{m_{,r}}{m}\right)^2 - \left(r\frac{f_{,r}}{f}\right)^2\right)
-\left( \left(\frac{m_{,\theta }}{m}\right)^2 -\left(\frac{f_{,\theta }}{f}\right)^2\right)\right]
\nonumber \\ & &
\nonumber\\
 & =& -R_{r\theta \theta r}= - R_{\theta r r\theta}= R_{\theta r \theta r} \ \ ,
\nonumber \\ & &
\nonumber\\
R_{r\phi r\phi} & = & -\frac{l}{4 f} \sin^2 \theta 
\left[ \left(\frac{l_{,\theta}}{l} - \frac{f_{,\theta}}{f} + 2 \cot \theta\right)
\left(\frac{m_{,\theta}}{m} - \frac{f_{,\theta}}{f}\right)
-\left(r \frac{l_{,r}}{l} - r \frac{f_{,r}}{f}+2 \right)
\left(r \frac{m_{,r}}{m} +3 r \frac{f_{,r}}{f}\right) \right.
\nonumber\\
& & 
\left.
+2\left( r^2 \frac{l_{,r,r}}{l} + 4 r \frac{l_{,r}}{l}- r^2 \frac{f_{,r,r}}{f} + 2\right)
-\left( r \frac{l_{,r}}{l} - r \frac{f_{,r}}{f} + 2\right)^2\right]
\nonumber \\ & &
\nonumber\\
 & =&  - R_{r\phi\phi r}= - R_{\phi r r\phi} = R_{\phi r \phi r} \ \ ,
\nonumber \\ & &
\nonumber\\
R_{\theta \phi \theta \phi} & = & - \frac{l}{4 f} r^2 \sin^2 \theta
\left[ 2 \cot \theta 
\left( 2 \frac{l_{,\theta}}{l} -\frac{m_{,\theta}}{m} -\frac{f_{,\theta}}{f} \right)
+2 r \frac{l_{,r}}{l }+ 2 r \frac{m_{,r}}{m} -4 r \frac{f_{,r}}{f}  \right.
\nonumber\\
& & 
\left.
+ r \frac{m_{,r}}{m} r \frac{l_{,r}}{l}
- r \frac{f_{,r}}{f} r \frac{l_{,r}}{l}
- r \frac{m_{,r}}{m} r \frac{f_{,r}}{f}
+ \left( r \frac{f_{,r}}{f}\right)^2 
+ 2 \frac{l_{,\theta ,\theta}}{l}-\left(\frac{l_{,\theta}}{l}\right)^2
- 2 \frac{f_{,\theta ,\theta}}{f}+2 \left(\frac{f_{,\theta}}{f}\right)^2 \right.
\nonumber\\
& & 
\left.
-\frac{m_{,\theta}}{m} \frac{l_{,\theta}}{l} 
-\frac{f_{,\theta}}{f} \frac{l_{,\theta}}{l} 
+\frac{m_{,\theta}}{m} \frac{f_{,\theta}}{f} \right]
\nonumber \\ & &
\nonumber\\
 & = &- R_{\theta \phi \phi \theta } = - R_{ \phi \theta \theta \phi} = R_{\phi \theta \phi \theta } \ \ ,
\nonumber \\ & &
 \nonumber\\
R_{r \phi \theta \phi} & = & -\frac{l}{4f} r  \sin^2 \theta
\left[ 2 \cot \theta \left( r \frac{l_{,r}}{l} - r \frac{m_{,r}}{m}\right)
+2\left( r \frac{l_{,r, \theta}}{l}- r \frac{f_{,r, \theta}}{f}\right) 
-r \frac{l_{,r}}{l} \frac{l_{,\theta}}{l}
\right. 
\nonumber\\
& & 
\left.
-\left(r \frac{m_{,r}}{m} \left( \frac{l_{,\theta}}{l} - \frac{f_{,\theta}}{f}\right)
 +\frac{m_{,\theta}}{m}\left(r \frac{l_{,r}}{l}- r \frac{f_{,r}}{f}\right)\right)
-2 \left( \frac{m_{,\theta}}{m} - \frac{f_{,\theta}}{f}\right) 
+ r \frac{f_{,r}}{f} \frac{f_{,\theta}}{f} \right]
\nonumber \\ & &
\nonumber\\
 & = &- R_{r \phi  \phi \theta } = - R_{\phi r  \theta \phi}= R_{\phi r \phi \theta }
= R_{\theta \phi r \phi}= - R_{\theta \phi \phi r}= -R_{ \phi \theta  r \phi}= R_{ \phi \theta \phi r} \ \ .
\nonumber\\ \nonumber
\end{eqnarray}
The Kretschmann scalar $K=R_{\mu \nu \lambda \rho} R^{\mu \nu \lambda \rho}$ is then given by
\begin{eqnarray}
K & = &
4 \left[ 
   \left(R_{0r0r} g^{00} g^{rr} \right)^2 
+  \left(R_{0\theta 0\theta} g^{00} g^{\theta \theta } \right)^2
+  \left(R_{0\phi 0 \phi} g^{00} g^{\phi \phi } \right)^2
+  \left(R_{r\theta r\theta} g^{rr} g^{\theta \theta } \right)^2
+  \left(R_{r\phi r\phi } g^{rr} g^{\phi \phi } \right)^2
+  \left(R_{\theta \phi \theta \phi} g^{\theta \theta } g^{\phi \phi} \right)^2
\right]
\nonumber \\
& &
+ 8 \left[
     \left(R_{0r0\theta} g^{00} \right)^2 
   + \left(R_{r \phi \theta \phi } g^{\phi \phi } \right)^2 \right] g^{rr} g^{\theta \theta } \ .
\nonumber \\ \nonumber
\end{eqnarray}
In order to show that the Kretschmann scalar is finite at the horizon,
we insert the expansion of the metric
functions eq.~(\ref{exp_horz1}) into the Riemann tensor.
In terms like 
($\frac{f_{,r}}{f}-\frac{m_{,r}}{m}$) and
($\frac{f_{,r,r}}{f}-\frac{m_{,r,r}}{m}$)
the divergences cancel, because 
the lowest terms in the expansion are of the same order in $r-r_{\rm H}$
for all metric functions.
The terms 
$( 2 + r \frac{m_{,r}}{m} - r \frac{f_{,r}}{f} )$ and 
$( 2 + r \frac{l_{,r}}{l} - r \frac{f_{,r}}{f} )$ are proportional to 
$r-r_{\rm H}$, and the term
$( r^2 \frac{l_{,r,r}}{l} + 4 r \frac{l_{,r}}{l}
 - r^2 \frac{f_{,r,r}}{f} + 2)$ 
is finite at the horizon.
The only term left, which could be divergent at the horizon, is  
$(r \frac{f_{,r}}{f} \frac{m_{,\theta}}{m}-2 r \frac{f_{,r,\theta}}{f})$.
However, this becomes
$\frac{2 r}{r- r_{\rm H}} 
 (\frac{m_{2,\theta}}{m_2}-2\frac{f_{2,\theta}}{f_2})$, 
which does not diverge because of relation (\ref{exp_horz2}).

\section{Appendix B}

Here we discuss our final choice of functions for the numerical
integration.

At the horizon the expansion coefficients
$f_2(\theta)$, $m_2(\theta)$ and $l_2(\theta)$
are of particular interest, because they enter into the expressions for the
temperature, the area and the circumferences. 
In order to obtain these functions directly
we have introduced the new functions $\bar{f}(\bar{x},\theta)$, 
$\bar{m}(\bar{x},\theta)$ and $\bar{l}(\bar{x},\theta)$,
\begin{equation}
f(\bar{x},\theta) = \bar{x}^2 \ \bar{f}(\bar{x},\theta)\ ,\ \ 
m(\bar{x},\theta) = \bar{x}^2 \ \bar{m}(\bar{x},\theta)\ ,\ \  
l(\bar{x},\theta) = \bar{x}^2 \ \bar{l}(\bar{x},\theta) \ ,
\end{equation}
where $\bar{x}= (1 -\frac{ x_{\rm H}}{x})$ 
is the compactified coordinate, eq.~(\ref{barx}).
The functions  $f_2(\theta)$, $m_2(\theta)$ and $l_2(\theta)$
are then given by 
\begin{equation}
f_2(\theta)  = \frac{1}{x_{\rm H}^2}\ \bar{f}(0,\theta)\ ,\ \   
m_2(\theta)  = \frac{1}{x_{\rm H}^2}\ \bar{m}(0,\theta)\ ,\ \  
l_2(\theta)  = \frac{1}{x_{\rm H}^2}\ \bar{l}(0,\theta)\ .  
\end{equation}

In the limit $x \rightarrow \infty$ the variable $\bar{x}$ approaches the 
value $1$. Consequently, at infinity
the boundary conditions for the new functions
$\bar{f}$, $\bar{m}$ and $\bar{l}$ coincide with the boundary conditions
of the functions $f$, $m$ and $l$, respectively.
At the horizon the boundary conditions for the new functions can be 
obtained from the expansion of the metric functions. They are given by
\begin{equation}
\left. (\bar{f}  -\partial_{\bar{x}} \bar{f})\right|_{\bar{x}=0}  =  0 \nonumber\ ,\ \ 
\left. (\bar{m}  +\partial_{\bar{x}} \bar{m})\right|_{\bar{x}=0}  =  0 \nonumber\ ,\ \ 
\left. (\bar{l}  +\partial_{\bar{x}} \bar{l})\right|_{\bar{x}=0}  =  0 \nonumber \ . 
\end{equation}

To satisfy the regularity condition (\ref{lm}) exactly
in the numerical calculations, we have introduced
the function $g(\bar{x},\theta)$,
\begin{equation}
g(\bar{x},\theta)=\frac{\bar{m}(\bar{x},\theta)}{\bar{l}(\bar{x},\theta)}
\ . \end{equation}
On the symmetry axis this function satisfies the boundary condition 
\begin{equation}
g|_{\theta=0}=1 
\   \end{equation}
and at the horizon
\begin{equation}
\partial_{\bar{x}} {g}|_{\bar{x}=0} =  0
\ . \end{equation}

The numerical calculations were mostly performed with the 
functions $\bar{f}(\bar{x},\theta)$, $g(\bar{x},\theta)$ and 
$\bar{l}(\bar{x},\theta)$.

\newpage
\begin{table}[p!]
\begin{center}
\begin{tabular}{|c|ccc|} \hline
\multicolumn{1}{|c|} { $ $ }&
\multicolumn{3}{ c|} {  EYMD$(\gamma=1)$ }\\
 \hline
 $x_{\rm H}$      &  k=$1$,  n=$2$ &   k=$1$, n=$4$  &  k=$2$,  n=$2$ \\   
 \hline
\multicolumn{1}{|c|} { $ $ }&
\multicolumn{3}{ c|}  {$\mu$ } \\
 \hline 
 $0.01$        &  $0.9766$ &   $1.6227$ &  $1.2762$ \\ 
 $0.1 $        &  $1.1183$ &   $1.7700$ &  $1.4081$ \\  
 $1.  $        &  $2.5649$ &   $3.2193$ &  $2.5649$ \\  
 $10  $        &  $20.079$ &   $20.244$ &  $20.079$ \\
 \hline
\multicolumn{1}{|c|} { $ $ }&
\multicolumn{3}{ c|}  {$T$ } \\
 \hline 
 $0.01$        &  $0.583$ &   $0.989$ &  $0.230$ \\ 
 $0.1 $        &  $0.727$ &   $0.076$ &  $0.039$ \\  
 $1.  $        &  $0.016$ &   $0.013$ &  $0.016$ \\  
 $10. $        &  $0.002$ &   $0.002$ &  $0.002$ \\
\hline
\multicolumn{1}{|c|} { $ $ }&
\multicolumn{3}{ c|}  {$S$ } \\
 \hline 
 $0.01$        &  $0.014$ &   $0.012$ &  $0.033$ \\ 
 $0.1 $        &  $1.167$ &   $1.116$ &  $2.243$ \\  
 $1.  $        &  $62.47$ &   $71.25$ &  $62.47$ \\  
 $10. $        &  $5045.$ &   $5081.$ &  $5045.$ \\
 \hline 
\multicolumn{1}{|c|} { $ $ }&
\multicolumn{3}{ c|}  {$D$ } \\
 \hline 
 $0.01$        &  $0.9609$ &   $1.6063$ &  $1.2619$ \\ 
 $0.1 $        &  $0.9487$ &   $1.6011$ &  $1.2319$ \\  
 $1.  $        &  $0.6022$ &   $1.3188$ &  $0.6023$ \\  
 $10. $        &  $0.0836$ &   $0.2682$ &  $0.0836$ \\
 \hline 
\multicolumn{1}{|c|} { $ $ }&
\multicolumn{3}{ c|}  {$L_e/L_p$ } \\
 \hline 
 $0.01$        &  $0.9999$ &   $1.0000$ &  $0.9993$ \\ 
 $0.1 $        &  $0.9974$ &   $0.9997$ &  $0.9910$ \\  
 $1.  $        &  $0.9980$ &   $0.9968$ &  $0.9980$ \\  
 $10. $        &  $1.0000$ &   $0.9998$ &  $1.0000$ \\
 \hline 
\end{tabular}
\end{center} 
\vspace{1.cm} 
{\bf Table 1}\\
The dimensionless mass $\mu$,
the temperature $T$, the entropy $S$, the dilaton charge $D$
and the ratio $L_e/L_p$
of the EYMD black hole solutions 
with node number $k=1$ and winding
number $n=2$, $k=1$ and $n=4$, as well as $k=2$ and $n=2$ 
are shown for the values 
of the horizon radius $x_{\rm H}=0.01$, $0.1$, $1$ and $10.$ 

\end{table}

\newpage

\begin{figure}
\centering
\vspace{-1cm}
\mbox{  \epsfysize=12cm \epsffile{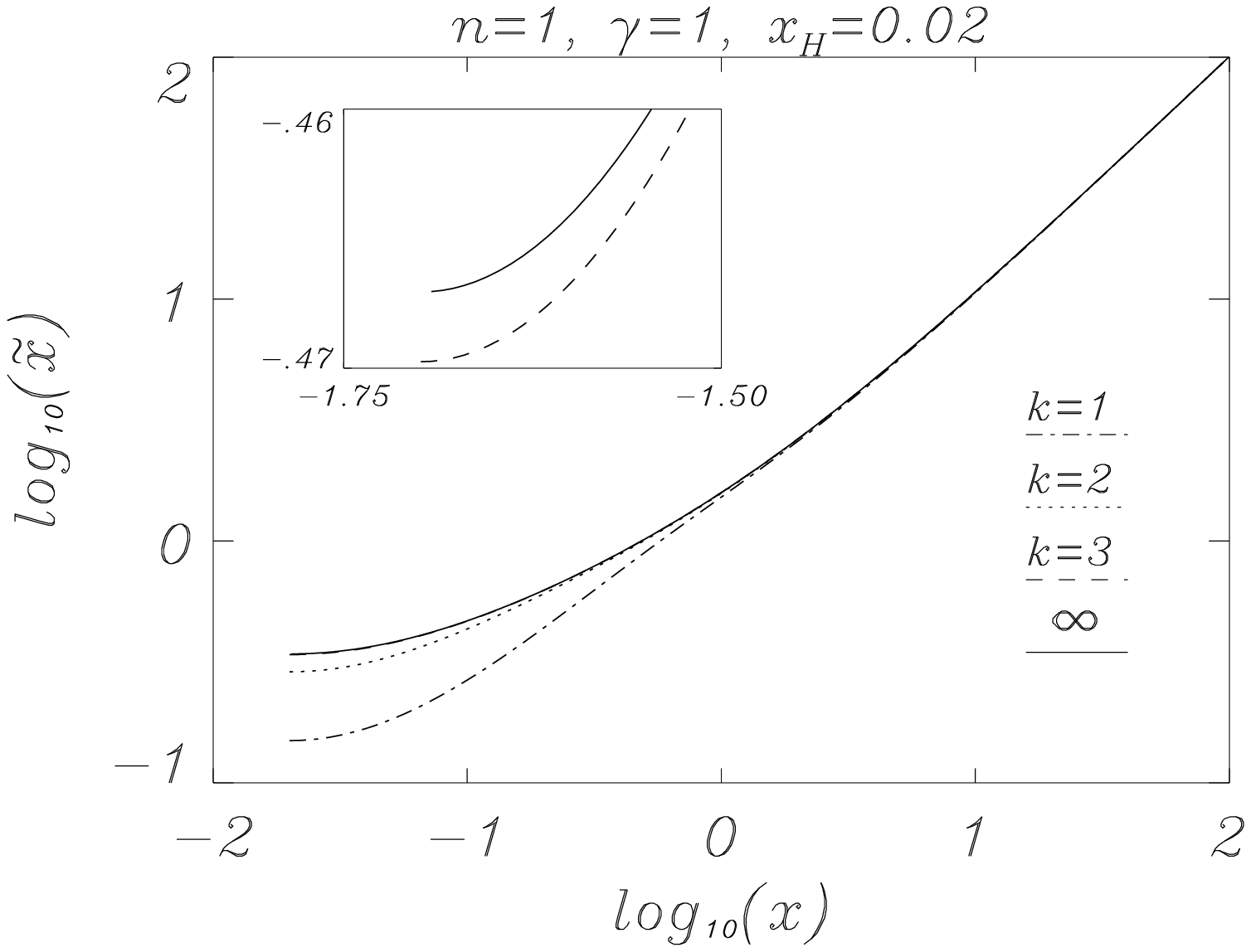}}\\
\end{figure}
\noindent Fig.~1\\
The coordinate transformation between the isotropic coordinate $x$
and the Schwarzschild-like coordinate $\tilde x$
is shown for the static spherically symmetric solutions ($n=1$)
of EYMD theory with $\gamma=1$ and $k=1-3$.
Also shown is the coordinate transformation for the limiting EMD solution.
\clearpage
\newpage
\begin{figure}
\centering
\vspace{-1cm}
\mbox{  \epsfysize=9cm \epsffile{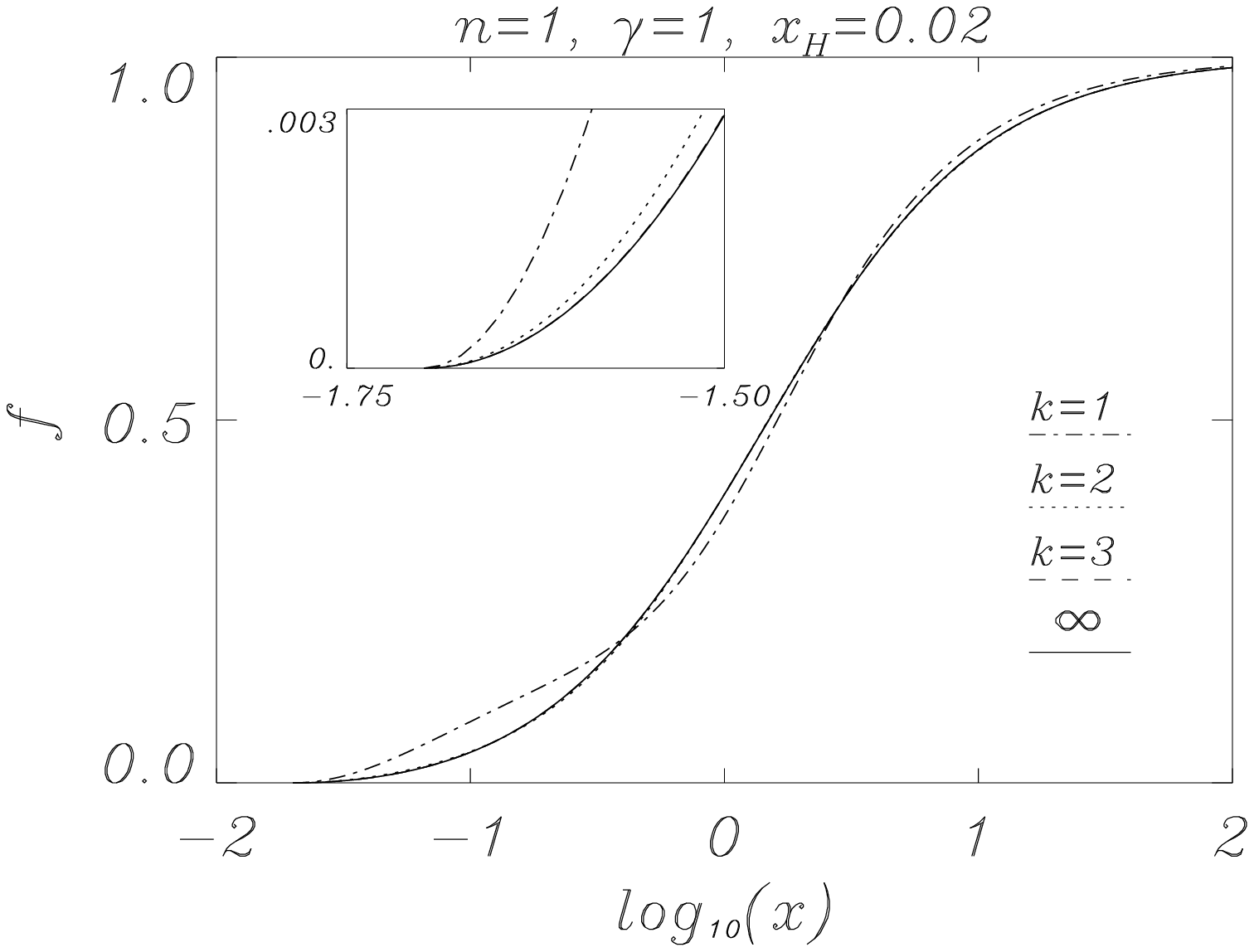}}\\
\end{figure}
\noindent Fig.~2a\\
The metric function $f$ is shown 
for the static spherically symmetric solutions ($n=1$)
of EYMD theory with $\gamma=1$ and $k=1-3$.
Also shown is the metric function of the limiting EMD solution.
\vspace{1cm}\\
\begin{figure}
\centering
\vspace{-1cm}
\mbox{  \epsfysize=9cm \epsffile{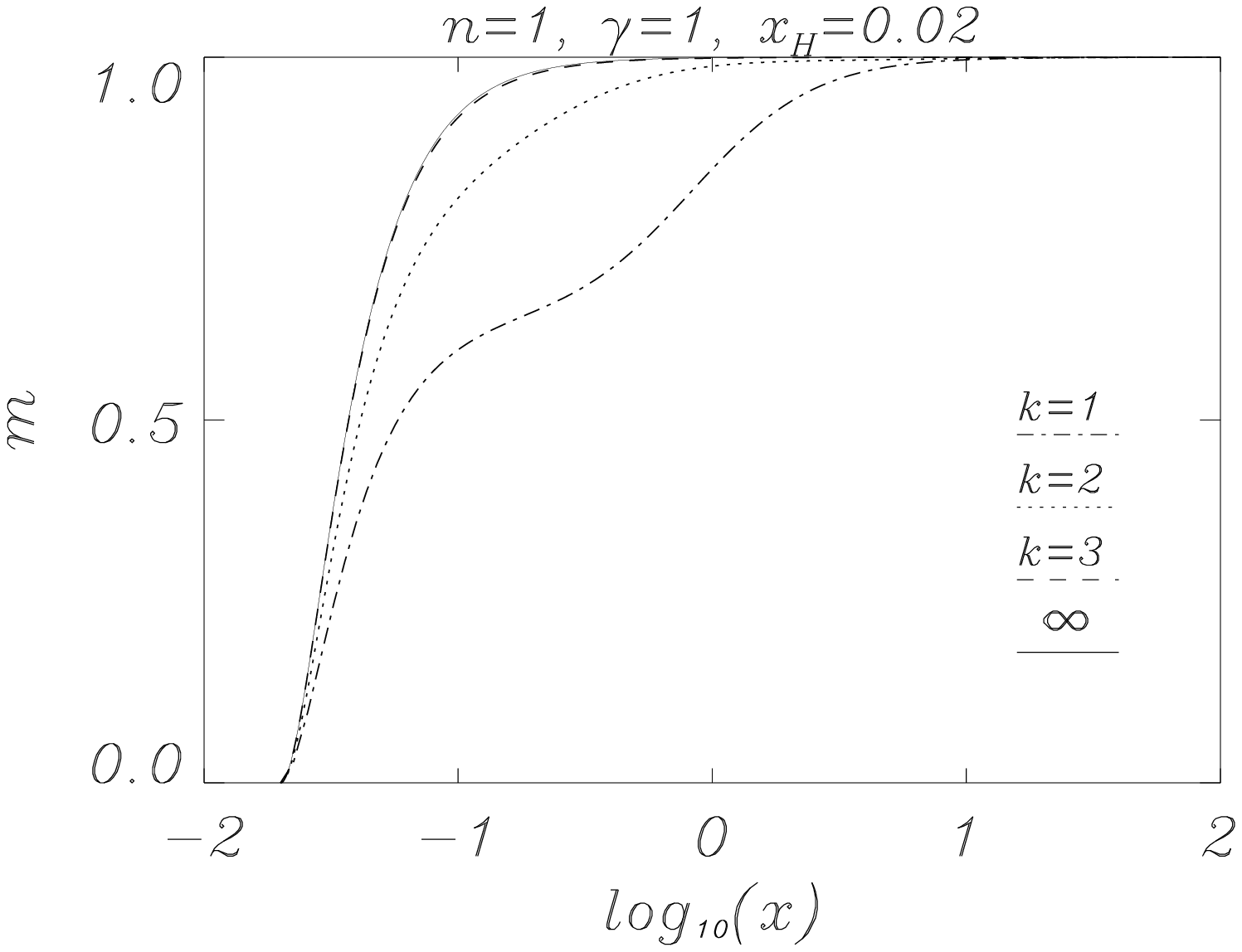}}\\
\end{figure}
\noindent Fig.~2b\\
Same as Fig.~2a for the metric function $m$.
\clearpage
\newpage
\begin{figure}
\centering
\vspace{-1cm}
\mbox{  \epsfysize=9cm \epsffile{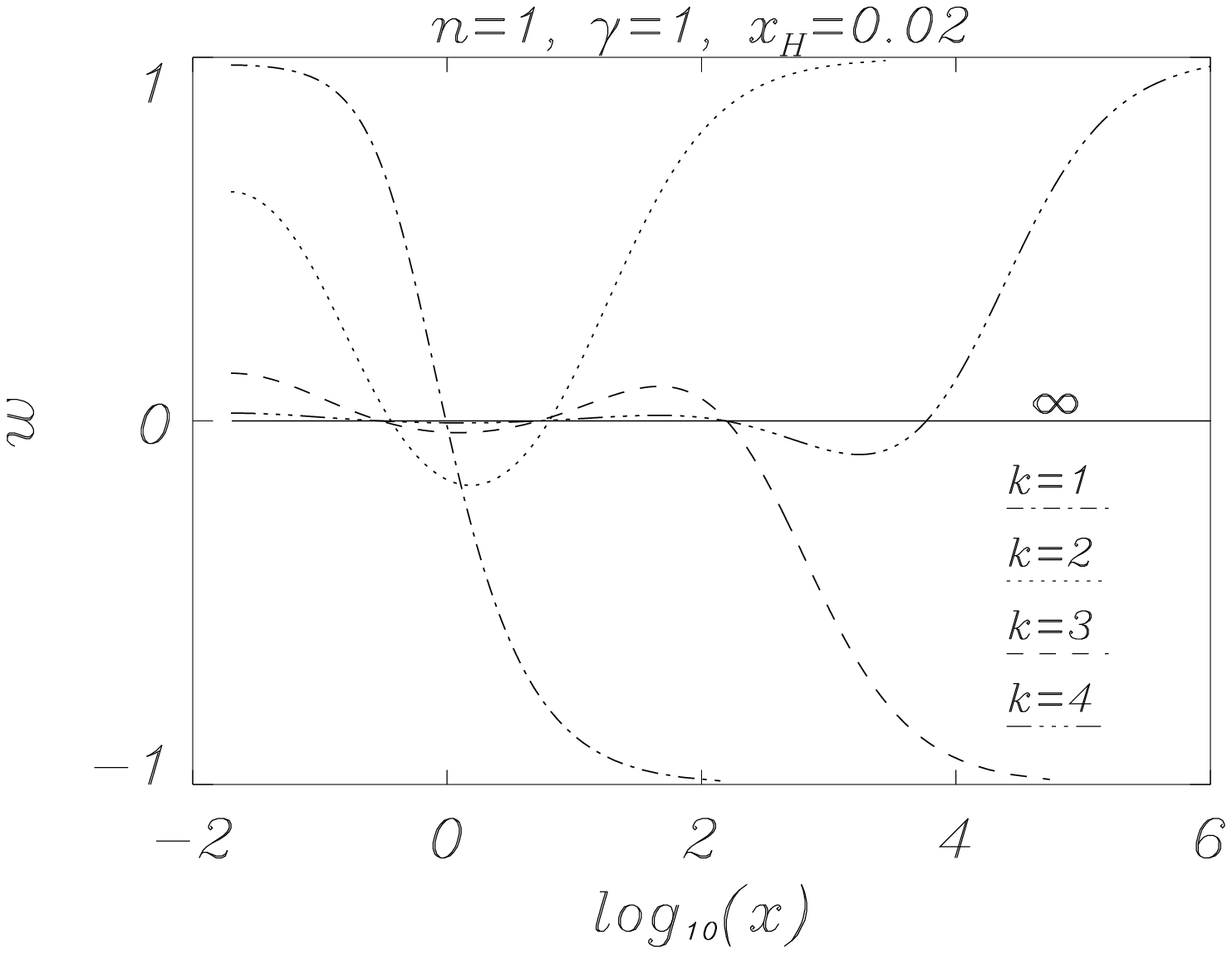}}\\
\end{figure}
\noindent Fig.~3a\\
Same as Fig.~2a for the gauge field function $w$, for $k=1-4$.
\vspace{1.cm}\\
\begin{figure}
\centering
\vspace{-1cm}
\mbox{  \epsfysize=9cm \epsffile{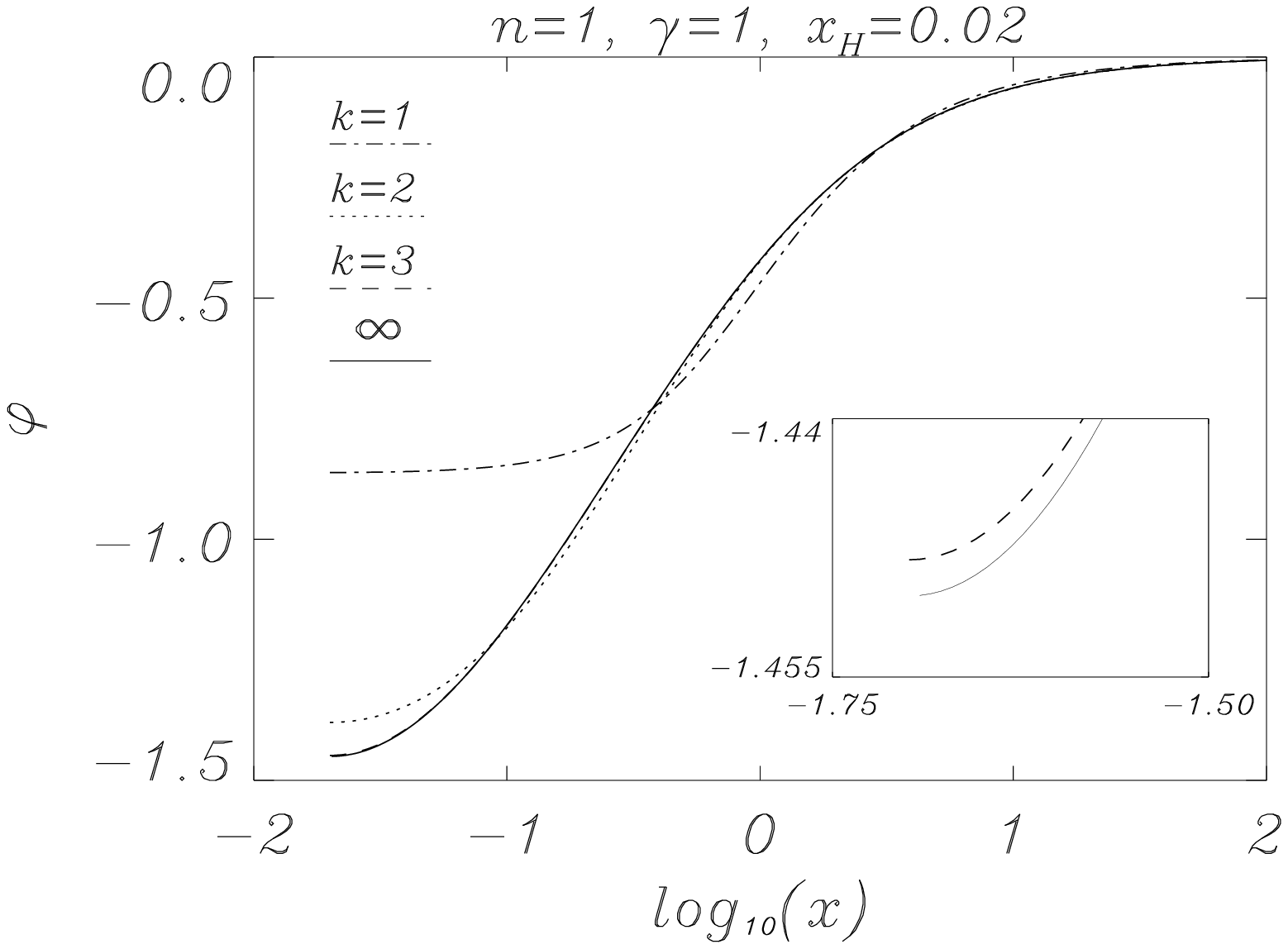}}\\
\end{figure}
\noindent Fig.~3b\\
Same as Fig.~2a for the dilaton function $\varphi$.
\clearpage
\newpage
\begin{figure}
\centering
\vspace{-1cm}
\mbox{  \epsfysize=12cm \epsffile{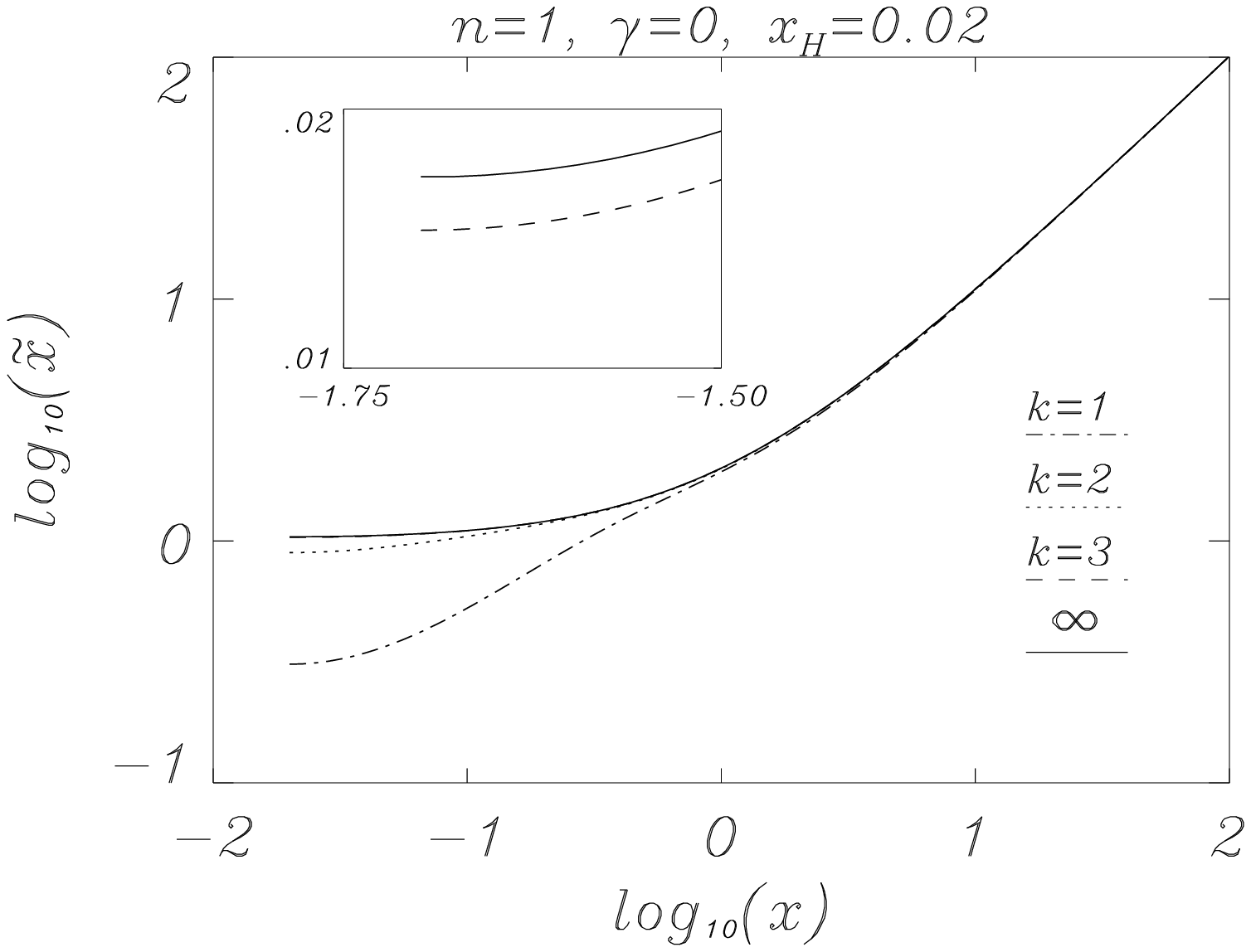}}\\
\end{figure}
\noindent Fig.~4\\
The coordinate transformation between the isotropic coordinate $x$
and the Schwarzschild-like coordinate $\tilde x$
is shown for the static spherically symmetric solutions ($n=1$)
of EYM theory with $k=1-3$.
Also shown is the coordinate transformation for the limiting RN solution.
\clearpage
\newpage
\begin{figure}
\centering
\vspace{-1cm}
\mbox{  \epsfysize=9cm \epsffile{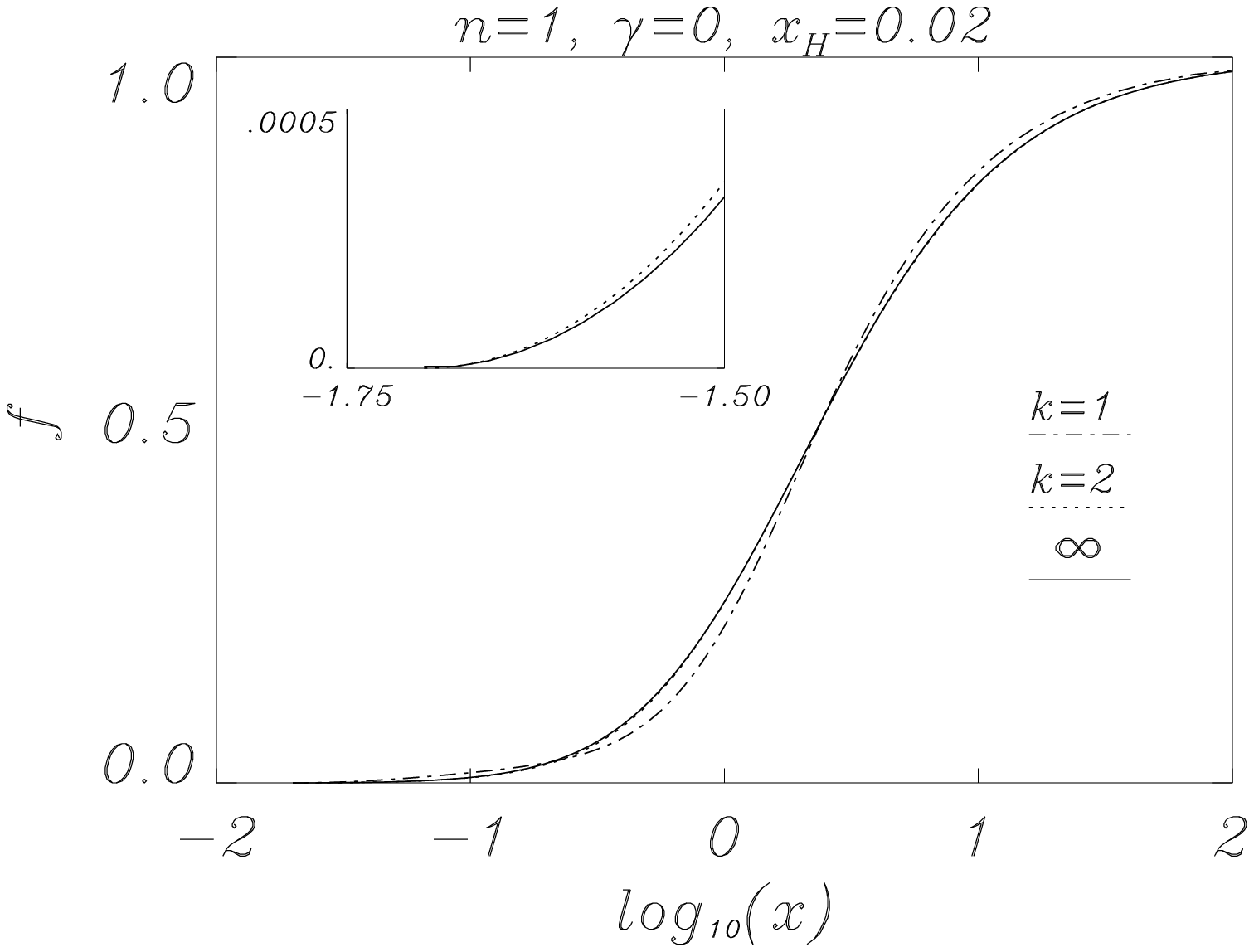}}\\
\end{figure}
\noindent Fig.~5a\\
The metric function $f$ is shown 
for the static spherically symmetric solutions ($n=1$)
of EYM theory with $k=1-2$.
Also shown is the metric function of the limiting RN solution.
\vspace{1.cm}\\ 
\begin{figure}
\centering
\vspace{-1cm}
\mbox{  \epsfysize=9cm \epsffile{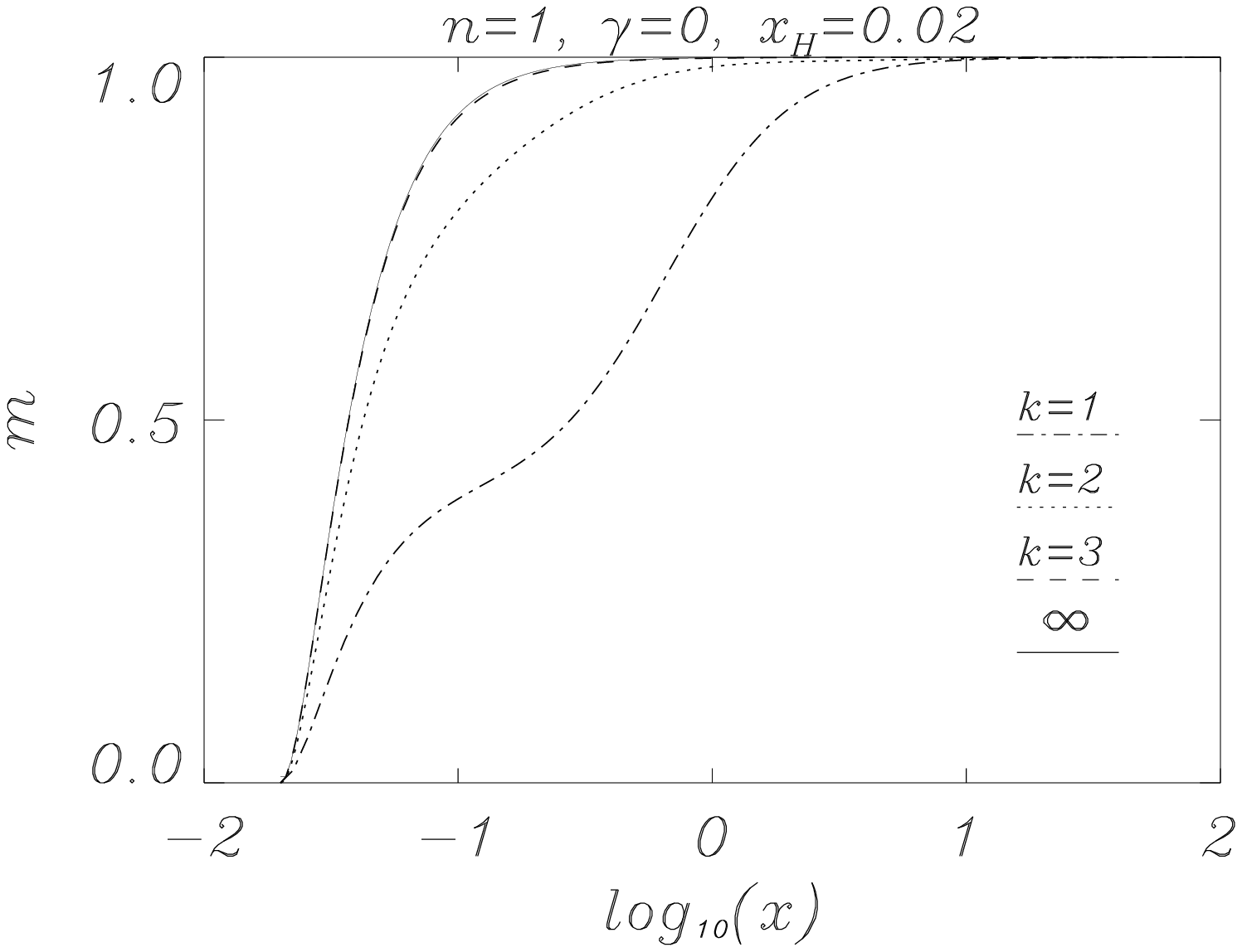}}\\
\end{figure}
\noindent Fig.~5b\\
Same as Fig.~5a for the metric function $m$ for $k=1-3$ .
\clearpage
\newpage
\begin{figure}
\centering
\vspace{-1cm}
\mbox{  \epsfysize=9cm \epsffile{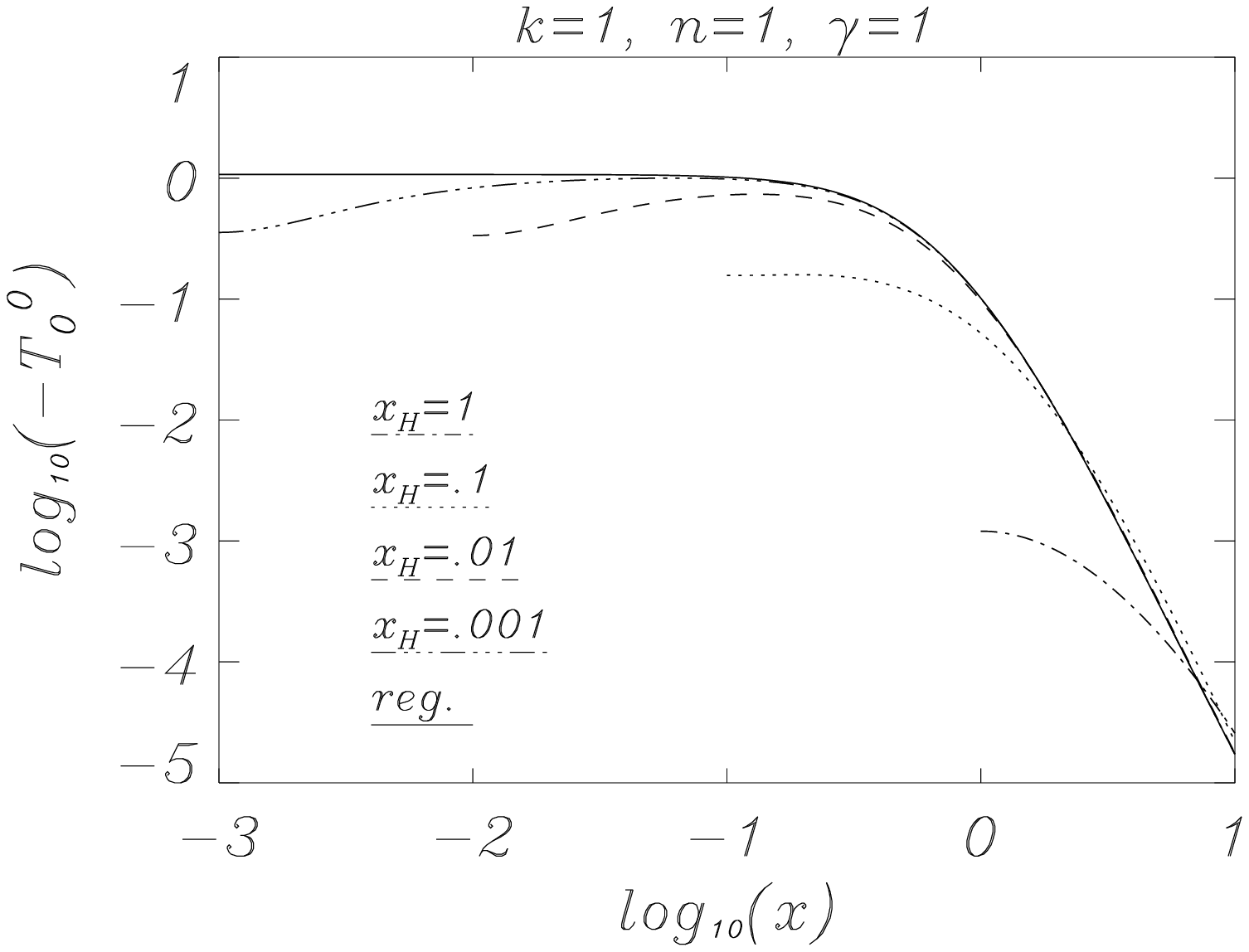}}\\
\end{figure}
\noindent Fig.~6a\\
The energy density $\epsilon=-T_0^0$ 
is shown for the static spherically symmetric black hole solutions ($n=1$)
of EYMD theory with $\gamma=1$ and $k=1$
and the horizon radii $x_{\rm H}=1$, 0.1, 0.01 and 0.001,
as well as for the corresponding globally regular solution.
\vspace{1.cm}\\
\begin{figure}
\centering
\vspace{-1cm}
\mbox{  \epsfysize=9cm \epsffile{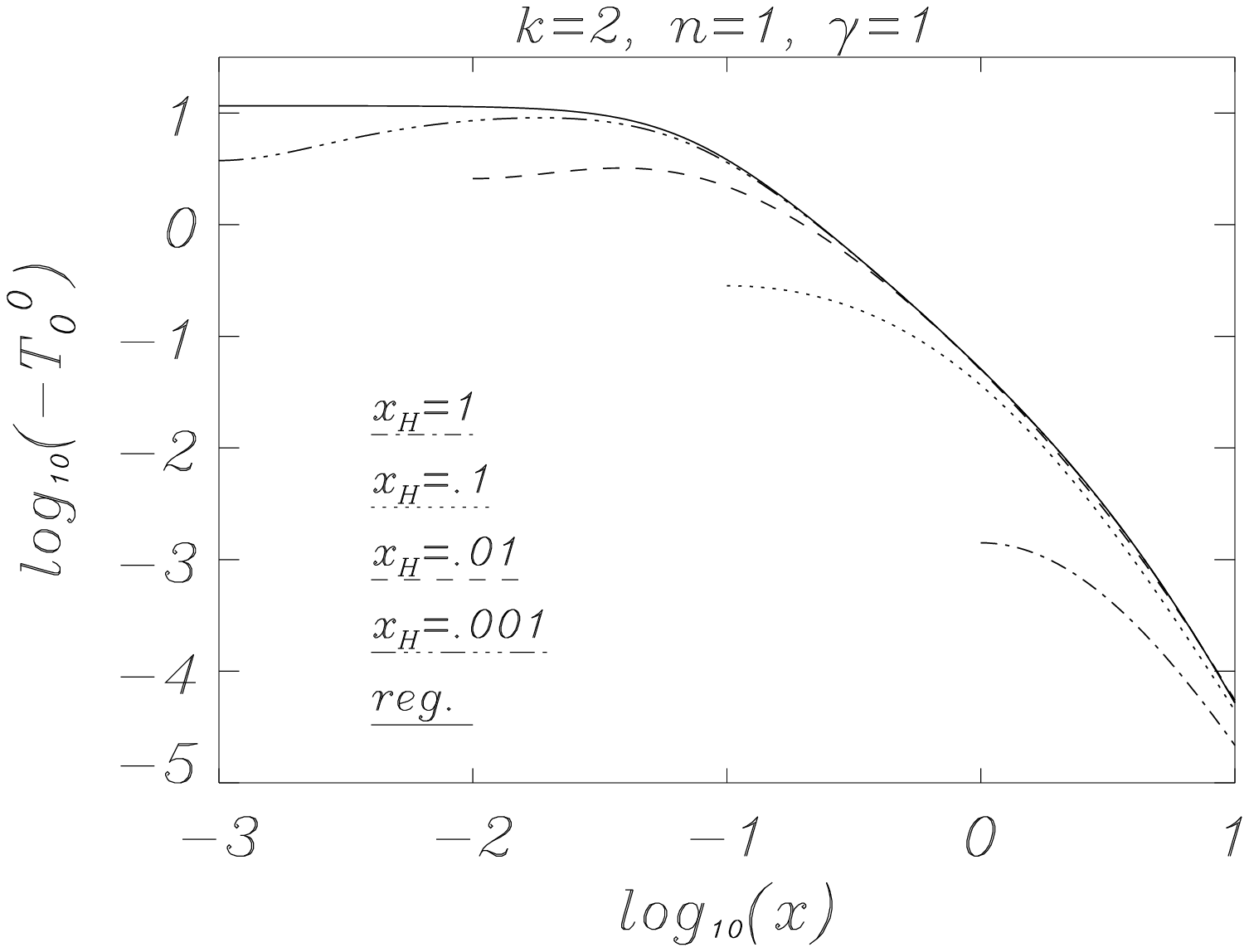}}\\
\end{figure}
\noindent Fig.~6b\\
Same as Fig.~6a for $k=2$.
\clearpage
\newpage
\begin{figure}
\centering
\vspace{-1cm}
\mbox{  \epsfysize=12cm \epsffile{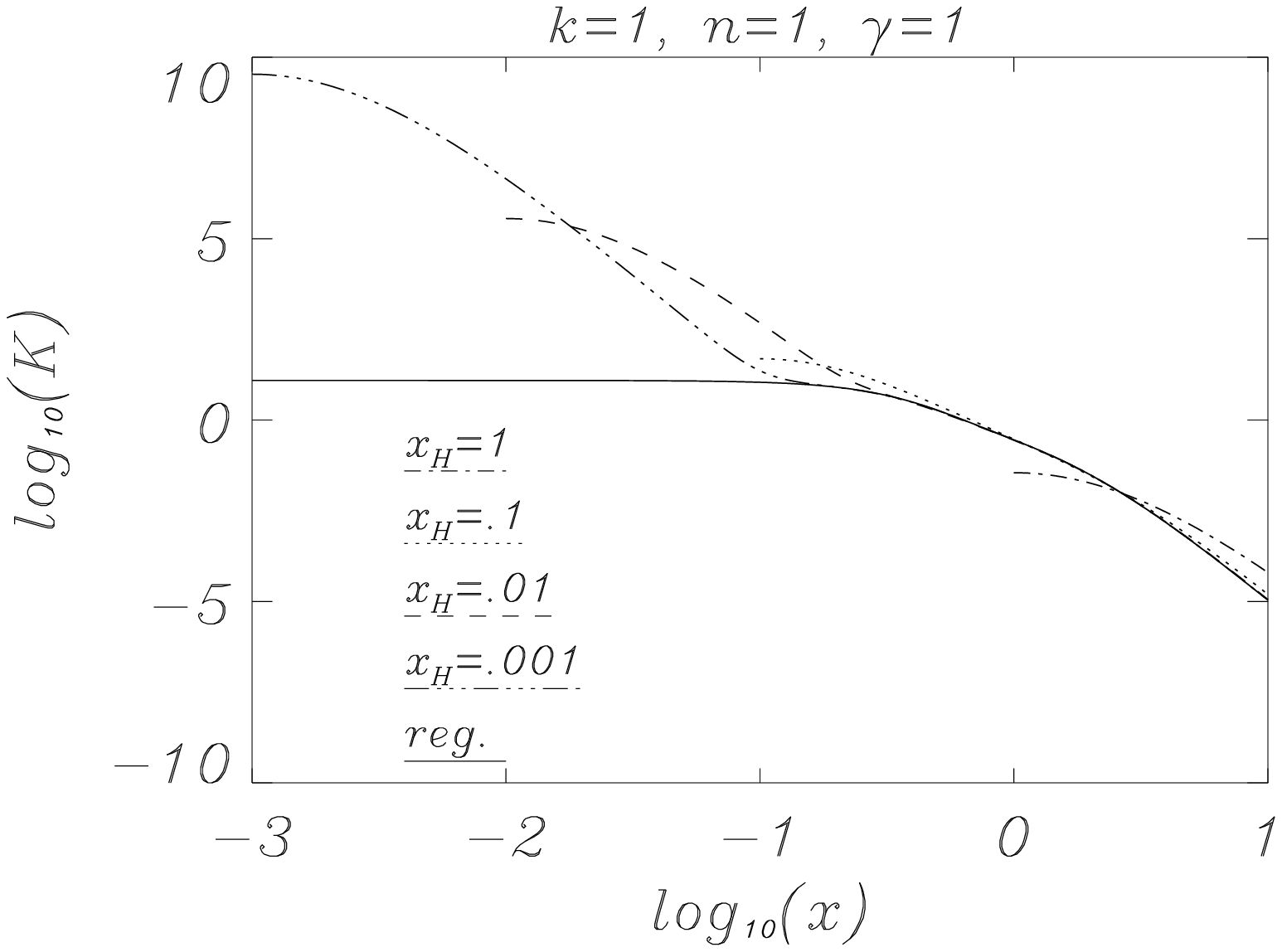}}\\
\end{figure}
\noindent Fig.~7\\
The Kretschmann scalar
is shown for the static spherically symmetric black hole solutions ($n=1$)
of EYMD theory with $\gamma=1$ and $k=1$
and the horizon radii $x_{\rm H}=1$, 0.1, 0.01 and 0.001,
as well as for the corresponding globally regular solution.
\clearpage
\newpage
\begin{figure}
\centering
\vspace{-1cm}
\mbox{  \epsfysize=10cm \epsffile{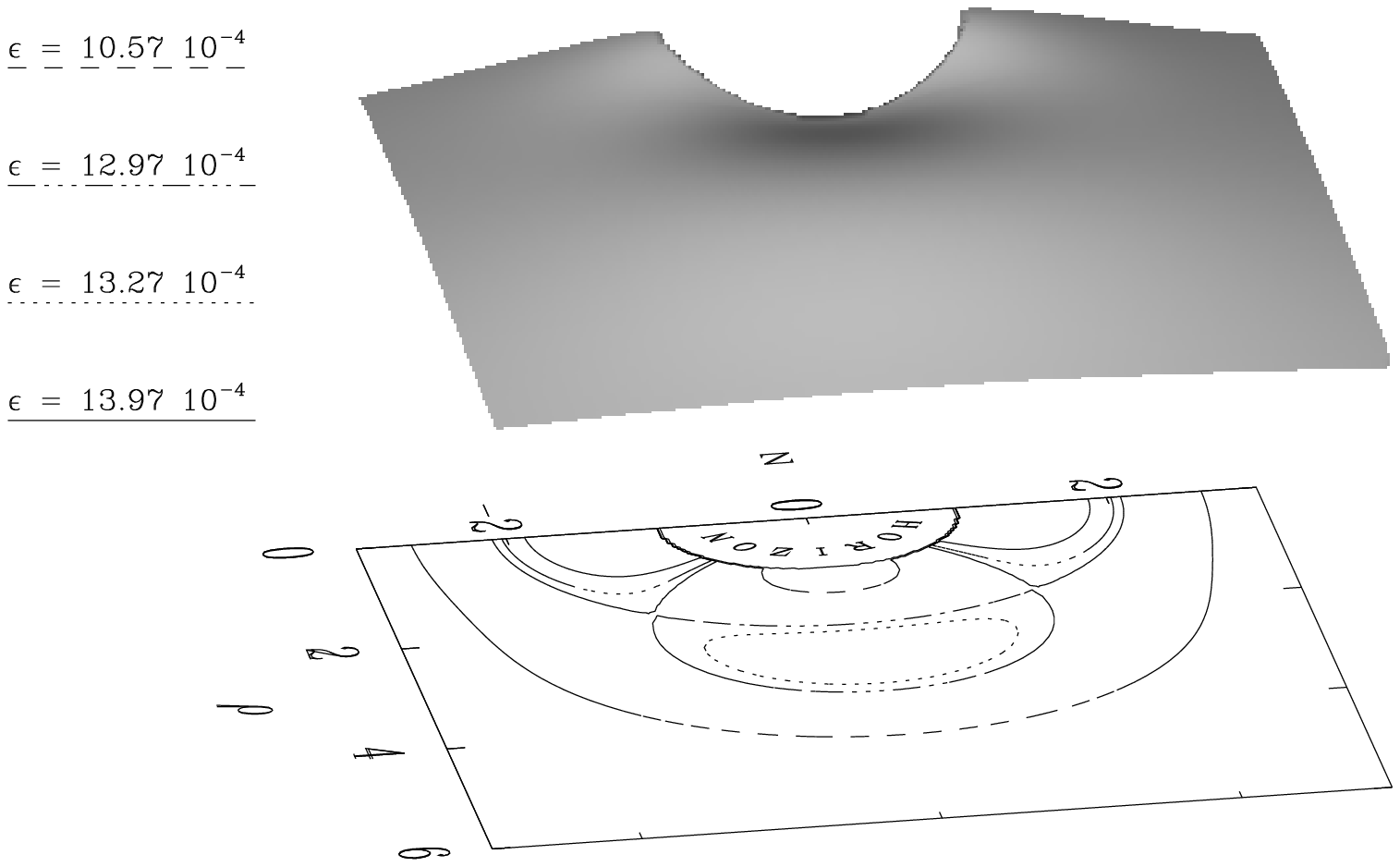}}
\end{figure}
\noindent Fig.~8a\\
The energy density $\epsilon=-T_0^0$ 
is shown for the static axially symmetric black hole solution 
of EYMD theory with $\gamma=1$, $n=2$ and $k=1$
and the horizon radius $x_{\rm H}=1$
in a 3-dimensional plot and a contour plot
with axes $\rho$ and $z$.
\vspace{1cm}\\
\begin{figure}
\centering
\vspace{-1cm}
\mbox{ 
\epsfysize=3cm \epsffile{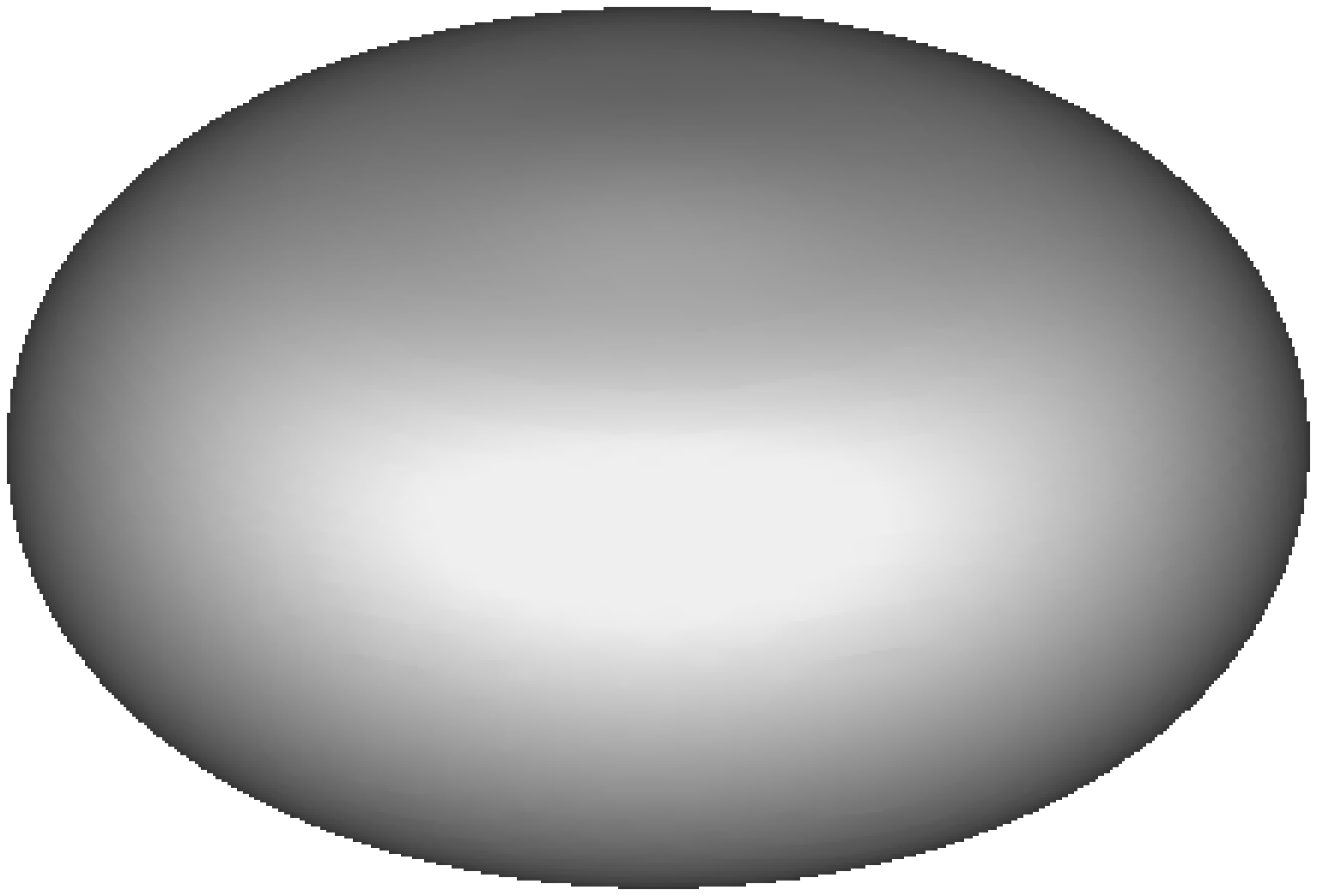}
\hspace{3.cm}\epsfysize=3cm \epsffile{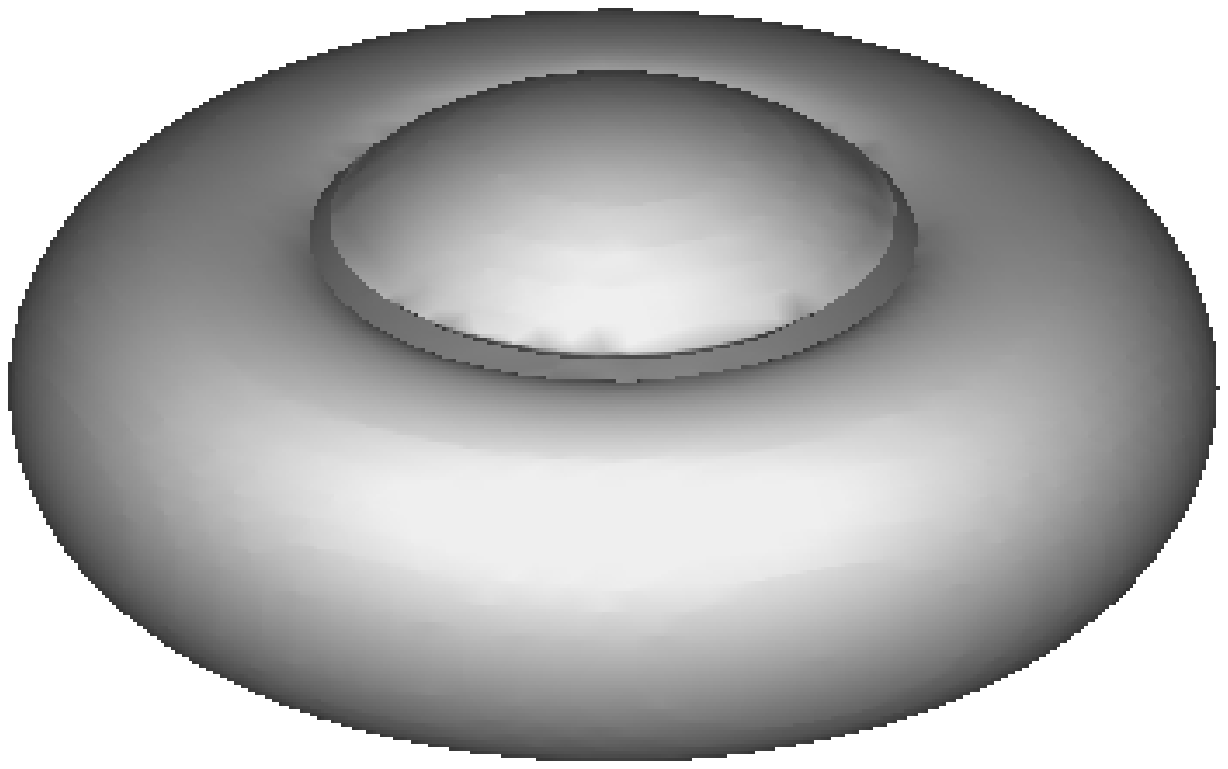}
} \\
\mbox{ $\varepsilon= 10.57 \ 10^{-4}$}\hspace{5cm}
\mbox{ $\varepsilon= 12.97 \ 10^{-4}$} 
\vspace{1cm}\\
\mbox{ 
\hspace*{-1cm} \epsfysize=3cm \epsffile{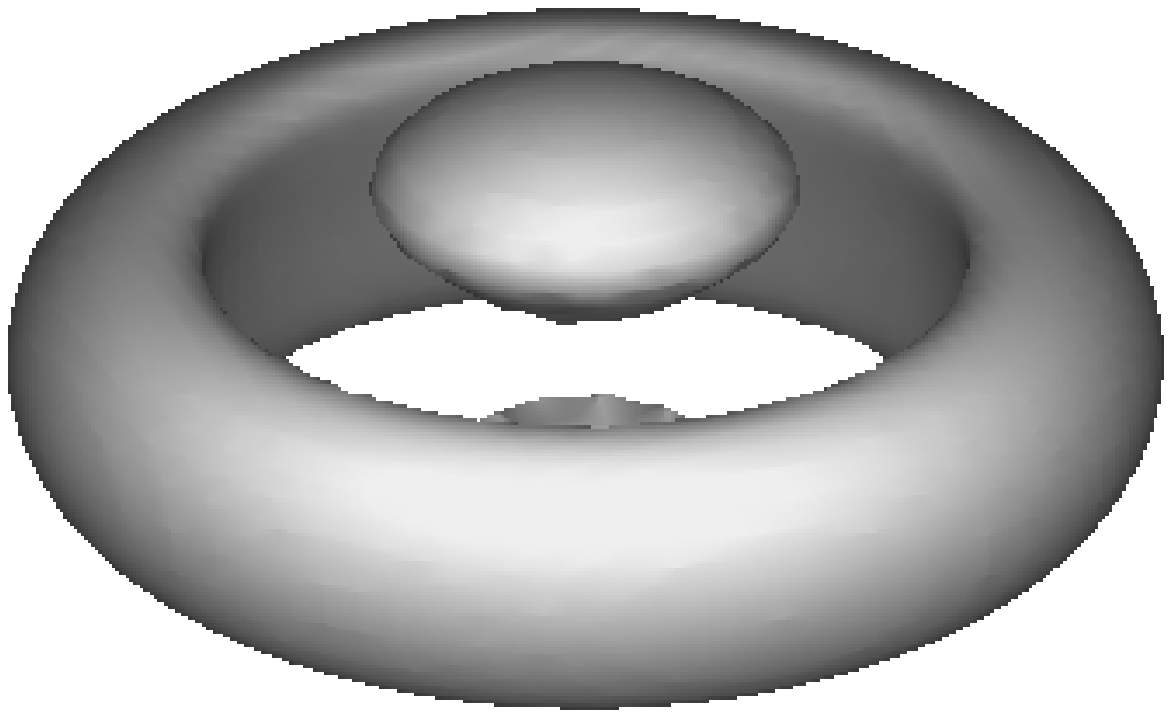}
\hspace{4cm}\epsfysize=3cm \epsffile{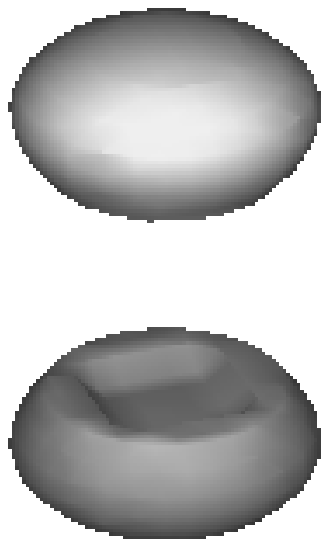}
} \\
\mbox{ $\varepsilon= 13.27 \ 10^{-4}$} \hspace{5cm}
\mbox{ $\varepsilon= 13.97 \ 10^{-4}$}
\end{figure}
\noindent Figs.~8b-e\\
Surfaces of constant energy density $\epsilon=-T_0^0$ 
are shown for the solution of Fig.~8a.
\clearpage
\newpage
\begin{figure}
\centering
\vspace{-1cm}
\mbox{  \epsfysize=12cm \epsffile{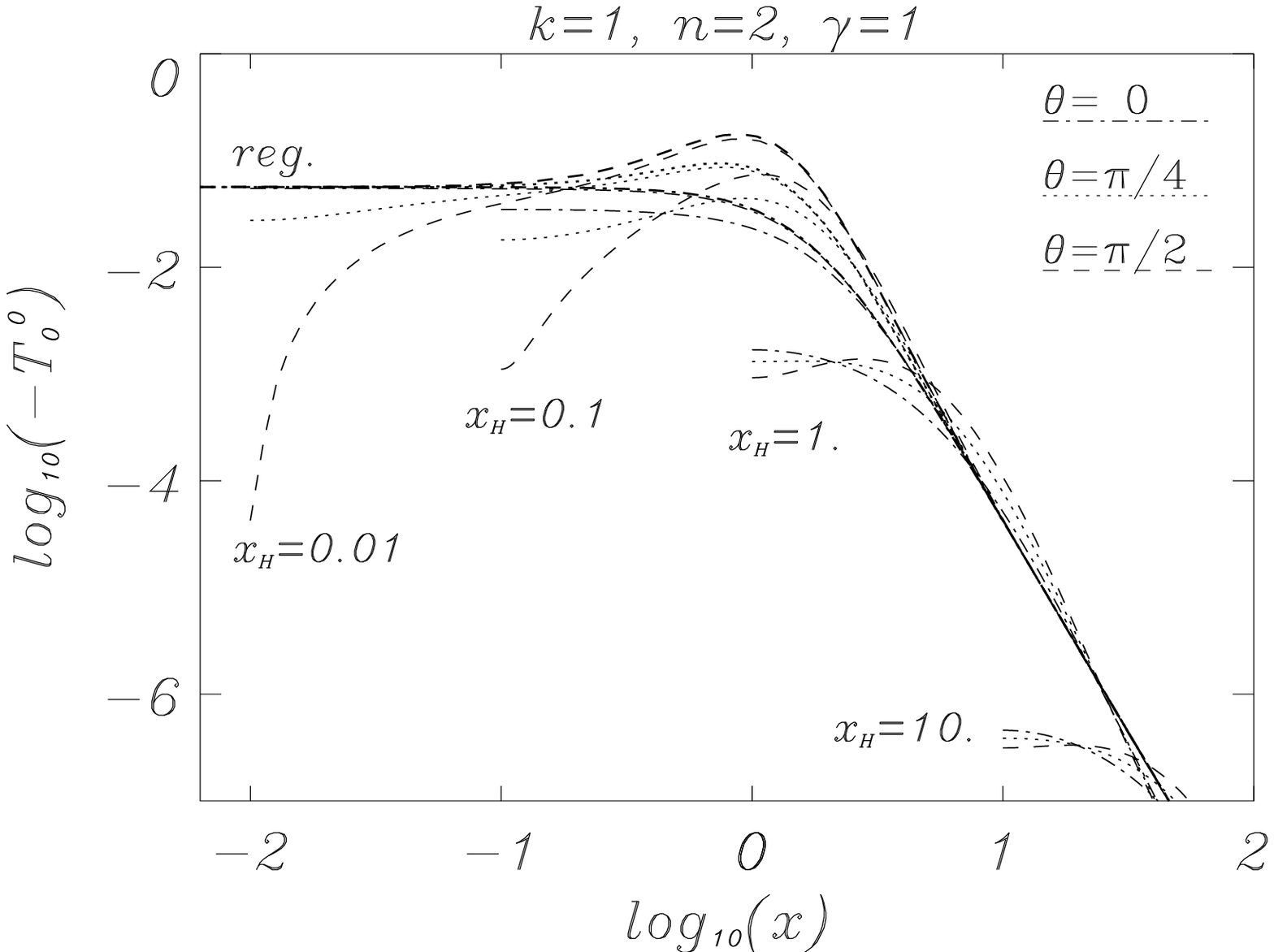}}\\
\end{figure}
\noindent Fig.~9\\
The energy density $\epsilon=-T_0^0$ is shown as a function 
of the dimensionless coordinate $x$ for the angles 
$\theta=0$, $\theta=\pi/4$ and $\theta=\pi/2$ 
for the EYMD black hole solutions with $\gamma=1$,
winding number $n=2$, node number $k=1$ and 
horizon radii $x_{\rm H}=0.01$,
$x_{\rm H}=0.1$, $x_{\rm H}=1$,
and $x_{\rm H}=10$,
as well as for the corresponding globally regular solution.
\clearpage
\newpage
\begin{figure}
\centering
\vspace{-1cm}
\mbox{  \epsfysize=6.cm \epsffile{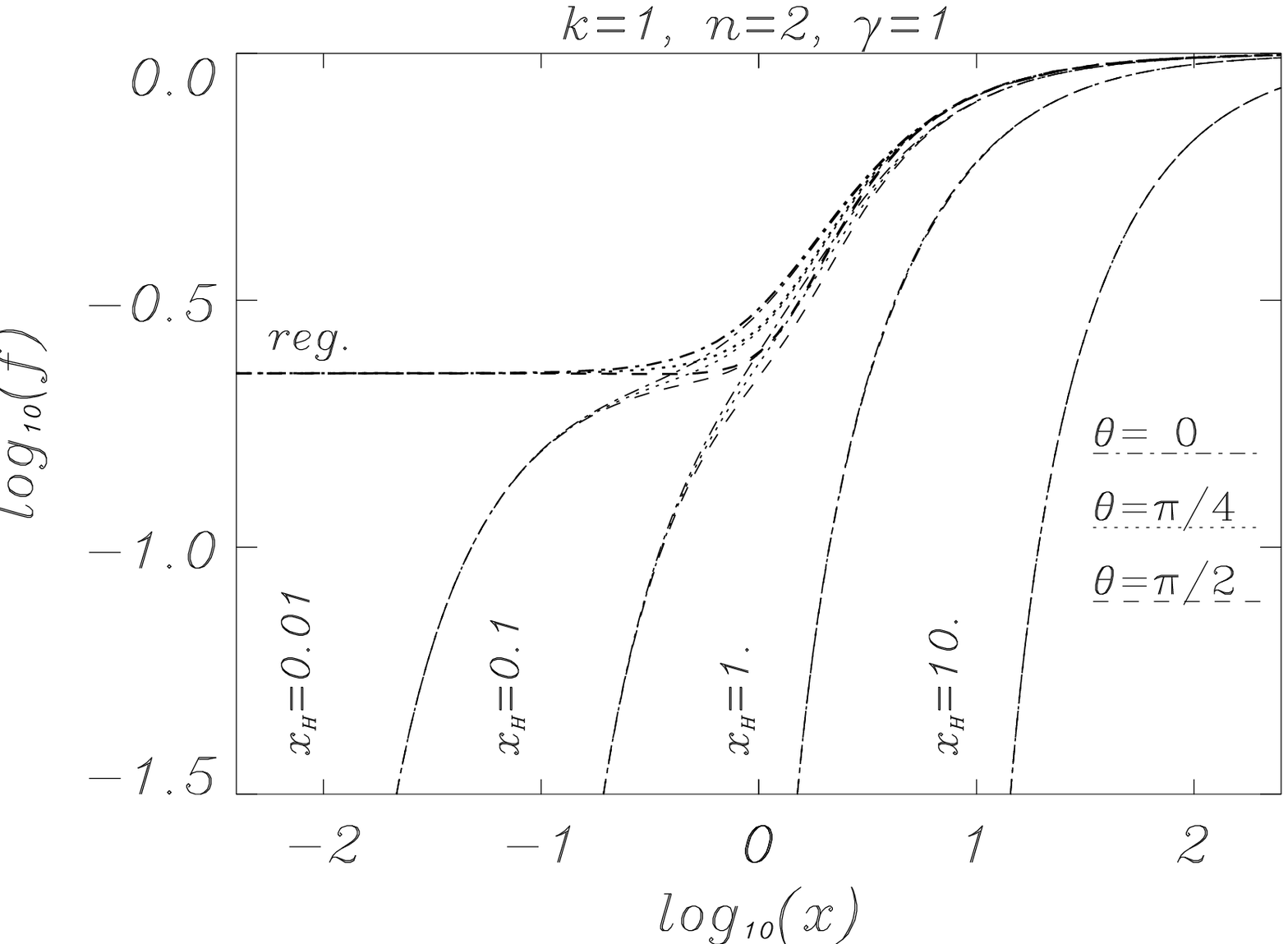}}\\
\end{figure}
\noindent Fig.~10a
Same as Fig.~9 for the metric function $f$.
\vspace{1.cm}\\
\begin{figure}
\centering
\vspace{-1cm}
\mbox{  \epsfysize=6.cm \epsffile{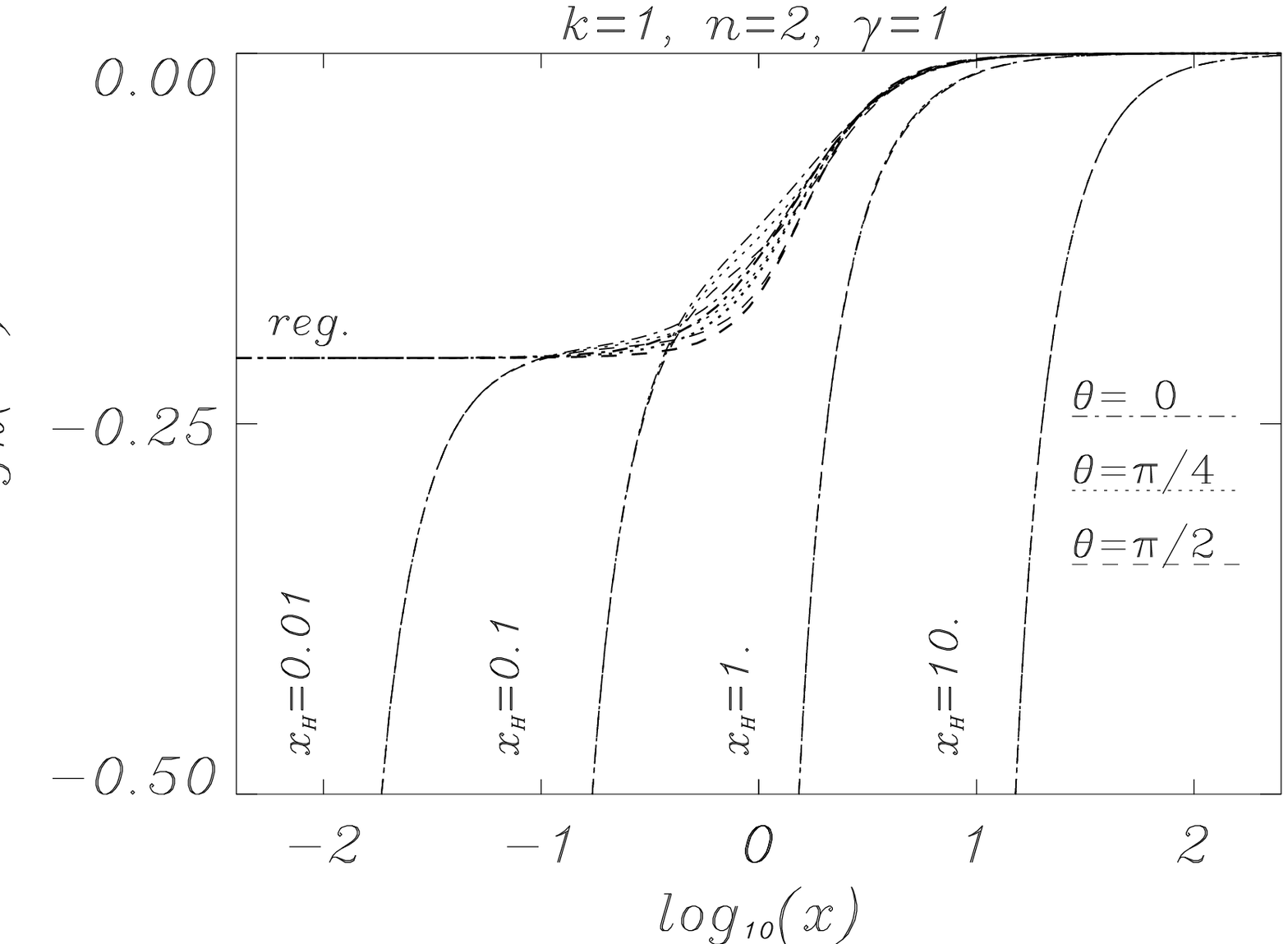}}\\
\end{figure}
\noindent Fig.~10b
Same as Fig.~9 for the metric function $m$.
\vspace{1.cm}\\
\begin{figure}
\centering
\vspace{-1cm}
\mbox{  \epsfysize=6.cm \epsffile{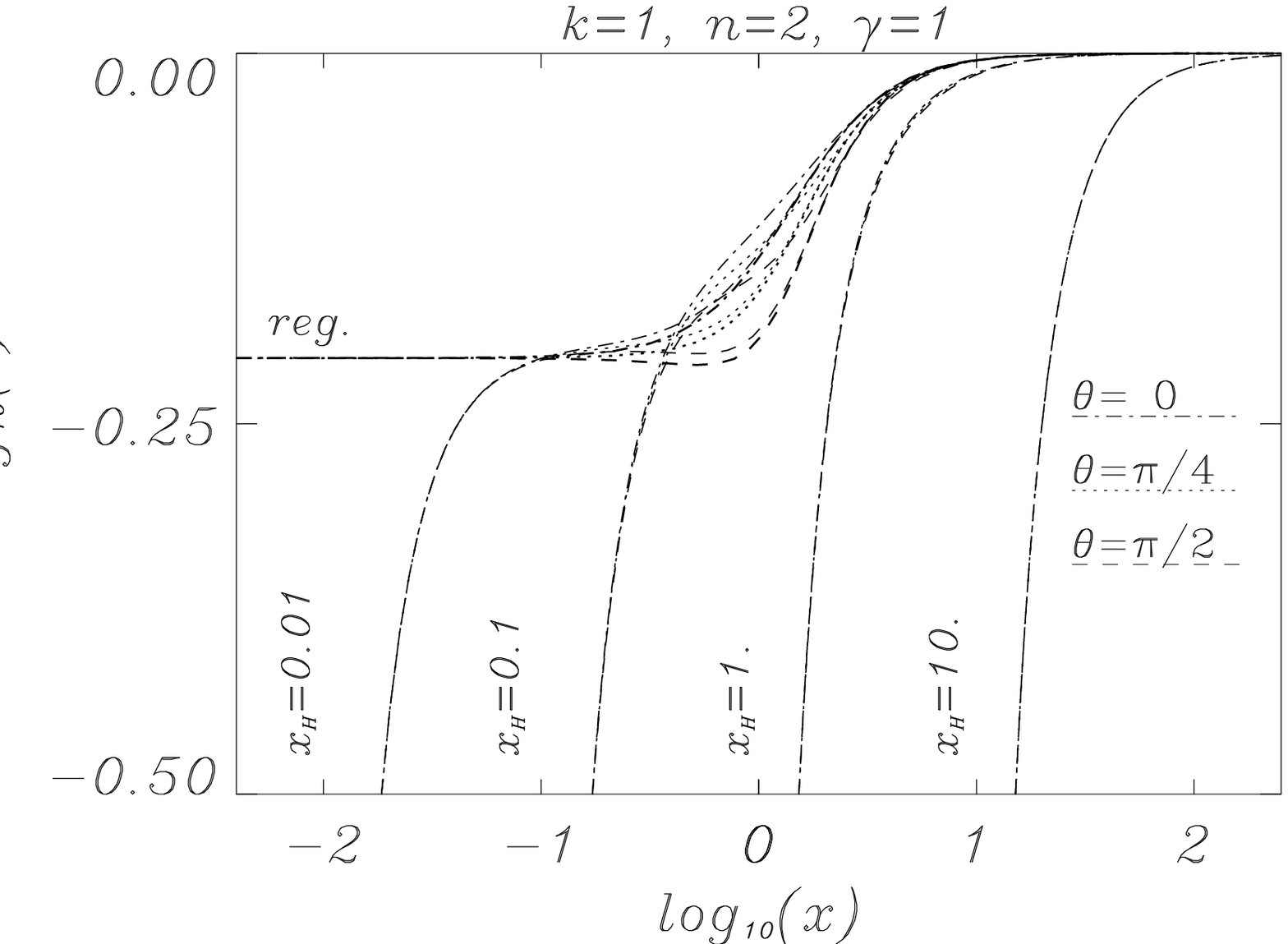}}\\
\end{figure}
\noindent Fig.~10c
Same as Fig.~9 for the metric function $l$.
\clearpage
\newpage
\begin{figure}
\centering
\vspace{-1cm}
\mbox{Fig.~11a  \epsfxsize=7.5cm \epsffile{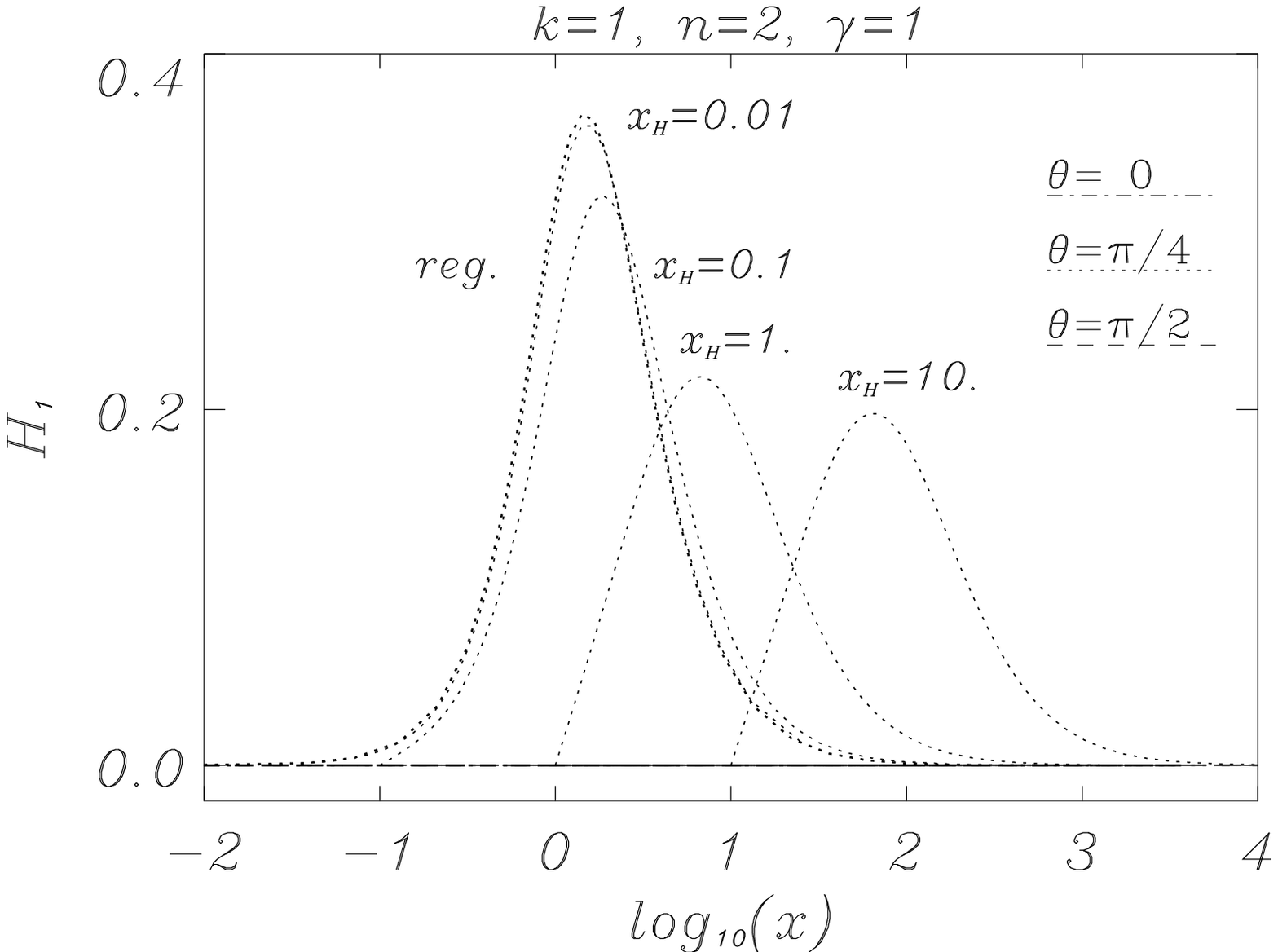}
       \epsfxsize=7.5cm \epsffile{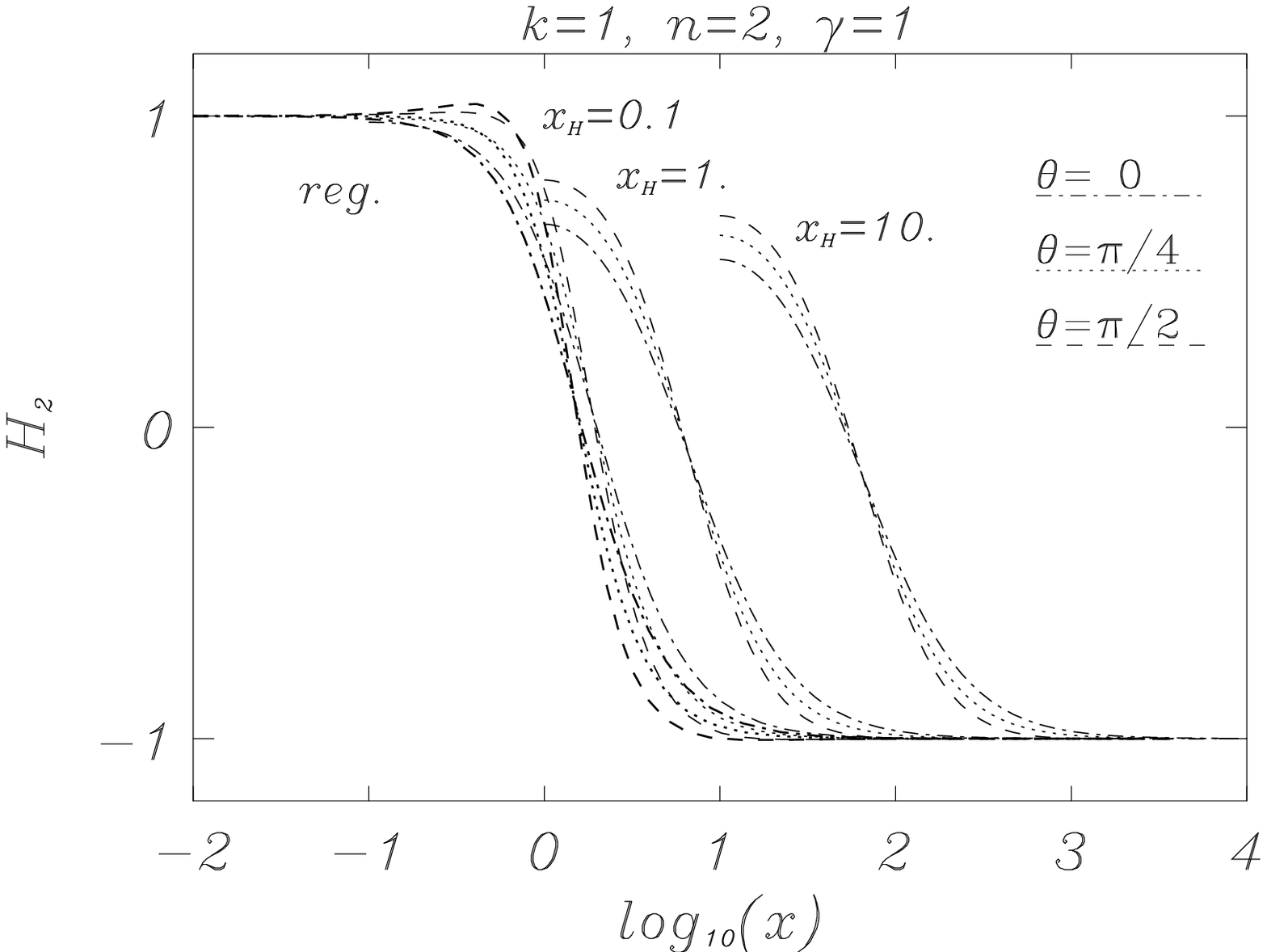} Fig.~11b}
\vspace{2cm}        \\
\mbox{Fig.~11c  \epsfxsize=7.5cm \epsffile{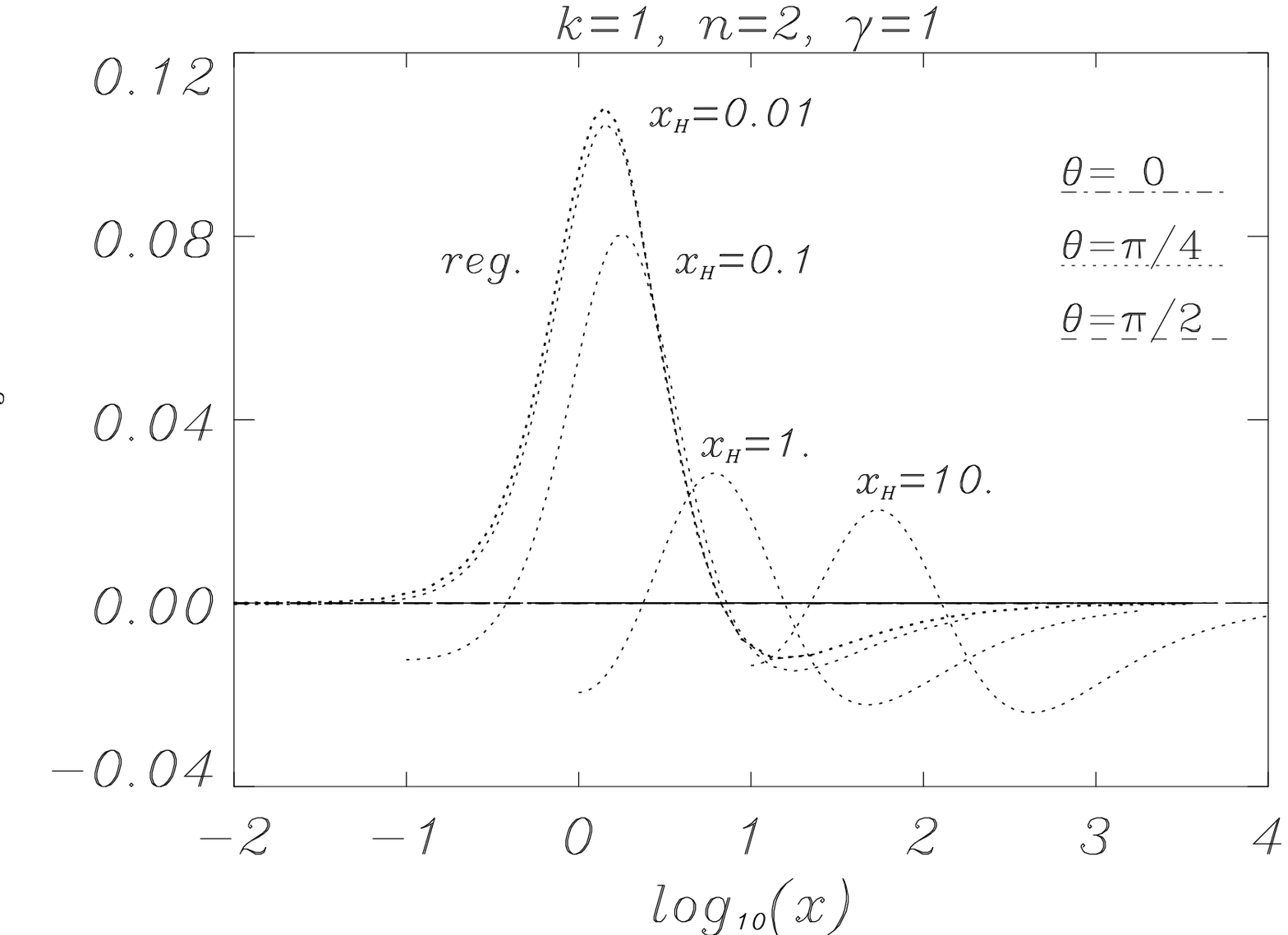}
       \epsfxsize=7.5cm \epsffile{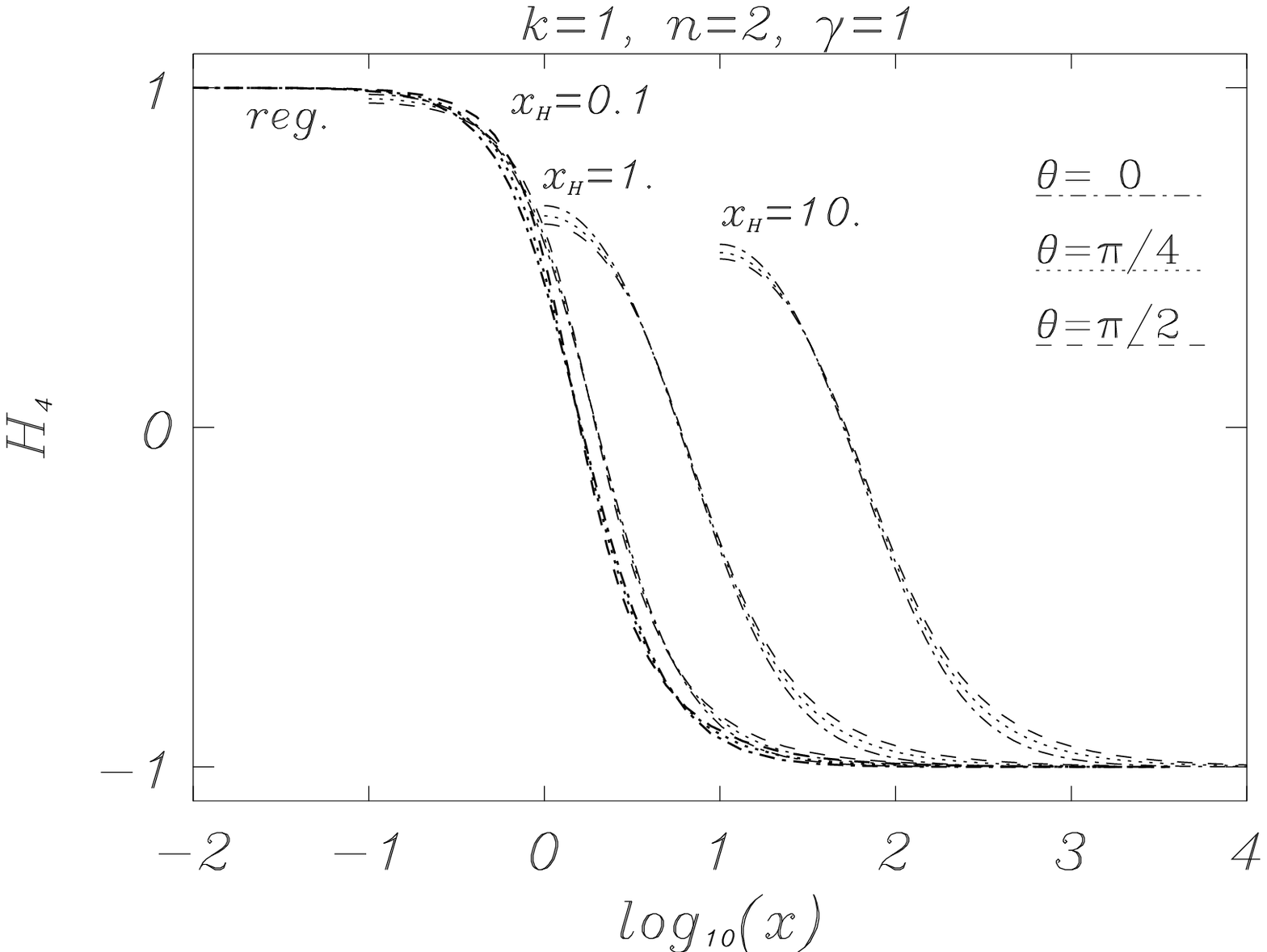} Fig.~11d}\vspace{1.cm}\\
\end{figure}
\noindent Fig.~11a
Same as Fig.~9 for the gauge field function $H_1$\\
\vspace{1.cm}\\
Fig.~11b
Same as Fig.~9 for the gauge field function $H_2$ for 
horizon radii 
$x_{\rm H}=0.1$, $x_{\rm H}=1$,
and $x_{\rm H}=10$.
\vspace{1.cm}\\
Fig.~11c
Same as Fig.~9 for the gauge field function $H_3$.
\vspace{1.cm}\\
Fig.~11d
Same as Fig.~11b for the gauge field function $H_4$.
\clearpage
\newpage
\begin{figure}
\centering
\vspace{-1cm}
\mbox{  \epsfysize=12cm \epsffile{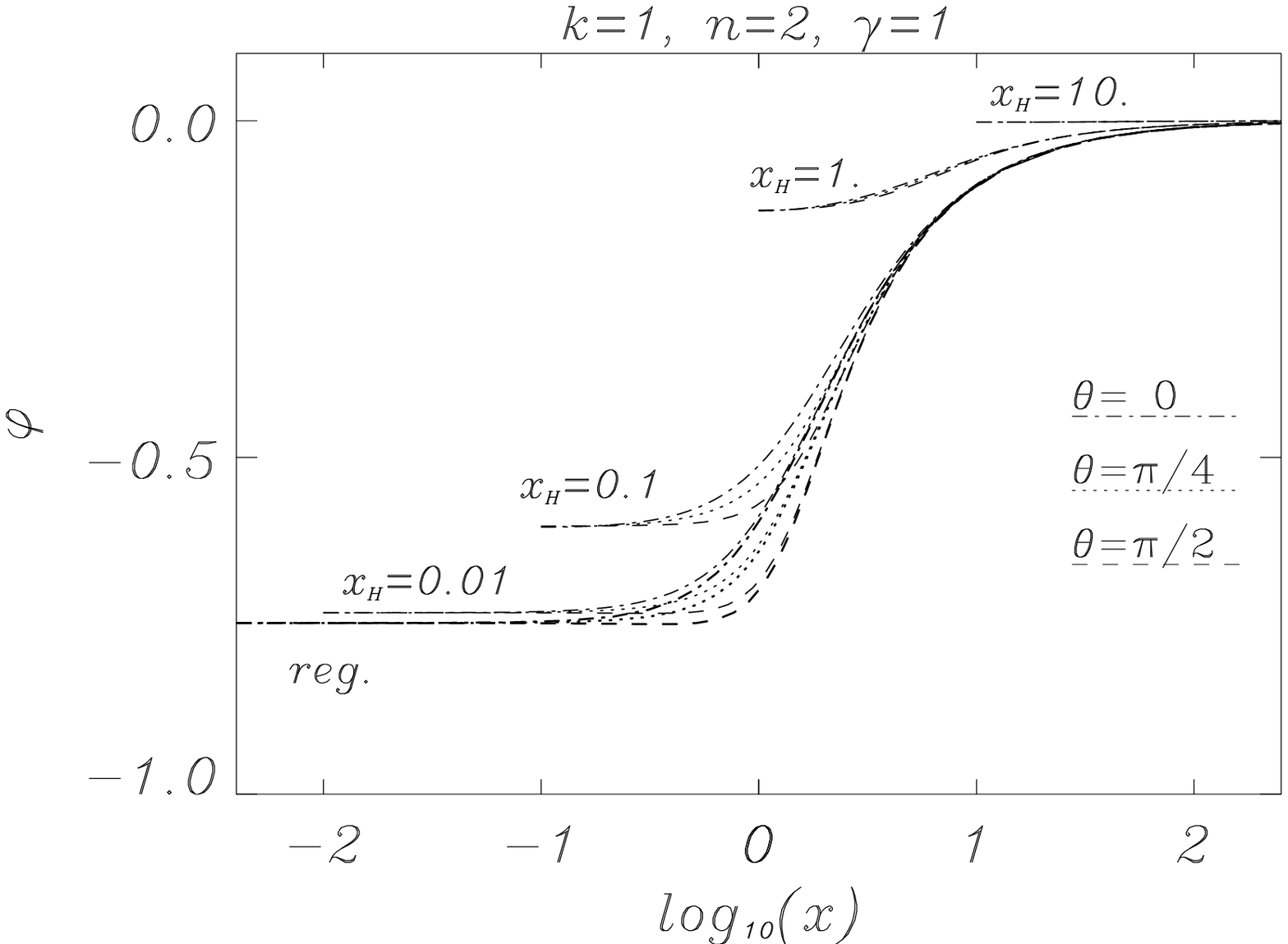}}\\
\end{figure}
\noindent Fig.~12\\
Same as Fig.~9 for the dilaton function $\varphi$.
\clearpage
\newpage
\begin{figure}
\centering
\vspace{-1cm}
\mbox{  \epsfysize=12cm \epsffile{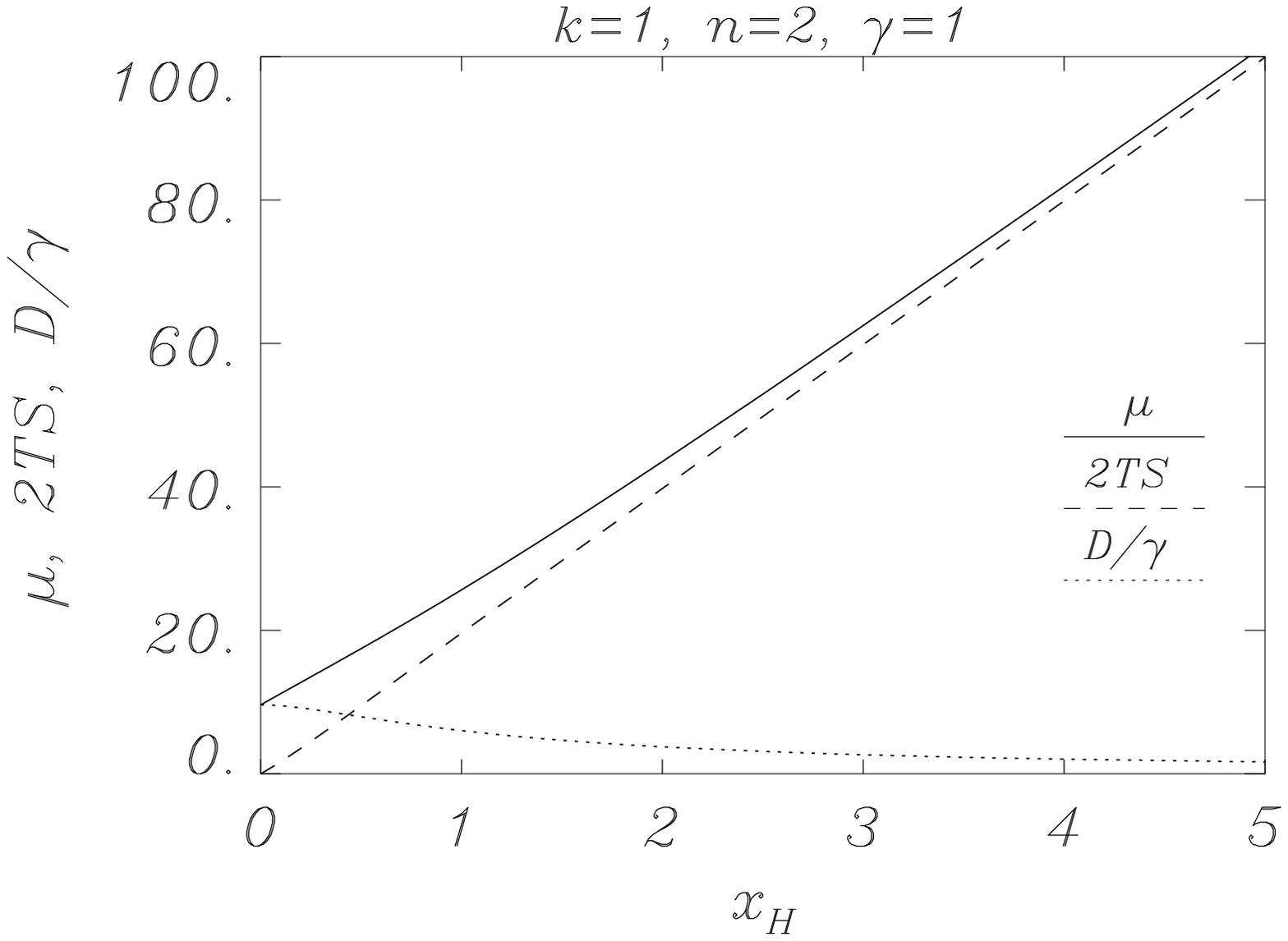}}\\
\end{figure}
\noindent Fig.~13\\
The dimensionless mass $\mu$ 
is shown as a function of the horizon radius $x_{\rm H}$
for the EYMD black hole solutions with $\gamma=1$,
winding number $n=2$, node number $k=1$.
Also shown are $2 TS$ and $D/\gamma$.
\clearpage
\newpage
\begin{figure}
\centering
\vspace{-1cm}
\mbox{  \epsfysize=12cm \epsffile{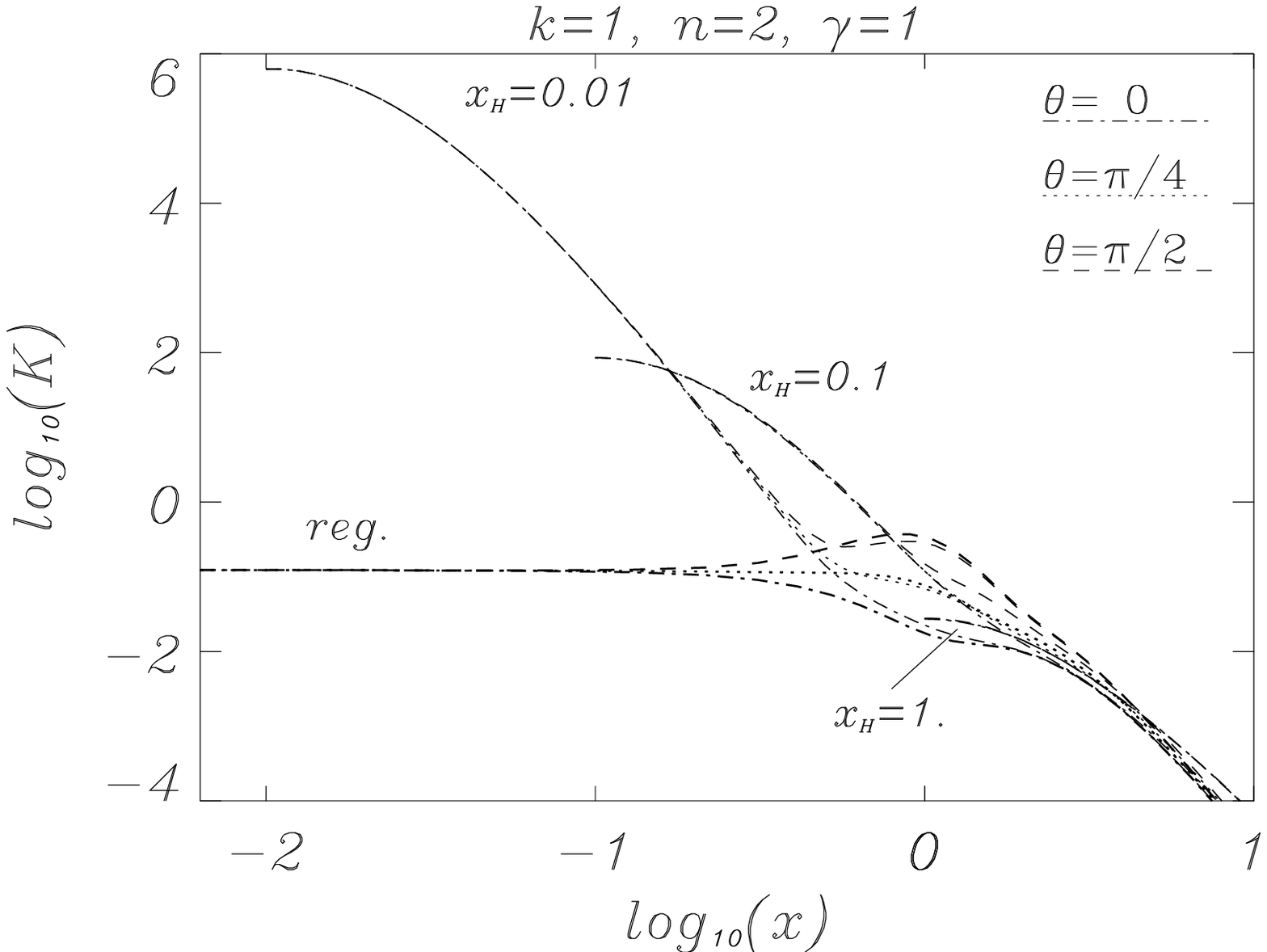}}\\
\end{figure}
\noindent Fig.~14\\
The Kretschmann scalar is shown as a function 
of the dimensionless coordinate $x$ for the angles 
$\theta=0$, $\theta=\pi/4$ and $\theta=\pi/2$ 
for the EYMD black hole solutions with $\gamma=1$, 
winding number $n=2$, node number $k=1$
and the horizon radii $x_{\rm H}=1$, 0.1, and 0.01,
as well as for the corresponding globally regular solution.
\clearpage
\newpage
\begin{figure}
\centering
\vspace{-1cm}
\mbox{  \epsfysize=6cm \epsffile{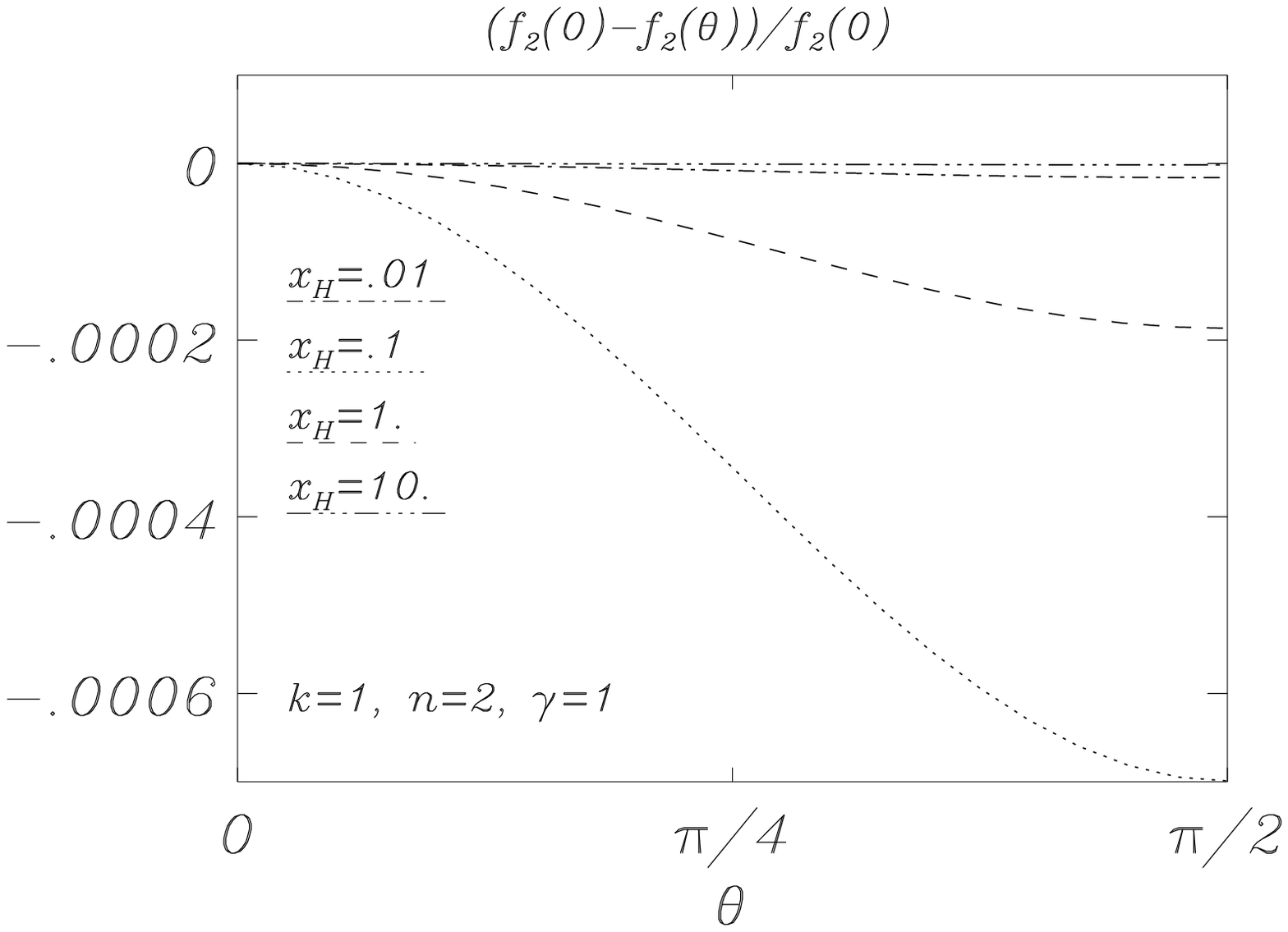}}\\
\end{figure}
\noindent Fig.~15a\\
The normalized expansion coefficient $f_2$ 
of the metric function $f$
is shown as a function of the angle $\theta$
for the EYMD black hole solutions with $\gamma=1$, 
winding number $n=2$, node number $k=1$
and the horizon radii $x_{\rm H}=10$, 1, 0.1, and 0.01.
\vspace{.5cm}\\
\begin{figure}
\centering
\vspace{-1cm}
\mbox{  \epsfysize=6cm \epsffile{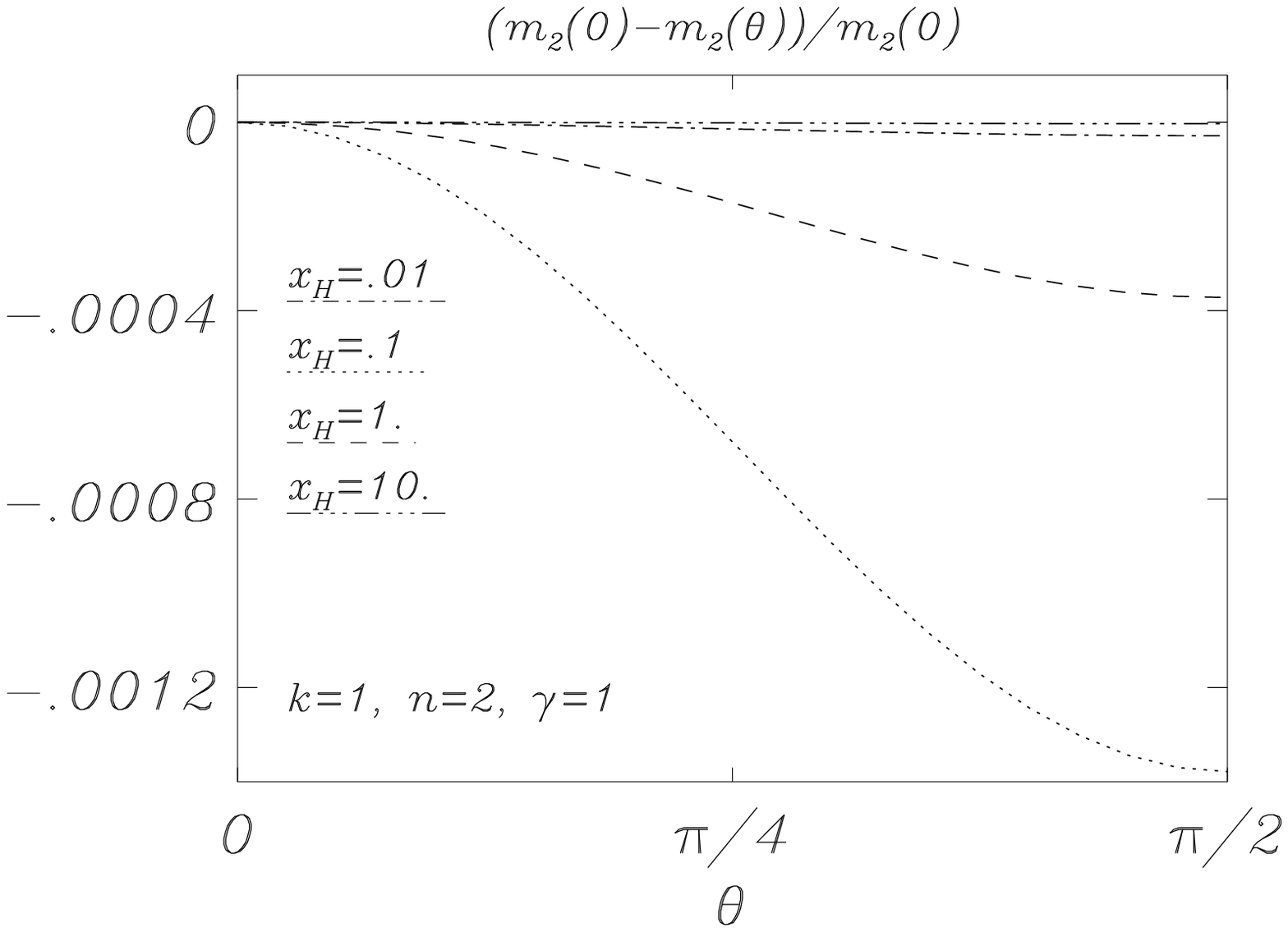}}\\
\end{figure}
\noindent Fig.~15b
Same as Fig.~15a for the expansion coefficient $m_2$ of the metric
function $m$.
\vspace{.5cm}\\
\begin{figure}
\centering
\vspace{-1cm}
\mbox{  \epsfysize=6cm \epsffile{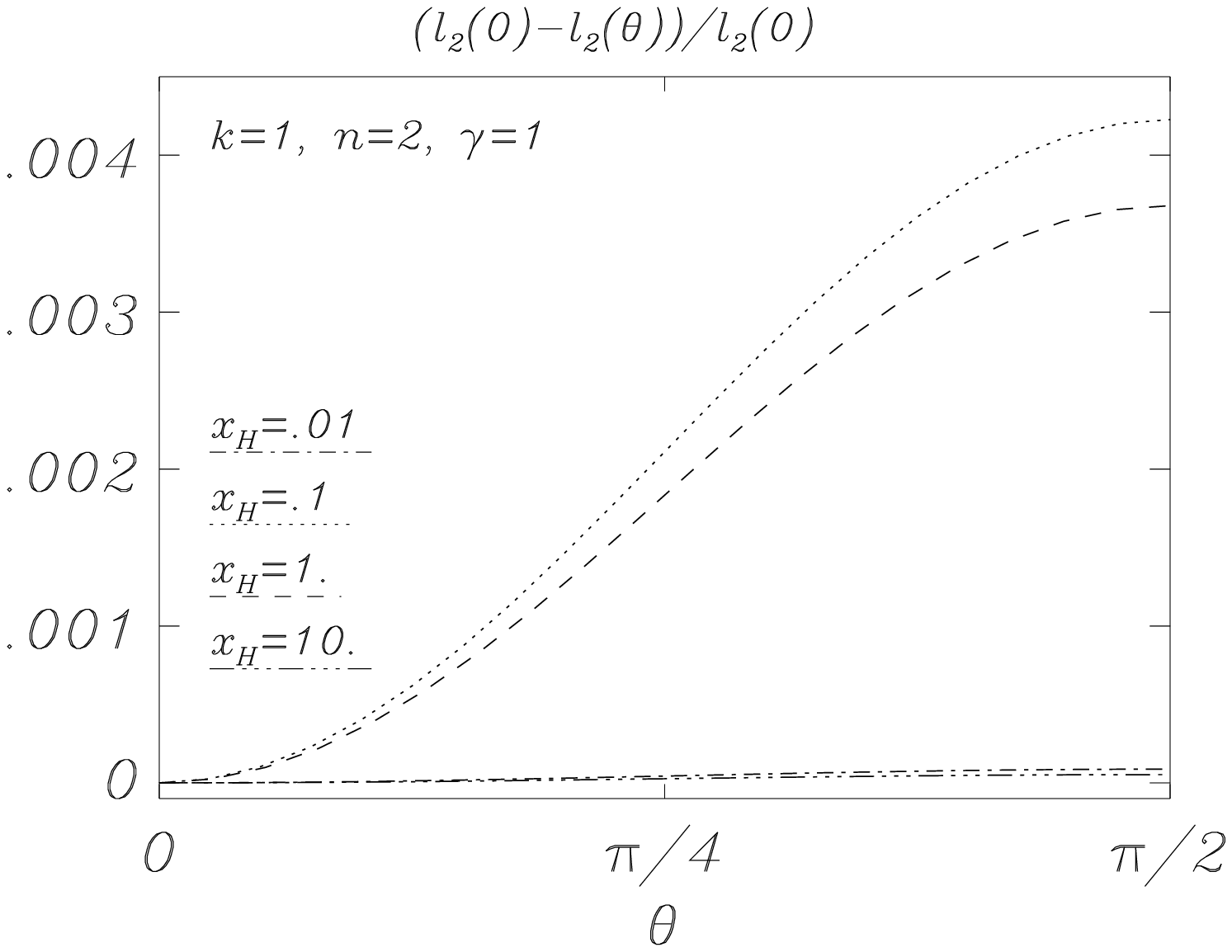}}\\
\end{figure}
\noindent Fig.~15c
Same as Fig.~15a for the expansion coefficient $l_2$ of the metric
function $l$.
\clearpage
\newpage
\begin{figure}
\centering
\vspace{-1cm}
\mbox{  \epsfysize=12cm \epsffile{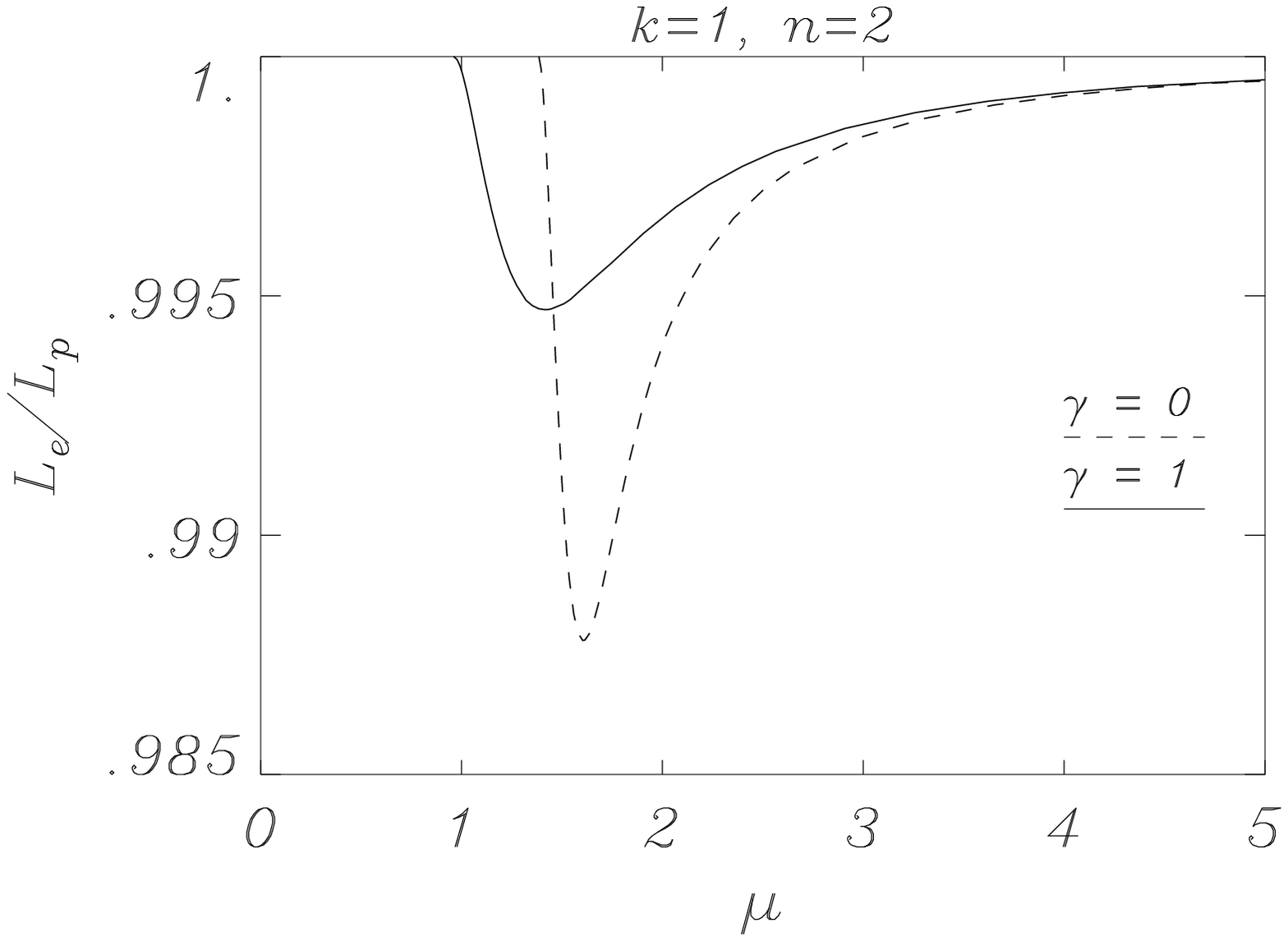}}\\
\end{figure}
\noindent Fig.~16\\
The ratio $L_e/L_p$ of the circumference of the horizon along the
equator $L_e$ to the circumference of the horizon along the poles
$L_p$ is shown as a function of the dimensionless mass $\mu$
for the black hole solutions with
winding number $n=2$ and node number $k=1$
of EYMD theory with $\gamma=1$ and EYM theory.
\clearpage
\newpage
\begin{figure}
\centering
\vspace{-1cm}
\mbox{  \epsfysize=12cm \epsffile{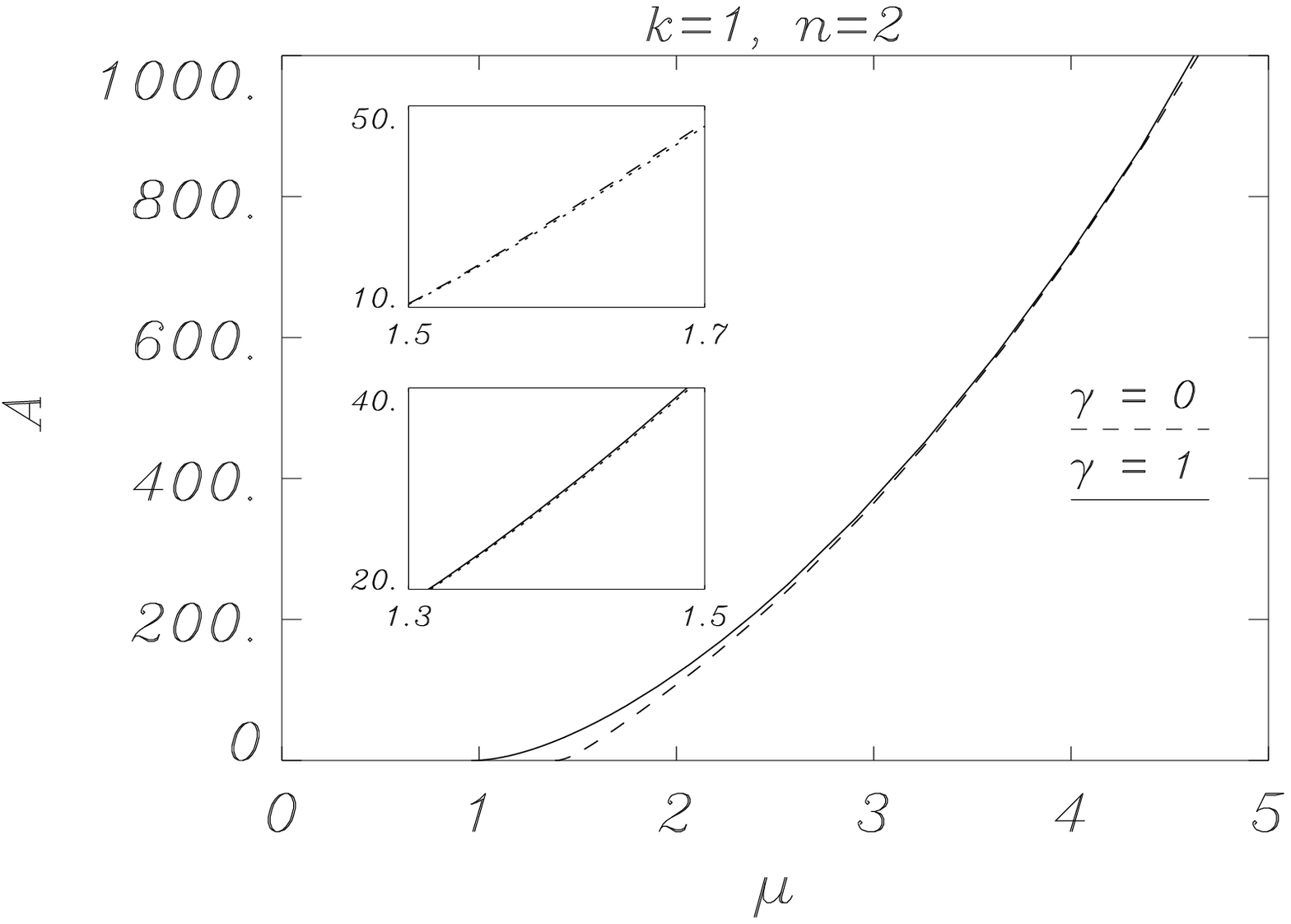}}\\
\end{figure}
\noindent Fig.~17\\
Same as Fig.~16 for the area $A$ of the horizon.
\clearpage
\newpage
\begin{figure}
\centering
\vspace{-1cm}
\mbox{  \epsfysize=12cm \epsffile{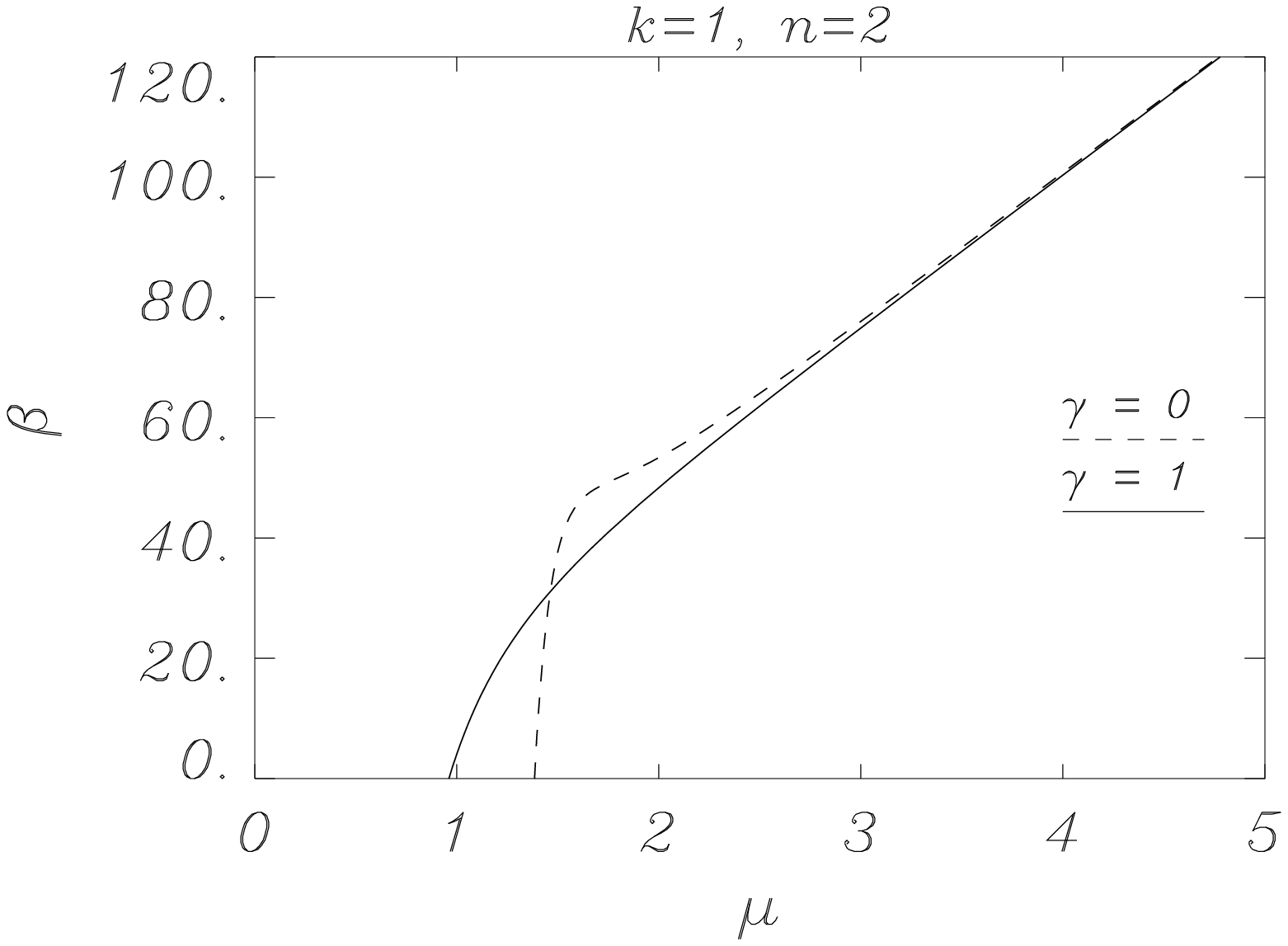}}\\
\end{figure}
\noindent Fig.~18\\
Same as Fig.~16 for the inverse temperature $\beta=T^{-1}$.
\clearpage
\newpage
\begin{figure}
\centering
\vspace{-1cm}
\mbox{ Fig.~19a \epsfxsize=7.5cm \epsffile{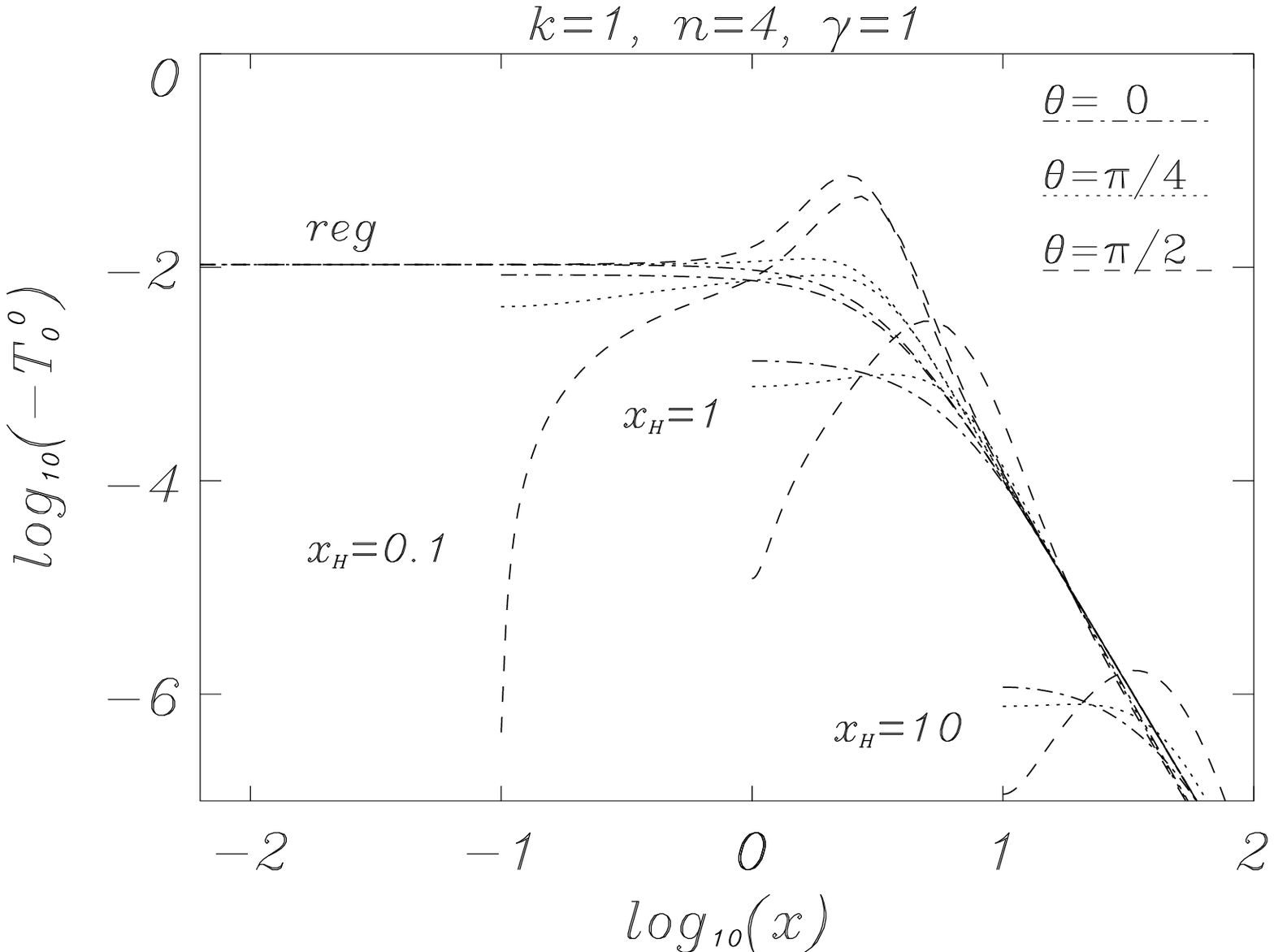}
        \epsfxsize=7.5cm \epsffile{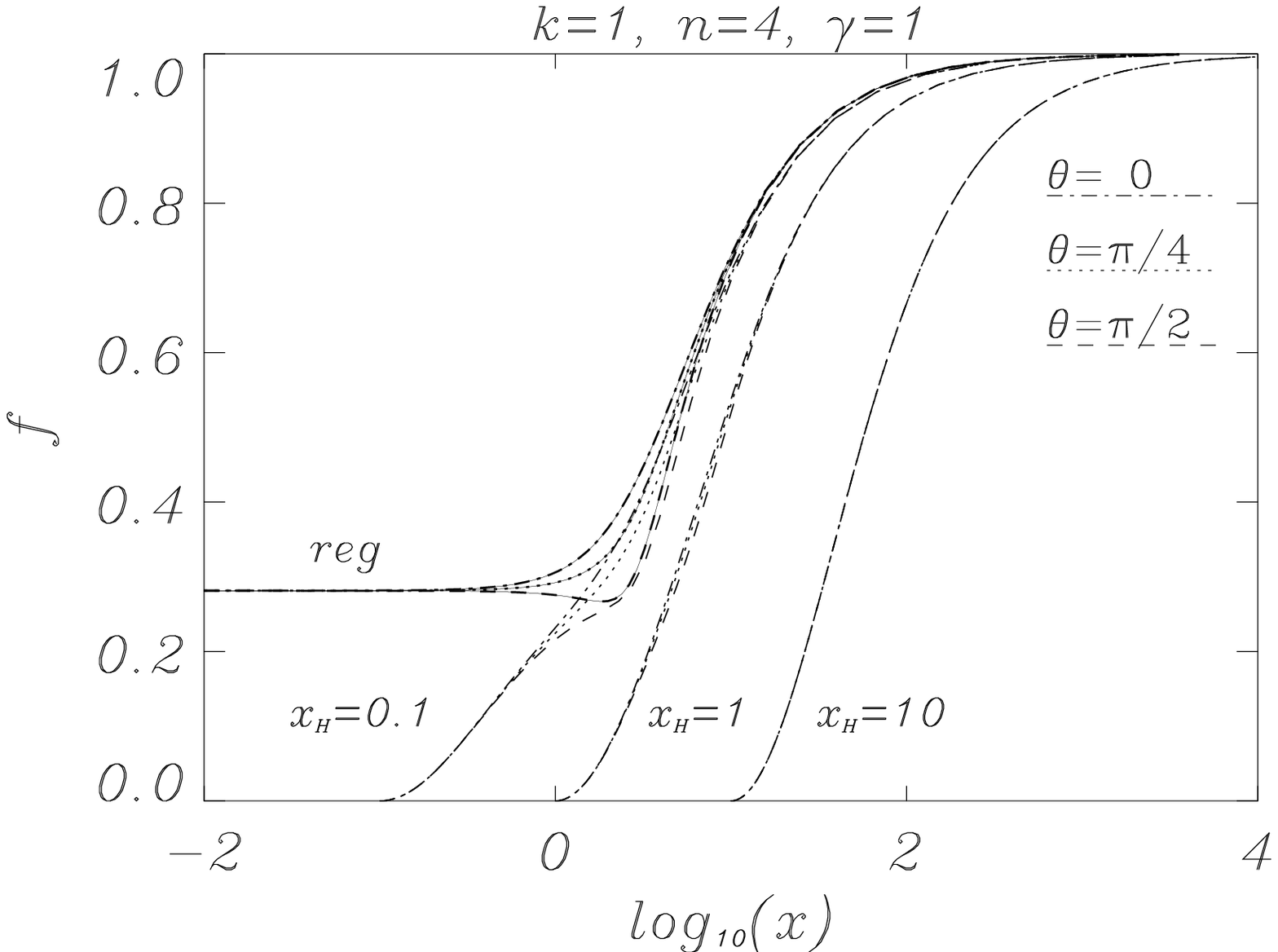}Fig.~19b}
\vspace{2cm}        \\
\mbox{ Fig.~19c \epsfxsize=7.5cm \epsffile{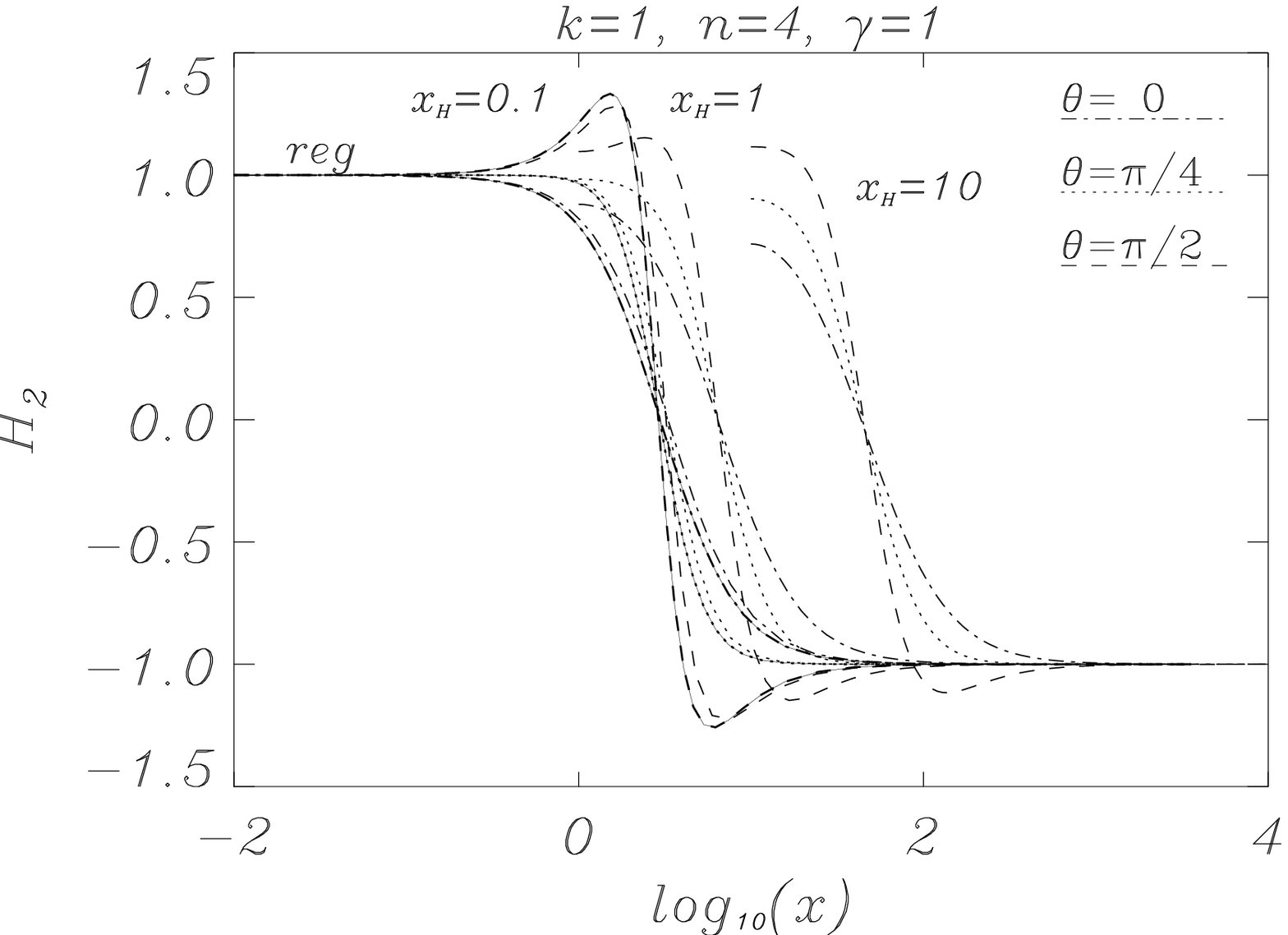}
        \epsfxsize=7.5cm \epsffile{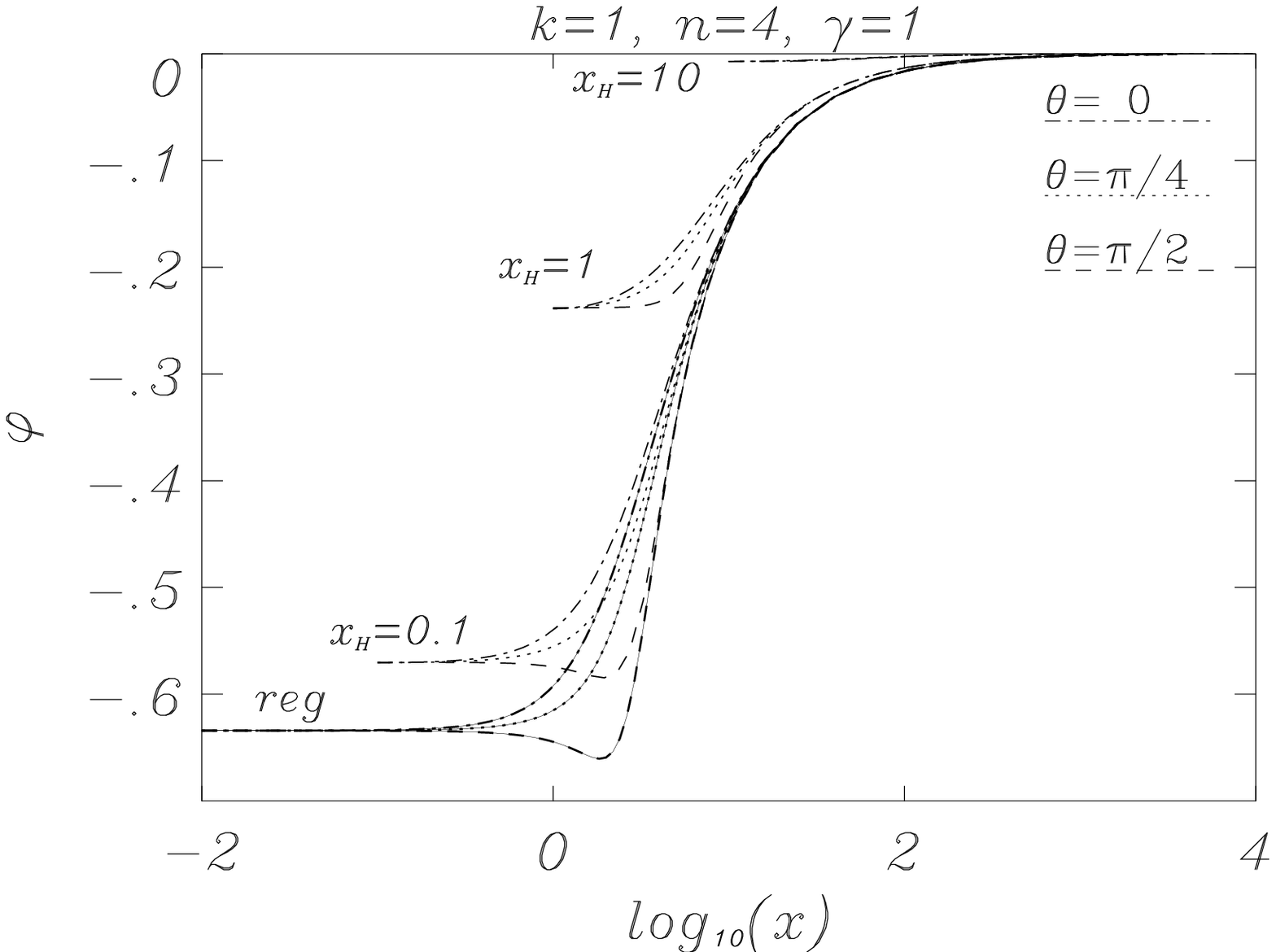}Fig.~19d}\vspace{1.cm}\\
\end{figure}
\noindent Fig.~19a\\
The energy density $\epsilon=-T_0^0$ is shown as a function 
of the dimensionless coordinate $x$ for the angles 
$\theta=0$, $\theta=\pi/4$ and $\theta=\pi/2$ 
for the EYMD black hole solutions with $\gamma=1$,
winding number $n=4$, node number $k=1$ and 
horizon radii $x_{\rm H}=0.1$, $x_{\rm H}=1$,
and $x_{\rm H}=10$,
as well as for the corresponding globally regular solution.
\vspace{1.cm}\\
Fig.~19b\\
Same as Fig.~19a for the metric function $f$.
\vspace{1.cm}\\
Fig.~19c\\
Same as Fig.~19a for the gauge field function $H_2$.
\vspace{1.cm}\\
Fig.~19d\\
Same as Fig.~19a for the dilaton function $\varphi$.
\clearpage
\newpage
\begin{figure}
\centering
\vspace{-1cm}
\mbox{  \epsfysize=12cm \epsffile{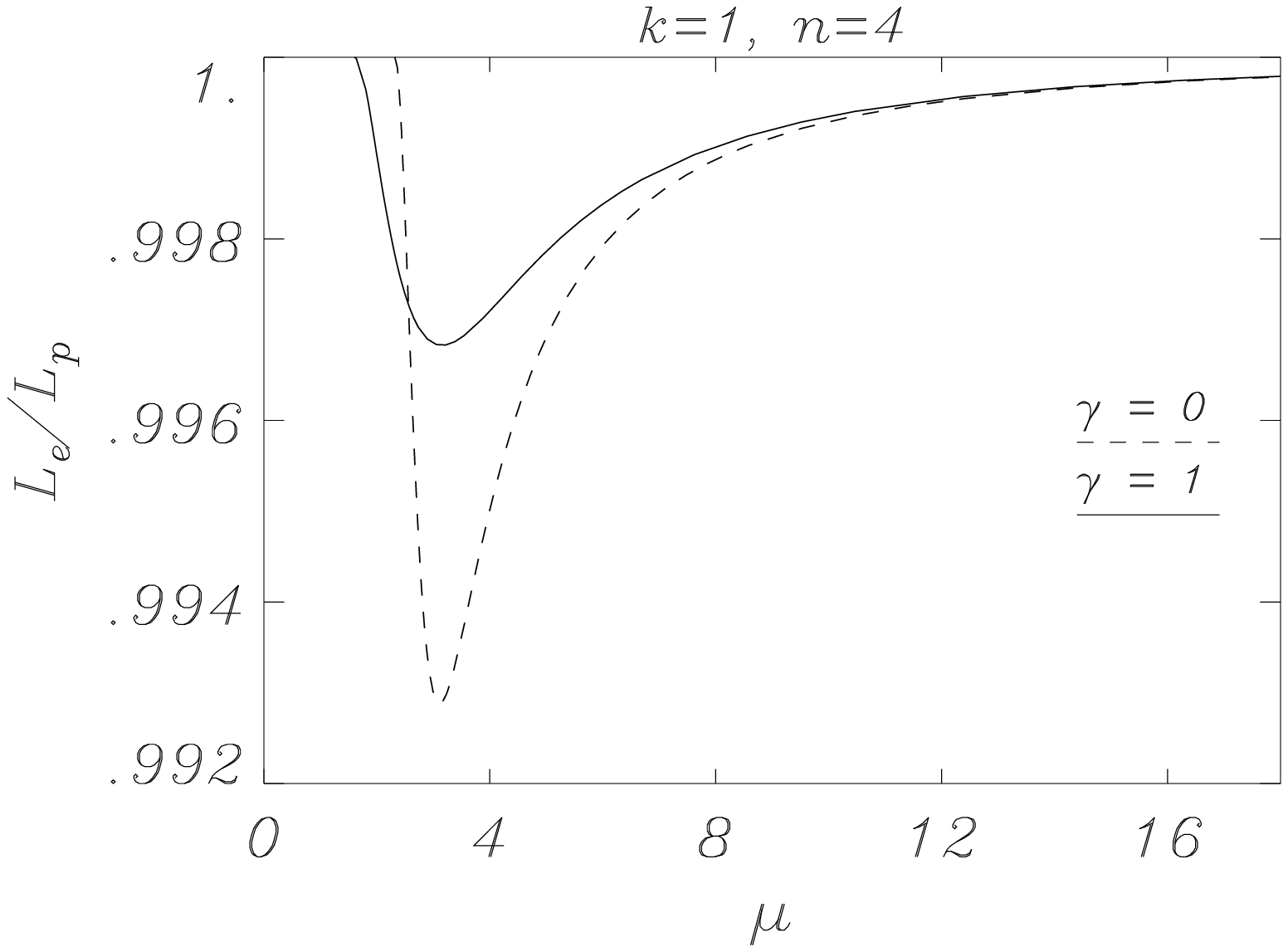}}\\
\end{figure}
\noindent Fig.~20\\
The ratio $L_e/L_p$ of the circumference of the horizon along the
equator $L_e$ to the circumference of the horizon along the poles
$L_p$ is shown as a function of the dimensionless mass $\mu$
for the black hole solutions with
winding number $n=4$ and node number $k=1$
of EYMD theory with $\gamma=1$ and EYM theory.
\clearpage
\newpage
\begin{figure}
\centering
\vspace{-1cm}
\mbox{ Fig.~21a \epsfxsize=7.5cm \epsffile{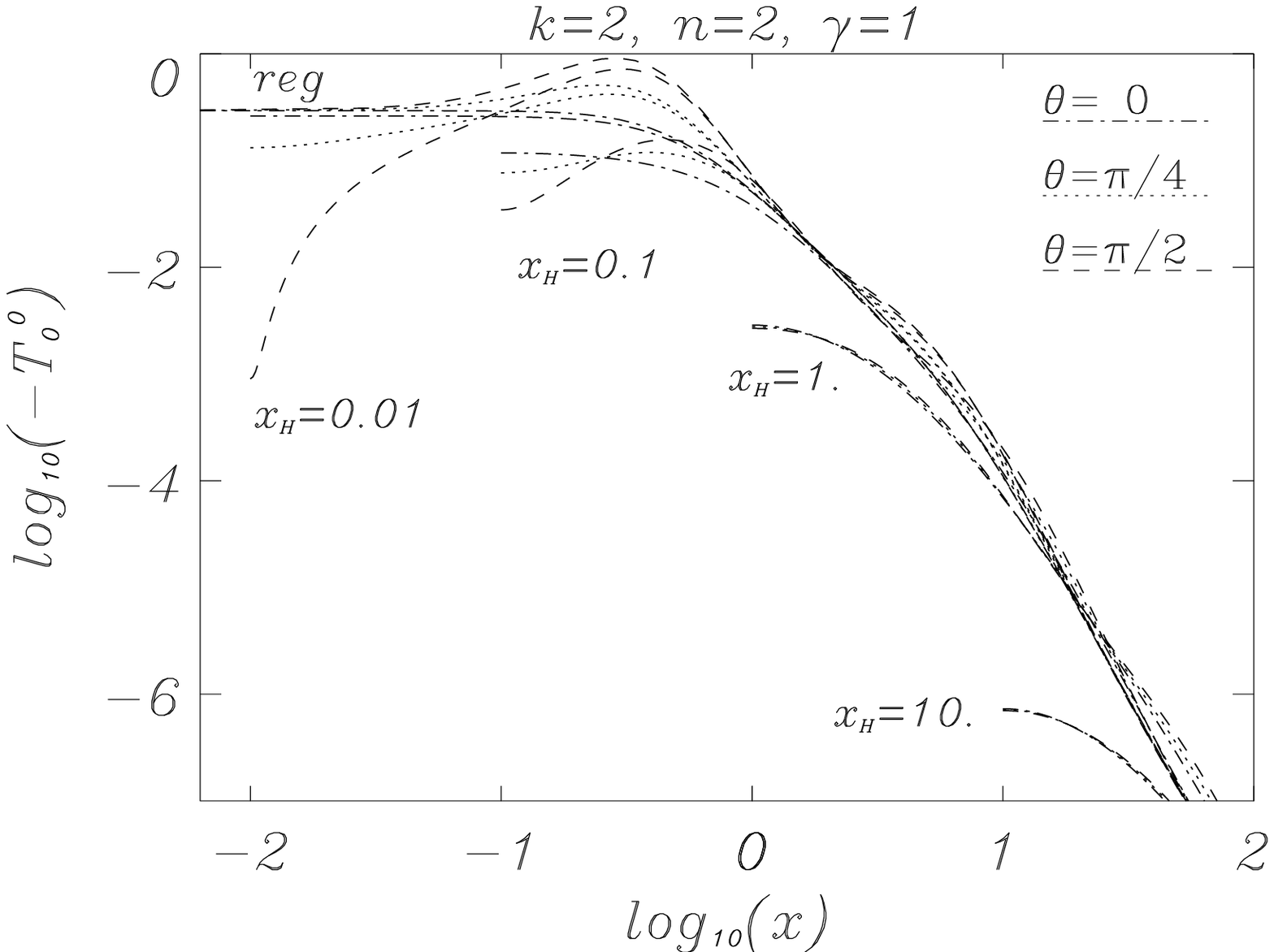}
        \epsfxsize=7.5cm \epsffile{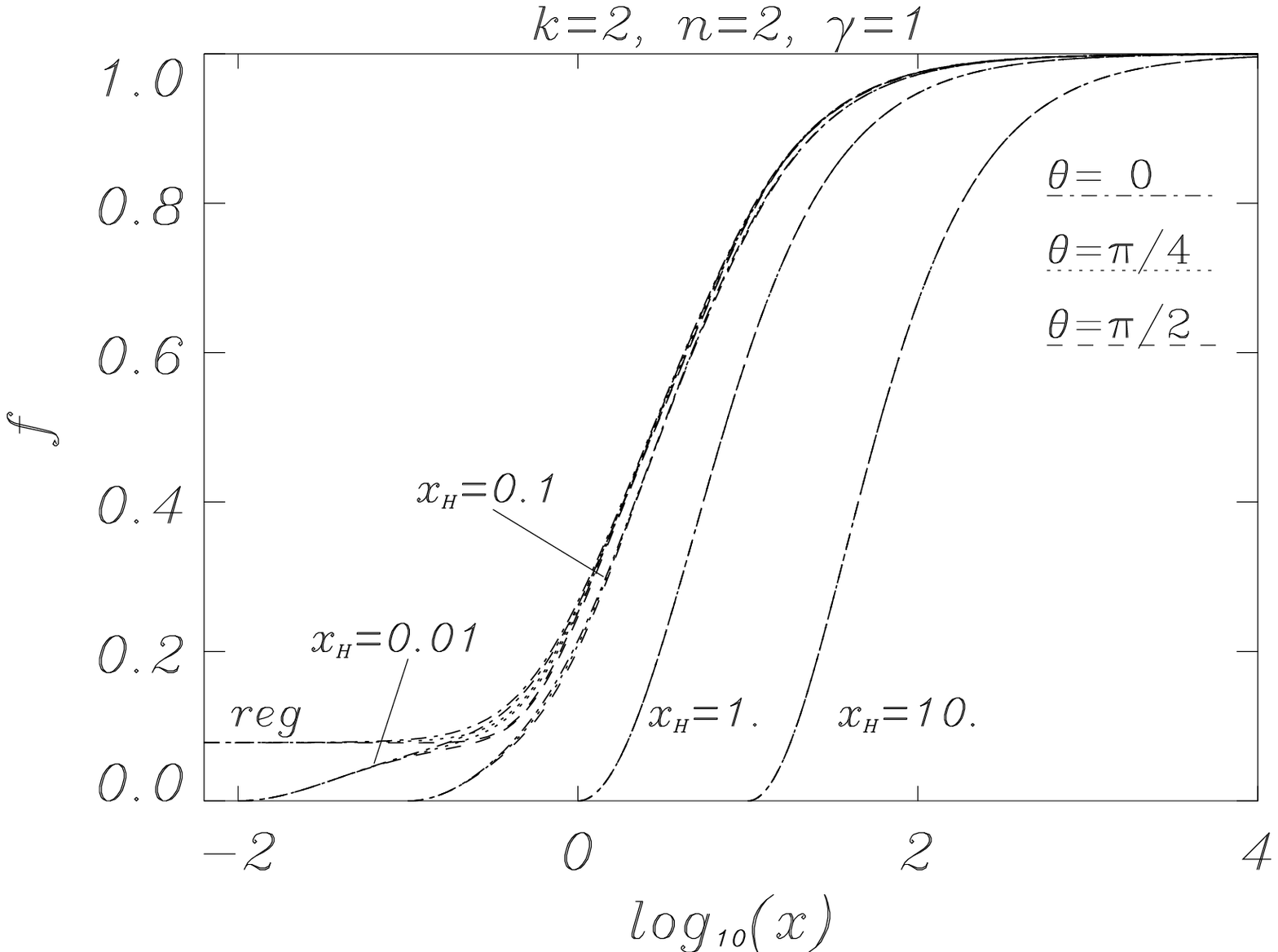}Fig.~21b}
\vspace{2cm}        \\
\mbox{  Fig.~21c\epsfxsize=7.5cm \epsffile{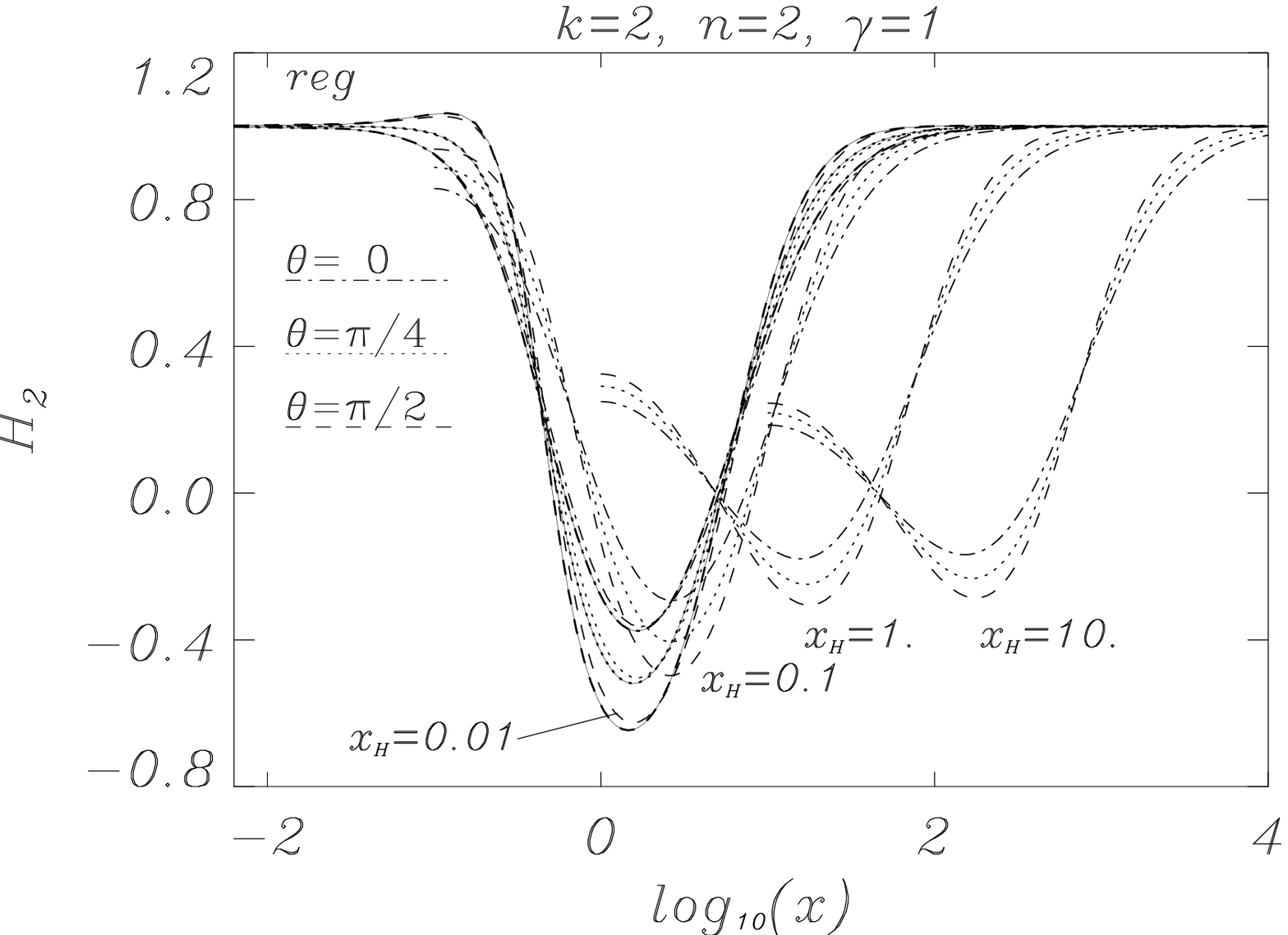}
        \epsfxsize=7.5cm \epsffile{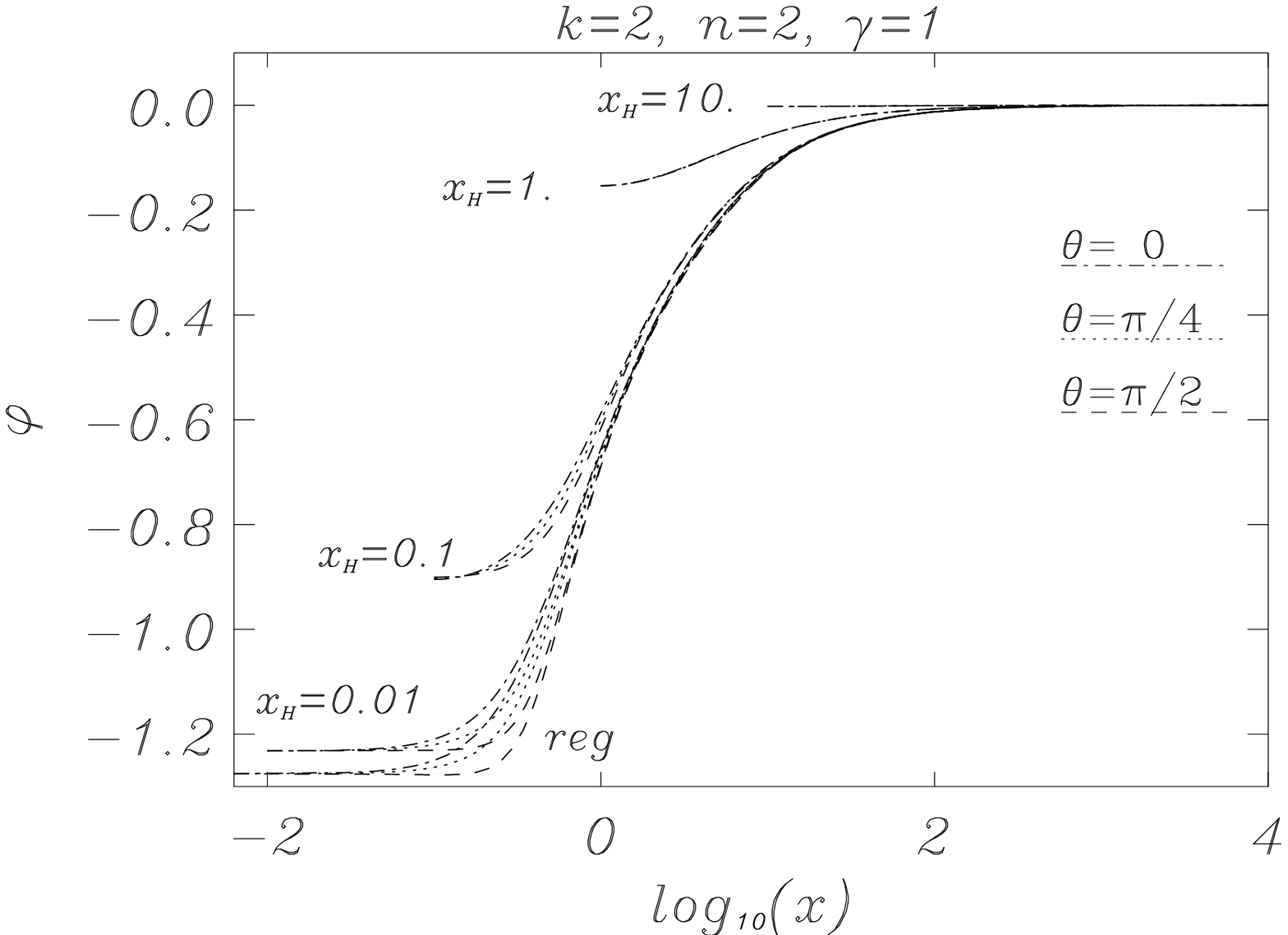}Fig.~21d}\vspace{1.cm}\\
\end{figure}
\noindent Fig.~21a\\
The energy density $\epsilon=-T_0^0$ is shown as a function 
of the dimensionless coordinate $x$ for the angles 
$\theta=0$, $\theta=\pi/4$ and $\theta=\pi/2$ 
for the EYMD black hole solutions with $\gamma=1$,
winding number $n=2$, node number $k=2$ and 
horizon radii $x_{\rm H}=0.01$, $x_{\rm H}=0.1$,
$x_{\rm H}=1$, and $x_{\rm H}=10$,
as well as for the corresponding globally regular solution.
\vspace{1.cm}\\
Fig.~21b\\
Same as Fig.~21a for the metric function $f$.
\vspace{1.cm}\\
Fig.~21c\\
Same as Fig.~21a for the gauge field function $H_2$.
\vspace{1.cm}\\
Fig.~21d\\
Same as Fig.~21a for the dilaton function $\varphi$.
\clearpage
\newpage
\begin{figure}
\centering
\vspace{-1cm}
\mbox{  \epsfysize=6cm \epsffile{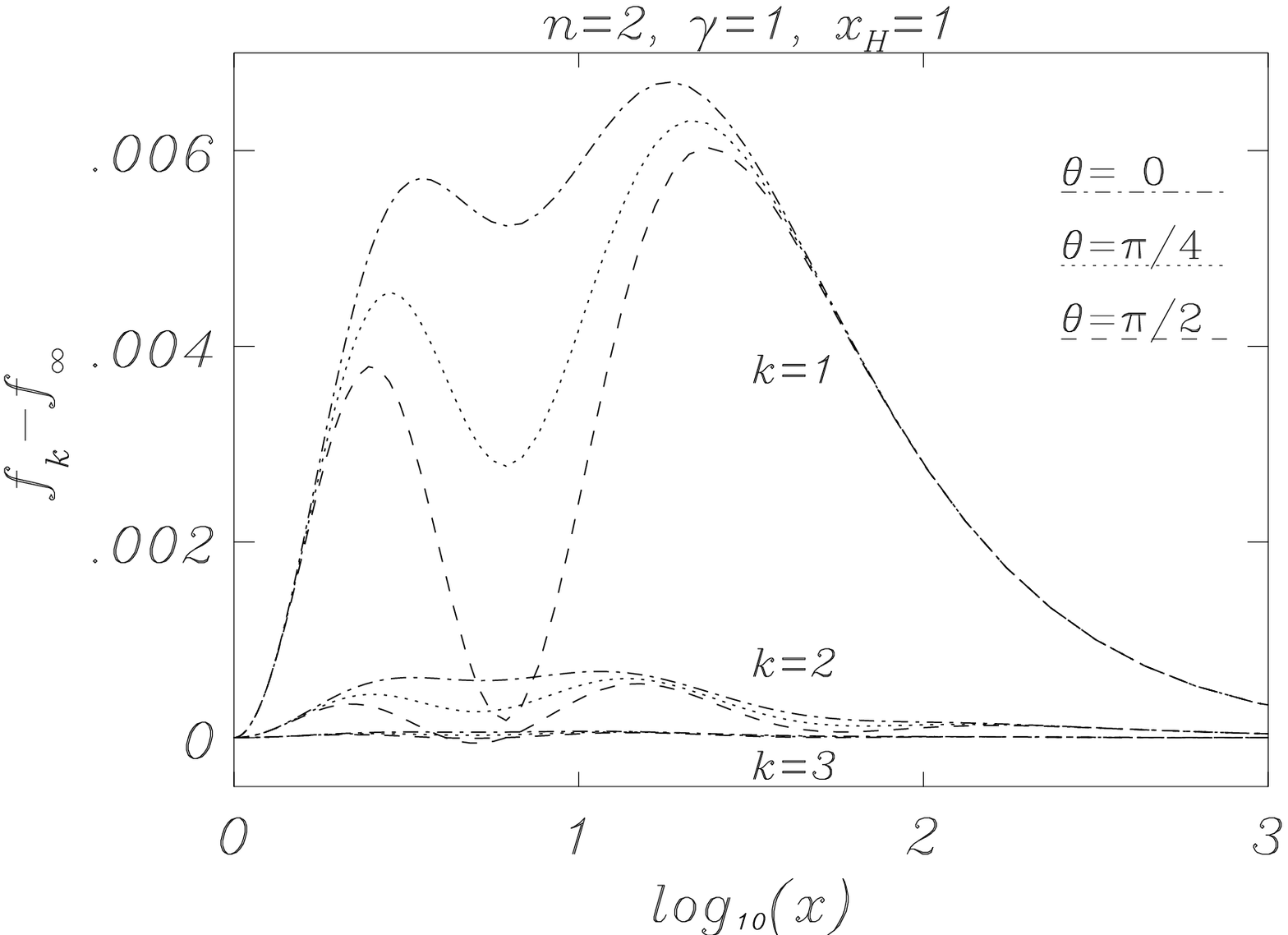}}\\
\end{figure}
\noindent Fig.~22a
The difference of the metric function $f_k$ and the 
metric function $f_\infty$ of the limiting solution
is shown as a function
of the dimensionless coordinate $x$ for the angles 
$\theta=0$, $\theta=\pi/4$ and $\theta=\pi/2$ 
for the EYMD black hole solutions with $\gamma=1$,
horizon radius $x_{\rm H}=1$, winding number $n=2$, 
and node numbers $k=1-3$.
\vspace{.5cm}\\
\begin{figure}
\centering
\vspace{-1cm}
\mbox{  \epsfysize=6cm \epsffile{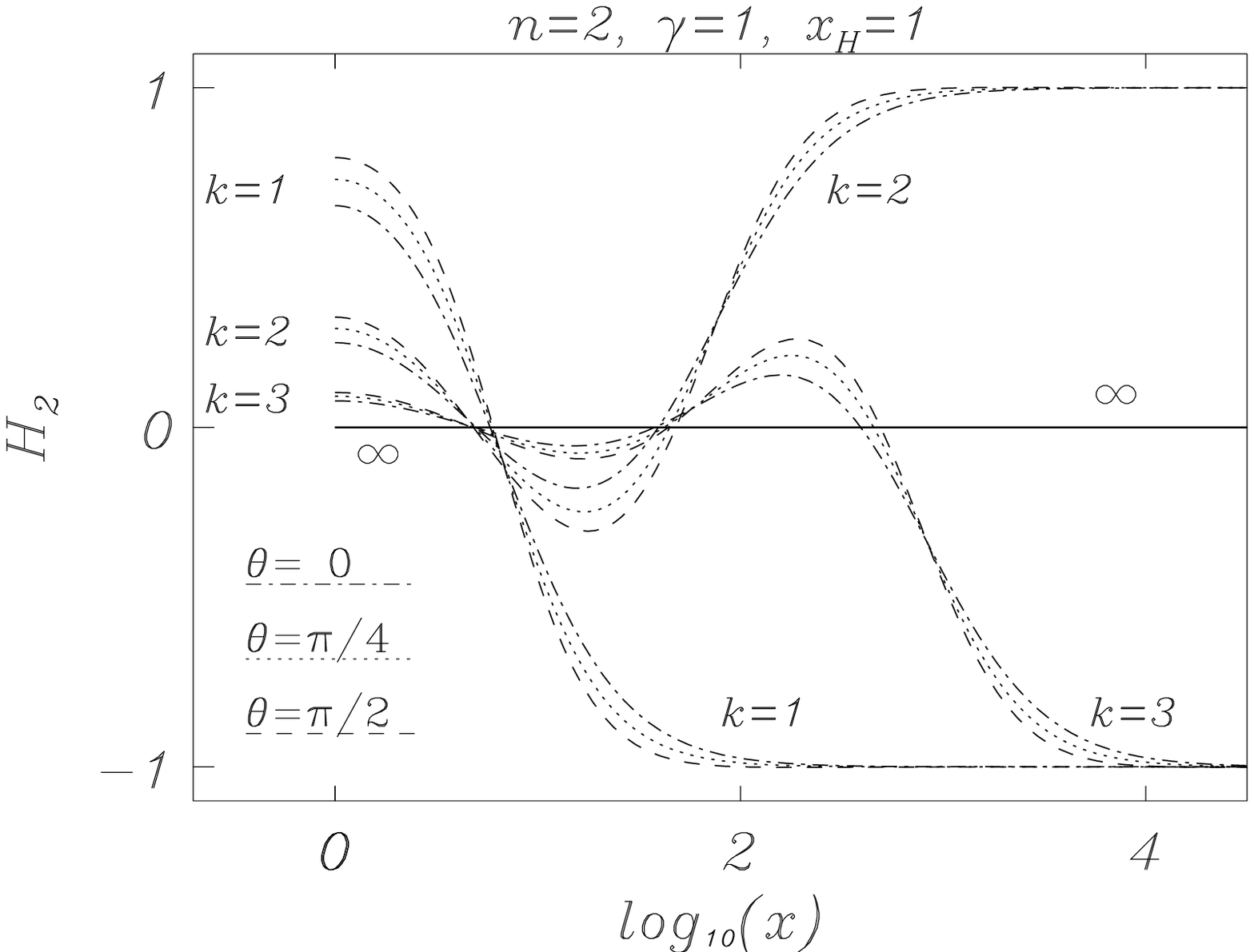}}\\
\end{figure}
\noindent Fig.~22b
The gauge field function $H_2$ is shown as a function
of the dimensionless coordinate $x$ for the angles 
$\theta=0$, $\theta=\pi/4$ and $\theta=\pi/2$ 
for the EYMD black hole solutions with $\gamma=1$,
horizon radius $x_{\rm H}=1$, winding number $n=2$, 
and node numbers $k=1-3$.
Also shown is the gauge field function of the limiting EMD solution.
\vspace{.5cm}\\
\begin{figure}
\centering
\vspace{-1cm}
\mbox{  \epsfysize=6cm \epsffile{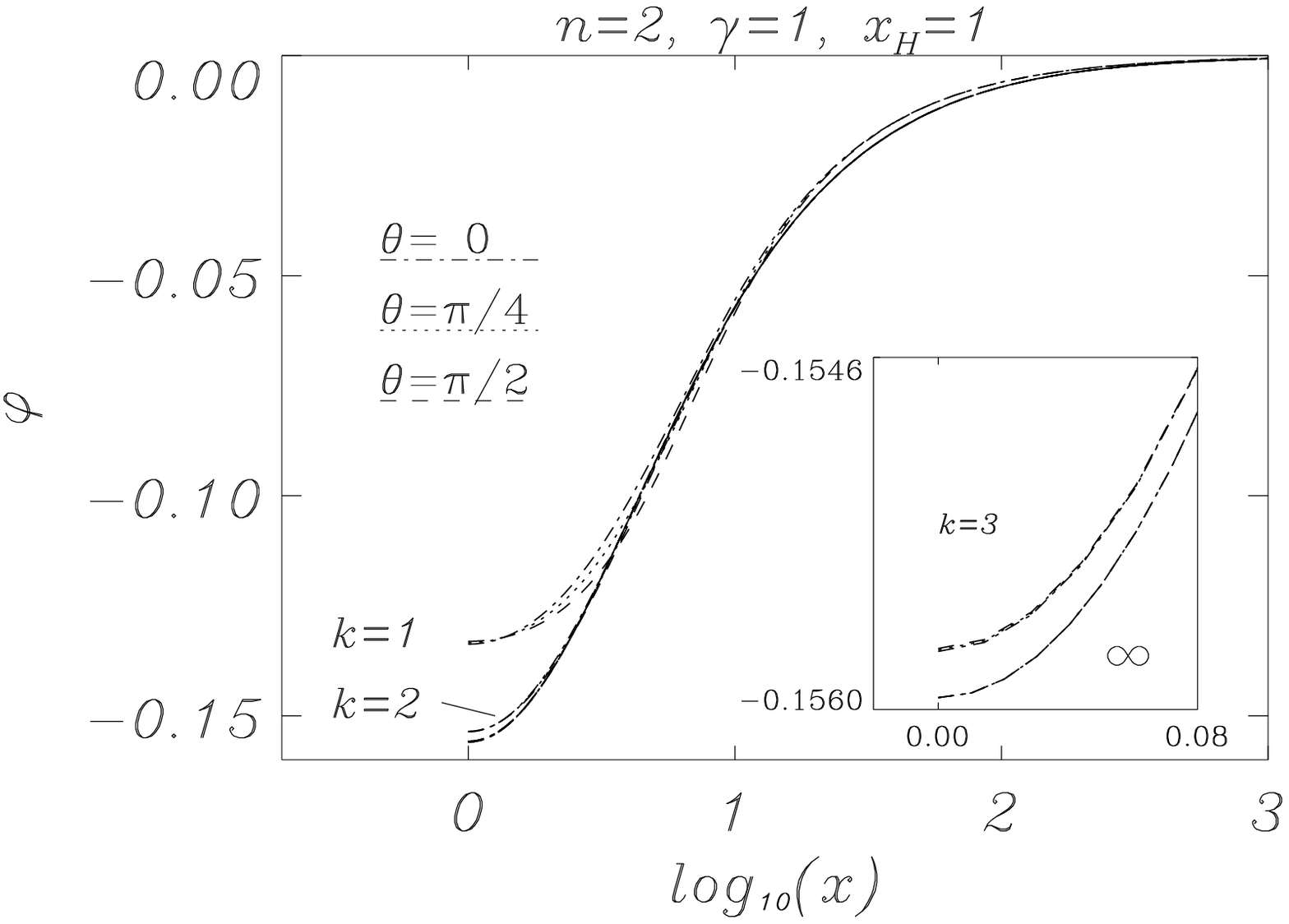}}\\
\end{figure}
\noindent Fig.~22c
Same as Fig.~22b for the dilaton function $\varphi$.


\begin{thebibliography}{000}
\bibitem{bek}
 see e.g.~A.~E. Mayo and J.~D. Bekenstein,
 No hair for spherical black holes: charged and nonminimally coupled
 scalar field with self--interaction,
 Phys. Rev. D54 (1996) 5059.
\bibitem{hair-em}
 W. Israel,
 Event horizons in static electrovac space-times,
 Commun. Math. Phys. 8 (1968) 245;
 D.~C. Robinson,
 Uniqueness of the Kerr black hole,
 Phys. Rev. Lett. 34 (1975) 905;
 P. Mazur,
 Proof of uniqueness of the Kerr-Newman black hole solution,
 J. Phys. A 15 (1982) 3173.
\bibitem{review}
 N. Straumann,
 Black holes with hair,
 Class. Quantum Grav. 16 (1993) S155;\\
 P. Bizon,
 Gravitating solitons and hairy black holes,
 Acta Phys. Polon. B25 (1994) 877.
\bibitem{kk3}
 B. Kleihaus and J. Kunz,
 Static black hole solutions with axial symmetry,
 Phys. Rev. Lett. 79 (1997) 1595.
\bibitem{ewein}
 S.~A. Ridgway and E.~J. Weinberg,
 Static black hole solutions without rotational symmetry,
 Phys.Rev. D52 (1995) 3440.
\bibitem{brod}
 O. Brodbeck, M. Heusler, N. Straumann and M.~S. Volkov,
 Rotating solitons and non-rotating, non-static black holes,
 gr-qc/9707057.
\bibitem{slomo}
 M.~S. Volkov and N. Straumann, 
 Slowly rotating nonabelian black holes,
 Phys. Rev. Lett. 79 (1997) 1428;\\
 O. Brodbeck and M. Heusler,
 Stationary perturbations and infinitesimal rotations
 of static Einstein-Yang-Mills configurations
 with bosonic matter,
 Phys. Rev. D56 (1997) 6278.
\bibitem{kk2}
 B. Kleihaus and J. Kunz,
 Static axially symmetric solutions of Einstein-Yang-Mills-dilaton
 theory, 
 Phys. Rev. Lett. 78 (1997) 2527.
\bibitem{kk4}
 B. Kleihaus and J. Kunz,
 Static axially symmetric Einstein-Yang-Mills-dilaton solutions:
 I.Regular solutions, Phys. Rev. D57, in press,
 gr-qc/9707045.
\bibitem{bm}
 R. Bartnik and J. McKinnon,
 Particlelike solutions of the Einstein-Yang-Mills equations,
 Phys. Rev. Lett. 61 (1988) 141.
\bibitem{su2}
 M.~S. Volkov and D.~V. Galt'sov,
 Black holes in Einstein-Yang-Mills theory,
 Sov. J. Nucl. Phys. 51 (1990) 747;\\
 P. Bizon,
 Colored black holes,
 Phys. Rev. Lett. 64 (1990) 2844;\\
 H.~P. K\"unzle and A.~K.~M. Masoud-ul-Alam,
 Spherically symmetric static SU(2) Einstein-Yang-Mills fields,
 J. Math. Phys. 31 (1990) 928.
\bibitem{eymd}
 E.~E. Donets and D.~V. Gal'tsov,
 Stringy sphalerons and non-abelian black holes,
 Phys. Lett. B302 (1993) 411;\\
 G. Lavrelashvili and D. Maison,
 Regular and black hole solutions of Einstein-Yang-Mills
 dilaton theory,
 Nucl. Phys. B410 (1993) 407.
\bibitem{emd}
 G.~W. Gibbons and K. Maeda,
 Black holes and membranes in higher-dimensional
 theories with dilaton fields,
 Nucl. Phys. B298 (1988) 741;\\
 D. Garfinkle, G.~T. Horowitz and A. Strominger,
 Charged black holes in string theory,
 Phys. Rev. D43 (1991) 3140.
\bibitem{book}
 see e.~g.~D. Kramer, H. Stephani, E. Herlt, and M. MacCallum,
 Exact Solutions of Einstein's Field Equations, Ch.~17
 (Cambridge University Press, Cambridge, 1980)
\bibitem{rr} 
 C. Rebbi and P. Rossi, 
 Multimonopole solutions in the Prasad-Sommerfield limit,
 Phys. Rev. D22 (1980) 2010.
\bibitem{kk}
 B. Kleihaus and J. Kunz, 
 Multisphalerons in the weak interactions,
 Phys. Lett. B329 (1994) 61;\\
 B. Kleihaus and J. Kunz, 
 Multisphalerons in the Weinberg-Salam theory,
 Phys. Rev. D50 (1994) 5343.
\bibitem{kk1}
 B. Kleihaus and J. Kunz,
 Axially symmetric multisphalerons in Yang-Mills-dilaton theory,
 Phys. Lett. B392 (1997) 135.
\bibitem{kkb}
 B. Kleihaus, J. Kunz and Y. Brihaye,
 The electroweak sphaleron at physical mixing angle,
 Phys. Lett. 273B (1991) 100;\\
 J. Kunz, B. Kleihaus, and Y. Brihaye,
 Sphalerons at finite mixing angle,
 Phys. Rev. D46 (1992) 3587.
\bibitem{bk}
 Y. Brihaye and J. Kunz,
 Axially symmetric solutions in the electroweak theory,
 Phys. Rev. D50 (1994) 4175.
\bibitem{rh2}
 The horizon of the axially symmetric Kerr solution 
 in Boyer-Lindquist coordinates
 also resides at a surface of constant $r$.
\bibitem{wein}
 S. Weinberg,
 Gravitation and Cosmology
 (Wiley, New York, 1972)
\bibitem{wald}
 R.~M. Wald, 
 General Relativity 
 (University of Chicago Press, Chicago, 1984)
\bibitem{kks3}
 B. Kleihaus, J. Kunz and A. Sood,
 SU(3) Einstein-Yang-Mills-dilaton sphalerons and black holes,
 Phys. Lett. B374 (1996) 289;
 B. Kleihaus, J. Kunz and A. Sood,
 Sequences of Einstein-Yang-Mills-dilaton black holes,
 Phys. Rev. D54 (1996) 5070.
\bibitem{lim}
 J.~A. Smoller and A.~G. Wasserman,
 An investigation of the limiting behavior of particle-like
 solutions to the Einstein-Yang/Mills equations
 and a new black hole solution,
 Commun. Math. Phys. 161 (1994) 365;\\
 P. Breitenlohner, P. Forgacs and D. Maison,
 Static spherically symmetric solutions of the
 Einstein-Yang-Mills equations,
 Commun. Math. Phys. 163 (1994) 141;\\
 P. Breitenlohner and D. Maison,
 On the limiting solution of the Bartnik-McKinnon family,
 Commun. Math. Phys. 171 (1995) 685.
\bibitem{fn1}
 The mass of the limiting solutions in Schwarzschild-like
 coordinates is constant for $0 \le \tilde x_{\rm H} \le 1$,
 $\mu = 1$.
\bibitem{schoen}
 W. Sch\"onauer and R. Wei\ss , 
 J. Comput. Appl. Math. 27, 279 (1989) 279;
 M. Schauder, R. Wei\ss\ and W. Sch\"onauer, 
 The CADSOL Program Package,
 Universit\"at Karlsruhe, Interner Bericht Nr. 46/92 (1992).
\bibitem{strau}
 N. Straumann and Z.~H. Zhou,
 Instability of the Bartnik-McKinnon solutions
 of the Einstein-Yang-Mills equations,
 Phys. Lett. B237 (1990) 353;\\
 N. Straumann and Z.~H. Zhou,
 Instability of colored black hole solutions,
 Phys. Lett. B243 (1990) 33.
\bibitem{mon}
 N.~J. Hitchin, N.~S. Manton and M.~K. Murray, 
 Symmetric monopoles,
 Nonlinearity 8 (1995) 661;\\
 C.~J. Houghton and P.~M. Sutcliffe,
 Tetrahedral and cubic monopoles,
 Commun. Math.Phys. 180 (1996)343;\\
 C.~J. Houghton and P.~M. Sutcliffe,
 Octahedral and dodecahedral monopoles, 
 Nonlinearity 9 (1996) 385.
\bibitem{bc}
 E. Braaten, S. Townsend and L. Carson,
 Novel structure of static multisoliton solutions
 in the Skyrme model,
 Phys. Lett. B235 (1990) 147;\\
 R.~A. Battye and P.~M. Sutcliffe,
 Symmetric Skyrmions,
 Phys. Rev. Lett. 79 (1997) 363;\\
 C.~J. Houghton, N.~S. Manton, P.~M. Sutcliffe,
 Rational Maps, Monopoles and Skyrmions,
 Nucl. Phys. B, in press, hep-th/9705151.
 
\end{thebibliography}
\end{document}